\documentclass[useAMS,usenatbib]{mn2e}
\usepackage{times}

\usepackage{amssymb}
\usepackage{./aas_macros}
\usepackage{graphicx}

\begin{document}

\title{Magneto-elastic oscillations of neutron stars with dipolar magnetic
fields}
\author[Michael Gabler, Pablo Cerd\'a-Dur\'an, Nikolaos Stergioulas, Jos\'e
A.~Font and Ewald M\"uller]
{Michael Gabler$^{1,2,3}$, 
Pablo Cerd\'a-Dur\'an$^1$, 
Nikolaos Stergioulas$^2$, 
Jos\'e A.~Font$^3$,
\and and Ewald M\"uller$^1$ 
\\
  $^1$Max-Planck-Institut f\"ur Astrophysik,
  Karl-Schwarzschild-Str.~1, 85741 Garching, Germany \\
  $^2$Department of Physics, Aristotle University of Thessaloniki,
  Thessaloniki 54124, Greece\\
$^3$Departamento de Astronom\'{\i}a y Astrof\'{\i}sica,
  Universitat de Val\`encia, 46100 Burjassot (Valencia), Spain 
}
\date{\today}
\maketitle
\begin{abstract}
By means of two dimensional, general-relativistic, magneto-hydrodynamical 
simulations we investigate the oscillations of magnetized neutron star models
(magnetars) for one particular dipolar magnetic field configuration
including the description of an extended solid crust. The aim of
this study is to understand the origin of the quasi-periodic oscillations
(QPOs) observed in the giant flares of soft gamma-ray repeaters (SGRs). We
confirm our previous findings which showed the existence of three different
regimes in the evolution depending on the magnetic field strength:
(a) a weak
magnetic field regime $B<5\times10^{13}\,$G, where crustal shear modes dominate
the evolution; (b) a regime of intermediate magnetic fields
$5\times10^{13}\,$G$\,<B<10^{15}\,$G, where Alfv\'en QPOs are mainly
 confined to the core of the neutron star and the crustal shear modes are damped
very efficiently; and (c) a strong field regime  $B>10^{15}\,$G, where
magneto-elastic oscillations reach the surface and approach the behavior of
purely Alfv\'en QPOs. When the Alfv\'en QPOs are confined to the core of the
neutron star, we find qualitatively similar QPOs as in the absence of a crust.
The lower QPOs associated with the closed field lines of the magnetic
field configuration are reproduced as in our previous simulations without
crust,
while the upper QPOs connected to the open field lines are displaced from the
polar axis. The position of these upper QPOs strongly depends on the magnetic
field strength. Additionally, we observe a family of edge QPOs and one new upper
QPO, which was not previously found in the absence of a crust. We extend our
semi-analytic model to obtain estimates for the continuum of the Alfv\'en
oscillations. Our results do not leave much room for a crustal-mode
interpretation of observed QPOs 
in SGR giant flares, but can accommodate an interpretation of these observations
as originating from Alfv\'en-like, global, turning-point QPOs 
(which can reach the surface of the star) in models with mean
surface magnetic field strengths in the narrow range of $3.8\times
10^{15}\,$G$\,\lesssim B\lesssim1.1\times{10^{16}}\,$G (for a sample of two
stiff EoS and various 
masses). This range is somewhat larger than estimates for magnetic
field strengths in known magnetars. The discrepancy may be resolved in
models including a more complicated magnetic field structure or
with models taking superfluidity of the neutrons and superconductivity of
the protons in the core into account.
\end{abstract}
%
\section{Introduction}
The detection of quasi-periodic oscillations (QPOs) in the giant flares of Soft
Gamma-ray Repeaters (SGRs) has raised hopes of drawing conclusions about their
interior structure, and thus to increase the understanding of the equation of
state (EoS) of matter at
supranuclear densities. SGRs are a class of magnetars, i.e.~very compact objects
with very strong magnetic fields~\citep{Duncan1992}. They show repeated
activity in the soft gamma-ray / hard X-ray spectrum, and as of today giant
flares with energies between $10^{44}-10^{46}\,$ erg/s have been detected in
three of these
objects. In the decaying X-ray tail of two of these events a number of
QPOs have been observed \citep[see][for recent reviews]{Israel2005, Watts2007}.
The frequencies of the QPOs are roughly $18$, $26$, $30$, $92$, $150$,
$625$, and $1840\,$Hz for the outburst of SGR 1806-20 and $28$, $53$, $84$, and
$155\,$Hz for SGR 1900+14. Recently, \cite{El-Mezeini2010} re-analyzed the
Rossi X-ray Timing Explorer (RXTE) observations and claimed that they discovered
$84$, $103$, and $648\,$Hz QPOs in the normal bursts of SGR 1806-20.
With a different method \cite{Hambaryan2011} have also found new QPOs in the
data of the SGR 1806-20 giant burst at frequencies $16.9$, $21.4$, $36.4$,
$59.0$, $116.3\,$Hz. If confirmed, these are truly fascinating discoveries
as the methods employed will allow one to increase the number of extracted QPO
frequencies.

The interpretation of the observed QPOs in terms of oscillations of
the magnetar itself seems very promising. It may allow for insight into the
properties of these objects and constrain the EoS
above nuclear matter density. The first step towards such magnetar
asteroseismology 
would be the identification of the modes of the star which have
frequencies in the right range and which could be observable during an outburst.
 For some time after the discovery of the QPOs in SGR 1900-16 in 1998 
the main focus was on the
torsional shear oscillations of the solid crust
because their frequencies in the range between $10$s of
Hz for nodeless modes and kHz for $n=1$ modes match the observed
frequencies cited above \citep[see][and references therein]{Duncan1998,
Strohmayer2005, Piro2005, Sotani2007, Samuelsson2007, Steiner2009}. From
energetic considerations it is very likely that
these oscillations are excited during a SGR outburst
\citep{Duncan1998,Levin2011}.
Moreover, torsional shear oscillations couple preferably to the exterior
magnetosphere \citep{Blaes1989},
where the emission in form of a trapped fireball is supposed to occur
\citep{Thompson2001}. Therefore, there is a natural channel of how these
oscillations may influence the X-ray signal emitted during the flare.
However, despite some recent improvements of the models \citep{Steiner2009},
the order of the frequencies of successive shear modes does not allow for a
complete interpretation of all observed QPOs.

Since the publications by \cite{Levin2006} and \cite{Glampedakis2006}
another possibility has been investigated. These authors show that the
crust-core coupling due to the extremely strong
magnetic field present in magnetars may have significant impact on the shear
oscillations of the crust. It was shown in 
simplified toy models that the
shear modes can be damped very efficiently into a magneto-hydrodynamic
(MHD) continuum of Alfv\'en
oscillations existing in the core of the neutron star.  This idea stimulated
further interest in the direction of magneto-elastic oscillations by 
 more detailed toy models \citep{Levin2007,Lee2007,Lee2008}. In
particular, \cite{Levin2007}
showed that due to the coupling through the
crust long-lived QPOs can be produced at the turning points or edges of the
continuum. 

\begin{figure}
\begin{center}	
 \includegraphics[width=.4\textwidth]{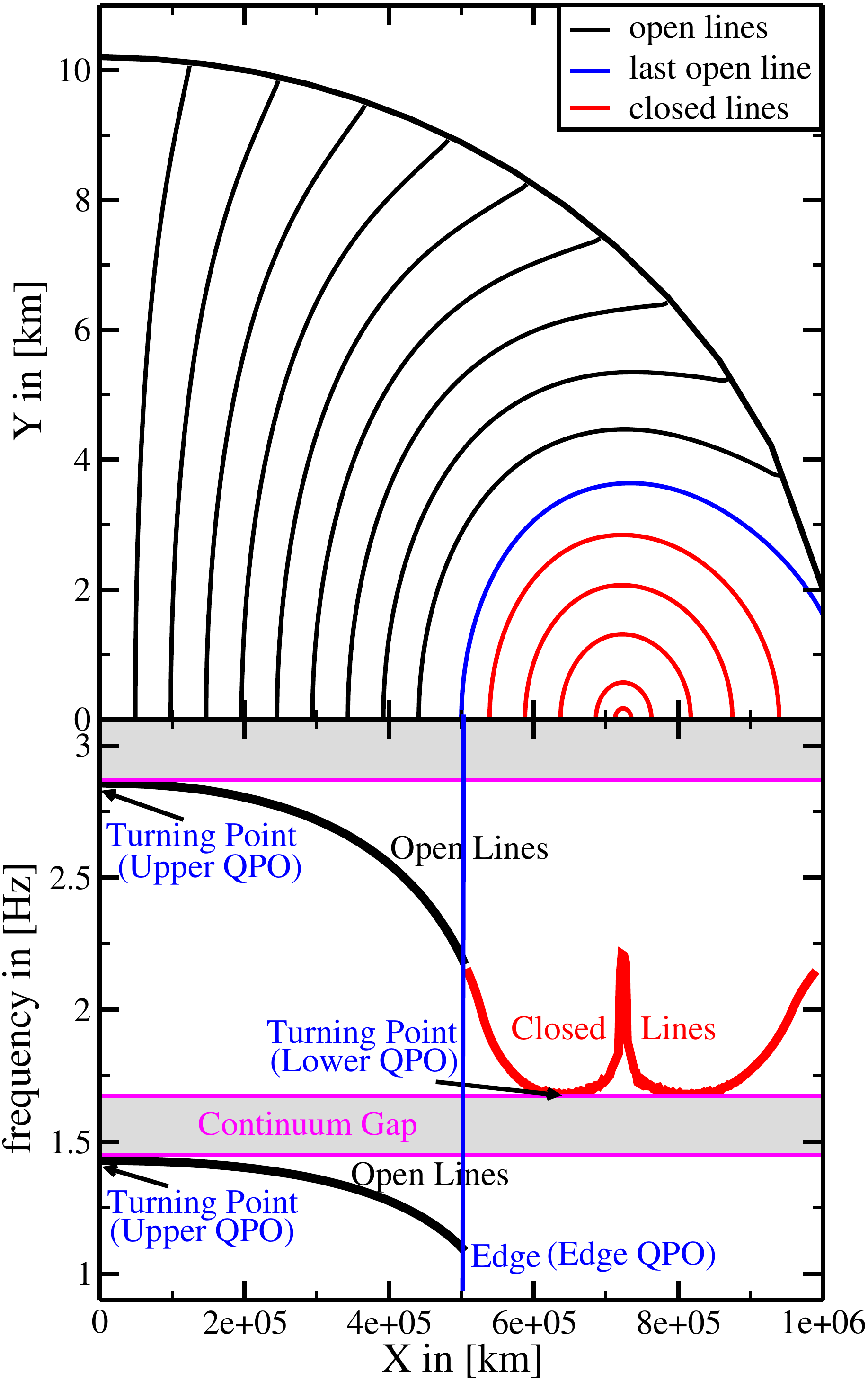}
\end{center}
\caption{{\it Upper panel:} Illustration of the dipolar magnetic field
configuration including field lines which close inside the neutron star (red
lines) and open lines which extend to the exterior (blue and black lines).
{\it Lower panel:} The frequencies of the corresponding field lines. The
ensemble of the frequencies of all field lines forms a continuum with edges and
turning points as indicated with the arrows. We call the QPOs related to the
turning point of the open field lines {\it upper} QPOs, the QPOs related to the
turning point of the closed field lines {\it lower} QPOs, and the QPOs related 
to the edge of the continuum at the last open field line edge QPOs.} 
\label{fig_intro}
\end{figure}

Before studying magneto-elastic oscillations in realistic scenarios it is
necessary to understand purely Alfv\'en oscillations of
neutron stars in General Relativity. Studies using general-relativistic
MHD models without taking an extended crust into account revealed two families
of long-lived QPOs in the continuum formed by Alfv\'en oscillations of
dipolar magnetic field configurations \citep{Sotani2008,Cerda2009,Colaiuda2009}.
\cite{Cerda2009} derived a semi-analytic model based on standing
waves in the short-wavelength approximation, which reproduces the MHD numerical 
results with very good agreement. The first family of QPOs, termed upper QPOs,
is
related to the turning point of the continuum at the open field lines close to
the polar axis (see Fig.\,\ref{fig_intro}).
The second family, the lower QPOs, can be found at the turning point in a region
of closed field lines near the equator. In this model the upper QPOs have their
maximum amplitudes at the surface of the star and are therefore candidates to
explain the observed QPOs. The different overtones have frequency ratios given
by integer numbers, making them very attractive to explain some of the observed
frequencies,  such as the $30$, $92$ and $150\,$Hz  QPOs
in SGR 1806-20.

Very recently a few groups have published results considering
magneto-elastic oscillations in magnetar models.
\cite{vanHoven2011} have presented a non-relativistic
model and shown that the oscillatory spectrum can
be influenced significantly by the physical properties of the model. For
example the presence of an entangled magnetic field allows for the existence
of discrete Alfv\'en modes, or the decoupling of neutrons in the core from
Alfv\'en waves can change the frequencies of the continuum.
In a different approach, based on the  coupling between a number of linear
one dimensional wave equations in the core and a two dimensional wave equation
in the crust, \cite{Colaiuda2011} 
suggested the existence of global, discrete Alfv\'en
modes in the gaps between continuous parts of the spectrum, as was
also suggested in \cite{vanHoven2011}.
In that case, the shear modes 
would not be damped resonantly and they could be observed as QPOs.
These oscillations are called gap modes.

In a previous paper \citep{Gabler2011letter}, 
we presented a realistic model of magneto-elastic oscillations in 
magnetars, finding that crustal shear oscillations, often invoked as 
an explanation of the QPOs seen after giant flares in SGRs, 
are damped by resonant absorption on timescales of at 
most 0.2s, for a lower limit on the dipole magnetic field strength of 
$5\times 10^{13}$G. At higher magnetic field strengths (typical in magnetars)
the damping timescale is even shorter. Our findings thus exclude torsional 
shear oscillations of the crust from explaining the observed low-frequency QPOs. 
In addition, we found that the Alfv\'en QPO model is a viable explanation of 
observed QPOs, if the dipole magnetic field strength exceeds a minimum 
strength  of about $10^{15}$G.  Then, predominantly Alfv\'en QPOs 
are no longer confined to the fluid core, but completely dominate over
shear oscillations in the crust region and have a maximum amplitude at the
surface of the star.

In the following, we present the details of the
extension of a previous model by \cite{Cerda2009}, where we performed two
dimensional MHD simulations of general-relativistic magnetar models with dipolar
magnetic fields. As sketched in \cite{Gabler2011letter} we extend this model by
including an extended solid crust, which is able to support shear stresses.
Here we complete the corresponding discussion with a larger set of numerical
simulations leading to quantitative estimates for the magnetic field strength.

Our results reveal 
the existence of different regimes depending on the magnetic field strength.
For weak magnetic fields $B<5\times10^{13}\,$G we are able to recover purely
crustal shear oscillations. At intermediate fields
$5\times10^{13}\,$G$<B<10^{15}\,$G
crustal modes are damped efficiently and the predominantly Alfv\'en QPOs are
limited to the fluid core of the neutron star. Finally, for strong fields
$B>10^{15}\,$G, the magneto-elastic oscillations are no longer confined to the
core by the interaction with the solid crust. They reach the surface with
significant amplitudes and may be responsible for the observed QPOs.
In the analysis presented below we will make substantial use of an 
extension of the semi-analytic
model presented in \cite{Cerda2009}, which is based on a short wave-length
approximation. We will further study the influence of the EoS on the properties
of
the equilibrium models, such as the change in frequency of the Alfv\'en
continuum or the dependence of the transition between the different regimes.

The paper is organized as follows: in Section \ref{sec_theory} we derive the
equations for general-relativistic, magneto-hydrodynamics of elastic objects in
the 3+1 split of general relativity. Those equations are then simplified to the
case of torsional oscillations in the small-amplitude limit. Furthermore we
discuss the boundary conditions we use, and the extension of the semi-analytic
model of \cite{Cerda2009}. Section\,\ref{sec_models} is concerned with the
equilibrium models we use as initial data for our simulations, and in
Section\,\ref{sec_numerics} we present the numerical methods employed.
Section\,\ref{sec_results} contains the results of our simulation and 
we discuss our findings in Section\,\ref{sec_discussion}.
Finally, the appendices are devoted to derive an alternative method to
compare our results with and to derive the eigenvalue problem of the shear
modes in our model.

In equations, we set $c=G=1$, with $c$ and $G$ being the speed of light
and Newton's gravitational constant, respectively. Latin indices
run over ($r, \theta, \varphi$) and Greek indices over ($t, r, \theta,
\varphi$). 
Partial and covariant derivatives are abbreviated by
comma and semicolon, respectively.
%
\section{Theoretical framework}\label{sec_theory}
In this section we present the equations governing the evolution of torsional
shear oscillations of magnetized neutron stars. First the equations
of general-relativistic, magneto-hydrodynamics (GRMHD) are revised and 
the effects of an elastic crust up to linear order in the perturbations are
included. Next, we apply a number of simplifications for 
 small-amplitude, axisymmetric torsional oscillations. It turns out that
these
simplifications are equivalent to the so-called anelastic approximation, see
\citet{Cerda2009}. This provides a natural framework that can be used to
 generalize the formalism in use. The next subsection is concerned with the
presentation of the boundary conditions applied in this work, and finally, we
sketch the derivation of a semi-analytic model to calculate the Alfv\'en
continuum.

Our numerical approach to study magneto-elastic oscillations of neutron stars
is based on an approximate Riemann solver. Additionally, we have obtained a
second method for the evolution of the crust coupled to the Riemann solver in
the core. In this alternative method we linearize the governing equations
and evolve the resulting wave equation numerically. The results are used as a
test to which we can compare our evolution with the Riemann solver method. We
give the technical details of this approach in Appendix \ref{appendix_linear}.
\subsection{General relativistic magneto-hydrodynamics
and elasticity}

We adopt the framework of 3+1 split of general
relativity. The general metric can thus be described by the following line
element
\begin{equation}
 ds^2 = - \alpha^2 dt^2 + \gamma_{ij} (dx^i + \beta^i dt)(dx^j + \beta^j dt)\,,
\end{equation}
where $\alpha$ and $\beta^i$ are called lapse function and shift vector,
respectively. For spherically symmetric space-times in isotropic coordinates
the 3-metric $\gamma_{ij}$ is conformally flat and 
 can be related to the spatial flat 3-metric $\hat\gamma_{ij}$ by the
conformal factor $\Phi$: 
\begin{equation}
 \gamma_{ij}=\Phi^4 \hat\gamma_{ij}\,.
\end{equation}
The description of matter in general relativity is based on the stress-energy
tensor $T^{\mu\nu}$. For a magnetized perfect fluid with elastic
properties, $T^{\mu\nu}$ can be written as a sum of different contributions:
\begin{equation}
 T^{\mu\nu} = T^{\mu\nu}_{\,\,\mathrm{fluid}} + T^{\mu\nu}_{\,\,\mathrm{mag}} +
T^{\mu\nu}_{\,\,\mathrm{elas}}\,.
\end{equation}
The different terms can be expressed as follows
\begin{equation}
 T^{\mu\nu}_{\,\,\mathrm{fluid}} = \rho h u^\mu u^\nu + P g^{\mu\nu}\,,
\end{equation}
where $\rho$ is the rest-mass density, $h=1+\epsilon+P/\rho$ the
specific enthalpy, $\epsilon$ the specific internal energy, $P$ the isotropic
fluid pressure, $u^\mu$ the 4-velocity of the fluid, and $g^{\mu\nu}$ the metric
tensor. Moreover, in the ideal MHD approximation, which we adopt here, 
the magnetic part is given by
\begin{equation}
 T^{\mu\nu}_{\,\,\mathrm{mag}} = b^2 u^\mu u^\nu + \frac{1}{2} b^2 g^{\mu\nu}-
b^\mu b^\nu\,,
\end{equation}
with $b^\mu$ being the magnetic 4-vector and $b^2 = b_\mu b^\mu$.

The theory of elasticity in general relativity is based on the fundamental work
of \citet{Carter1972} and was recently extended in a series of papers
by \citet{Karlovini2003,Karlovini2004b,Karlovini2007} and~\cite{Karlovini2004}. 
Our approach follows~\citet{Carter2006}. At the linear level in the
perturbations the
elastic effects of a crust enter in the form of the following contribution to
the
stress-energy tensor
\begin{equation}
  T^{\mu\nu}_{\,\,\mathrm{elas}} = - 2 \mu_\mathrm{S} \Sigma^{\mu\nu}\,.
\end{equation}
Here we have introduced the shear modulus
$\mu_\mathrm{S}$ and the shear tensor $\Sigma^{\mu\nu}$. The Lie derivative of
the latter
along the 4-velocity $\mathcal{L}_u \Sigma^{\mu\nu}$
is related to the strain tensor
$\sigma^{\mu\nu}$
\begin{equation}
\mathcal{L}_u \Sigma^{\mu\nu} =  \sigma^{\mu\nu}\,, 
\end{equation}
which is defined as
\begin{equation}
 \sigma^{\mu\nu} := \frac{1}{2} \left( u^\mu_{\,;\alpha} h^{\alpha\nu} +
u^\nu_{\,;\alpha} h^{\alpha\mu}\right) - \frac{1}{3} h^{\mu\nu}
u^\alpha_{\,;\alpha}\,,
\end{equation}
where 
\begin{equation}
h^{\mu\nu} := g^{\mu\nu} + u^\mu u^\nu 
\end{equation}
is a projection perpendicular to the 4-velocity.

The conservation of energy and momentum $\nabla_\nu T^{\mu\nu}=0$ and the
baryon
number conservation $\nabla_\nu J^\nu=0$, together with Maxwell's equations
$\nabla_\nu \hspace{.5mm} {}^{\ast} \hspace{-1mm} F^{\mu\nu}=0$ and
$\nabla_\nu
F^{\mu\nu}=4\pi \mathcal{J}^\mu$  (when expressed in the ideal MHD 
approximation) lead to the following form of a
flux-conservative hyperbolic system of equations
\begin{equation}
 \frac{1}{\sqrt{-g}} \left( \frac{\partial\sqrt{\gamma} U }{\partial t} +
\frac{\partial \sqrt{-g} F^i}{\partial x^i} \right) = S\,. 
\label{conservationlaw}
\end{equation}
The quantities just introduced are the Faraday electromagnetic
tensor field $F^{\mu\nu}$, and its dual ${}^{\ast} \hspace{-1mm} F^{\mu\nu}$,
the rest-mass current $J^{\mu} := \rho u^{\mu}$, and
the electric 4-current $\mathcal{J}^{\mu}$, while
$\sqrt{-g}=\alpha\sqrt{\gamma}$ and $\gamma = \det(\gamma_{ij})$. The state
vector $U$, the flux vector $F^i$, and the sources $S$ are given by
\begin{eqnarray}
 U &=& [D, S_j,\tau, B^k]\,,\label{eq_conserved}\\
 F^i &=& \left[ {\begin{array}{c}
D \hat v^i \\
S_j \hat v^i + \delta^i_j \left(P + \frac{1}{2}
b^2 \right) - \frac{b_j B^i}{W} - 2 \mu_\mathrm{S} \Sigma^i_{~j}\\
\tau \hat v^i +  \hat v^i \left(P + \frac{1}{2}
b^2\right) - \frac{\alpha b^t B^i}{W}\\
\hat v^i B^k - \hat v^k B^i
\end{array} } \right]\,, \label{flux_general}
\end{eqnarray}
\begin{equation}
S= \left[0,\frac{1}{2} T^{\mu\nu} \frac{\partial g_{\mu \nu}}{\partial
x^j},
\alpha T^{\mu\nu}\left( \delta^{t}_\nu \frac{\partial \ln\alpha}{\partial x^\mu}
-
\Gamma^t_{\nu\mu} \right),
0,0,0 \right]\,,
\end{equation}
where $\hat v^i \equiv v^i - \beta^i / \alpha$ and $W=\alpha u^t$ is the Lorentz
factor. Additionally, we introduce the
3-velocity of the fluid $v^i$, the generalized rest mass density $D$, the
momentum density $S_i$, and the energy density $\tau$
\begin{eqnarray}
 D &=& \rho W\,,\\
 S_i &=& (\rho h + b^2) W^2 v_i - \alpha b_i b^t\,,\\
 \tau&=& (\rho h + b^2) W^2 - \left(P + \frac{1}{2} b^2\right) - \alpha^2
(b^t)^2 - D\,.
\end{eqnarray}
The relation between $B^i$, the magnetic field measured by an Eulerian
observer, and the magnetic 4-vector $b^\mu$ 
is given by
\begin{equation}
 b^\mu = \left[ \frac{W B^i v_i }{\alpha}, \frac{B^i+ W^2 v^j B_j
\hat v^i}{W}\right]\,.
\end{equation}

The fluxes defined in Eq.\,(\ref{flux_general}) depend on the shear tensor
$\Sigma^{\mu\nu}$ which itself is not a function of the conserved variables
$U$. To complete the system (\ref{conservationlaw}) it is thus
necessary to describe how the components of the shear tensor evolve with time.
For simplicity, we will postpone the discussion of these terms to the next
section, where we first apply some simplifications.
\subsection{Torsional oscillations in the small-amplitude
limit and the anelastic
approximation}\label{sec:approximation}
This work is concerned with torsional oscillations of neutron stars. Therefore
we are allowed to perform a number of simplifications to the
full set of ideal GRMHD equations presented above.

(i) The temperature of an old neutron star is well below the Fermi temperature,
i.e. a barotropic EoS describes the matter with sufficient accuracy.
Hence, the equation for the energy density $\tau$ in system
(\ref{conservationlaw}) becomes redundant.

(ii) For purely poloidal background fields and in axisymmetry, the torsional
oscillations decouple at the linear level from the polar ones. For 
small-amplitude oscillations it is thus justified to evolve $B^\varphi$
and
$S_\varphi$ separately, and to keep $B^r$, $B^\theta$, $S_r$, and $S_\theta$
constant in time.

(iii) When setting the initial velocities in the
$r$ and $\theta$ 
 directions, and
therefore the polar perturbations, to zero, they remain zero during the
evolution (see (ii)). The continuity equation reduces to
\begin{equation}
 \frac{\partial \sqrt{\gamma} D}{\partial t} + \frac{\partial D
\hat v^\varphi}{\partial x^\varphi} =  \frac{\partial \sqrt{\gamma} D}{\partial
t} = 0\,,
\end{equation}
because we assume axisymmetry, i.e. the density $D$ stays constant
throughout the evolution.

(iv) Because we are dealing with torsional oscillations, which do not involve
large density changes, see (iii), and which only couple very weakly to the
spacetime evolution, the Cowling approximation should hold,
i.e. the metric can be regarded as fixed. Furthermore, the matter distribution
and thus the spacetime can be regarded as spherically symmetric, since the 
deviations from sphericity due to the magnetic field are small for the magnetic 
fields considered here  \citep[see]{Bocquet1995}. In the case of
isotropic coordinates and for non-rotating stars this implies that
the shift vector vanishes, $\beta^i  = 0$, and
consequently $\hat v^i = v^i$. 

In general, the eigenvalues of the flux-vector Jacobians of the system
(\ref{conservationlaw})
depend on the metric tensor, the conserved
variables, and the speed of sound. The latter dependence strongly limits the
 numerical time step, slowing down any time-explicit numerical simulation.
To overcome this limitation in a more general context, \cite{Bonazzola2007} 
suggested to remove all pressure dependencies from the fluxes $F^i$, which is
known as the anelastic approximation. This changes the properties of the
eigenvalues of the system making them independent of the speed of sound. 
 In our case, as a consequence of (iii), the eigenvalues of the flux-vector
Jacobians of the system (\ref{conservationlaw}) do no longer depend on 
the speed of sound (see below in this section) and the numerical time-step
is thus set by the speed of Alfv\'en waves, only.

The anelastic approximation provides a natural generalization
of the method applied in the present work. It allows one to describe situations
also involving perturbations in density and pressure, and to
study more complicated magnetic field configurations including the coupling
between axial and polar oscillations, as a first step towards full
3-dimensional simulations. This is, however, beyond the scope of the present
work.

To summarize, we only evolve the equations for
$S_\varphi$ and $B^\varphi$ and hold
all other conserved variables of Eq.\,(\ref{eq_conserved}) constant. Together
with Eq.\,(\ref{conservationlaw}) the following definitions describe the
evolution of torsional oscillations of magnetized neutron stars including an
elastic crust:
\begin{eqnarray}
 U &=& [S_\varphi, B^\varphi]  \label{reduced_withcrust1}\,,\\
 F^r &=& \left[ -
\frac{b_\varphi B^r}{W} - 2 \mu_\mathrm{S}
\Sigma^r_{~\varphi}, - v^\varphi B^r
\right]\,,  \label{flux_r}\\
 F^\theta &=& \left[ - \frac{b_\varphi B^\theta}{W}- 2 \mu_\mathrm{S}
\Sigma^\theta_{~\varphi},
-v^\varphi B^\theta \right]\,,\label{flux_theta}\\
S &=& [0, 0]\,. \label{reduced_withcrust2}
\end{eqnarray}

To derive the shear tensor $\Sigma^{\mu\nu}$ for torsional oscillations in the
small-amplitude limit, we follow \citet{Schumaker1983}. Since we assume
the Cowling approximation to hold, a diagonal metric $g^{\mu\nu}
=\mathrm{diag}(g^{\mu\nu})$, and
restrict our considerations to linear order in the displacements, the components
of $\Sigma^{\mu\nu}$ can be written as
\begin{equation}
\Sigma^{ij} = \frac{1}{2\alpha} \left[ g^{ik} \left( \xi^j \alpha
\right)_{,k}
+ g^{jk} \left( \xi^i \alpha \right)_{,k}\right] -
\frac{g^{ij}}{3\alpha}  \left( \xi^k \alpha
\right)_{,k}\,,\label{shear_tensor_general}
\end{equation}
while the other components vanish, $\Sigma^{\mu t} = \Sigma^{t \nu} = 0$.
Here, the displacement $\xi^j$ has been introduced. The corresponding
time derivative is related to the 3-velocity of the fluid as follows
\begin{equation}
\xi^j_{\,,t} = \alpha v^j = \frac{u^j}{u^t} \label{def_xidot}\,.
\end{equation}

For purely torsional oscillations the more general expression in
Eq.\,(\ref{shear_tensor_general}) simplifies to
\begin{equation}
 \Sigma^{ij} = \frac{1}{2}\left[ {\begin{array}{c c c}
0&0& g^{rr} \xi^\varphi_{\,,r}\\
0&0& g^{\theta\theta} \xi^\varphi_{\,,\theta}\\
g^{rr} \xi^\varphi_{\,,r}& g^{\theta\theta} \xi^\varphi_{\,,\theta}   &0\\
\end{array} } \right]\,.
\end{equation}

In its current form the system of equations (\ref{conservationlaw}) and
(\ref{reduced_withcrust1}) - (\ref{reduced_withcrust2}) is incomplete,
because the displacement appears in the fluxes, Eq.\,(\ref{flux_r})
and (\ref{flux_theta}). Hence, it is necessary to add (at least) one equation
for the evolution of $\xi^\varphi$. During the calculations it turned out that
the direct integration of $\xi^\varphi$ via Eq.\,(\ref{def_xidot}) is
numerically disadvantageous because of several reasons:

(i) The use of $\xi^\varphi$ requires an evaluation of its
 spatial derivatives when calculating the fluxes in
(\ref{reduced_withcrust1}) - (\ref{reduced_withcrust2}). This leads 
to numerical inaccuracies in regions where the displacement changes
strongly. 

(ii) Alternatively one could move the additional terms including the
derivatives of $\xi^\varphi$ to the source terms of the system of equations.
However, it is numerically more desirable to have a system of
conservation laws without any source terms. Furthermore, the spatial derivatives
of the shear modulus $\mu_\mathrm{S}$ would appear in the source
terms, which might give rise to spurious oscillations, because
$\mu_\mathrm{S}$ is given in tabulated form.

(iii) The interface conditions at the crust-core interface are given in terms
of $\xi^\varphi$ and $\xi^\varphi_{~,r}$. It 
 turns out that it is
numerically difficult to impose them directly on $\xi^\varphi$ ensuring a
sufficiently accurate corresponding condition for its radial derivative (see end
of Section\,\ref{sec_numerics}).

Therefore, instead of evolving $\xi^\varphi$, the system of
Eqs.\,(\ref{conservationlaw}) with Eqs.\,
(\ref{reduced_withcrust1}) - (\ref{reduced_withcrust2}) is supplemented with
the following equations for the spatial derivatives of $\xi^\varphi$
\begin{eqnarray} 
(\xi^\varphi_{\,,r})_{,t} - (v^\varphi \alpha)_{,r} &=&0\,,\label{eq_xi_dr}\\
(\xi^\varphi_{\,,\theta})_{,t} - (v^\varphi \alpha)_{,\theta} &=&
0\label{eq_xi_dtheta}\,,
\end{eqnarray}
which are based on Eq.\,(\ref{def_xidot}).

To obtain the eigenvalues we write the complete system in the form of a
conservation law like Eq.\,(\ref{conservationlaw}). This introduces the new
conserved variables $U$, fluxes $F^k$, and sources $S$:
\begin{eqnarray}
 U &=& [S_\varphi\,, B^\varphi\,, (\alpha \xi^\varphi_{\,,r})\,,
(\alpha\xi^\varphi_{\,,\theta})]\,,\label{eq_notusedI}\\
 F^k&=& \left[ {\begin{array}{c c c}
 - \frac{b_\varphi B^k}{W} - 2 \mu_\mathrm{S}
\Sigma^k_{\,\varphi} \\
- v^\varphi B^k\\
- \alpha v^\varphi \delta^k_{r}\\
- \alpha v^\varphi \delta^k_{\theta}
\end{array} } \right]\,, \label{eq_notusedII}\\
S &=& \left[0,0,- \delta^k_{r} \frac{\alpha v^\varphi}{\sqrt{-g}}\frac{\partial
\sqrt{-g}}{\partial x^k},- 
\delta^k_{\theta}
\frac{\alpha v^\varphi}{\sqrt{-g}}\frac{\partial \sqrt{-g}}{\partial
x^k} \right]\,,\label{eq_notusedIII}
\end{eqnarray}
where $k = \{r, \theta\}$.
The solution of the eigenvalue problem of the associated flux-vector Jacobian
leads to the following non-zero eigenvalues
\begin{eqnarray}
\lambda_{1/2}^{k} &=& \pm \sqrt{\frac{ (B^k)^2 + \mu_\mathrm{S} /
g_{kk} }{A}}\,, \label{eq:eigenvalues} \\
A &=& \frac{\partial S_\varphi}{\partial v_\varphi} = \rho h W^4 (1+v_\varphi
v^\varphi) + B^r B_r + B^\theta B_\theta\,.
\end{eqnarray} 
The set of equations (\ref{conservationlaw}) with
(\ref{eq_notusedI}), (\ref{eq_notusedII}), and (\ref{eq_notusedIII}) 
 is useful for computing the eigenvalues, but contains
non-zero sources in the conservation law. 
For the numerical time-evolutions, we
thus implement the original set Eq.\,(\ref{conservationlaw}) with Eqs.\,
(\ref{reduced_withcrust1}) - (\ref{reduced_withcrust2}) and
Eqs.\,(\ref{eq_xi_dr}) and (\ref{eq_xi_dtheta}) (see
Section \ref{sec_numerics} below for details).
\subsection{Boundary conditions at the surface and treatment of the crust-core
interface}\label{sec_bc}

The system of equations for magneto-elastic torsional oscillations contains two
degrees of freedom related to the two non-vanishing eigenvalues
(\ref{eq:eigenvalues}).  Therefore, we have to impose boundary
conditions at the surface of the star that mimic the incoming waves from the
magnetosphere, which 
are not included in our simulations. 
In the core, where $\mu_{\rm S}=0$, there are still two degrees of freedom.
Since we are simulating both regions (crust and core) having the same number of
degrees of freedom there is no need for boundary conditions at the
crust-core interface (although there is a need for a special treatment  
for numerical reasons, see below). 

In addition, in the case of ideal MHD without charges, the electric field is
continuous everywhere, and hence the velocity $v^\varphi$, too. This
implies continuity of the displacement, $\xi^\varphi$, 
and of its time derivative, $\xi^\varphi_{\,,t}$. At the surface of the star
and at the crust-core interface
the tangential derivative, $\xi^\varphi_{\,,\theta}$, is continuous, while no
restrictions apply to the continuity of $\xi^\varphi_{\,,r}$. 

{\it Boundary conditions at the surface}:
We assume that there are no current sheets at the surface of the star, i.e.
the tangential magnetic field components have to be continuous
\begin{equation}
b^\varphi_\mathrm{crust}=b^\varphi_\mathrm{atmosphere}
\label{eq:bcout1}
\end{equation}
at the surface. 

The conservation of momentum gives the continuity of the traction $t^\varphi$

\begin{equation}
 t^\varphi=T (\tilde \mathbf{n}, \tilde \mathbf{\varphi}) = T (\tilde
\mathbf{r}, \tilde \mathbf{\varphi}) =
T^{r\varphi}\,,
\end{equation}
i.e. the tangential stresses inside and outside the star have to
balance each other. Here the tilde indicates normalized vectors, $\tilde
\mathbf{n}$ is the normal to the surface of the star and thus $\tilde \mathbf{n}
= \tilde \mathbf{r}$.  

The continuous traction condition can be simplified in the case of continuous 
$b^{\varphi}$ and leads to a condition for $\xi^\varphi_{~,r}$: 
\begin{eqnarray}
T^{r\varphi}_\mathrm{crust}&=&T^{r\varphi}_\mathrm{atmosphere}\\
 b^r b^\varphi_\mathrm{crust}+\frac{\mu_\mathrm{S}}{\Phi^4}
\xi^\varphi_{\,\mathrm{crust},r}&=& b^r
b^\varphi_\mathrm{atmosphere}\\
 \xi^\varphi_{\,\mathrm{crust},r}&=&0 \,.
\label{eq:bcout2}
\end{eqnarray}
Eqs.\,(\ref{eq:bcout1}) and (\ref{eq:bcout2}) are the set of boundary
conditions that we apply at the surface of the star. We need the 
additional condition (\ref{eq:bcout2}), because we are evolving more variables 
than the degrees of freedom of the system.

The assumptions made here are motivated by the picture that the magnetospheric
field close to the surface will move with its footpoints in the crust, i.e. the
exterior solution relaxes to a force-free field on a much shorter
time scale than the interior evolves. This implies that currents can be
maintained in the magnetosphere, which is necessary to support more general
equilibrium configurations than considered here and hence to create a
twisted magnetospheric field.
A more detailed discussion of the coupling to the magnetosphere would exceed
the scope of this paper and will be purpose of further investigations.

Our boundary conditions are similar to those used in
previous work without the presence of a crust \citep{Sotani2008, Cerda2009,
Colaiuda2009} and in simulations with a crust
\citep{Gabler2011letter,Colaiuda2011}.
However, our boundary conditions differ from those of \cite{Lander2010} and
\cite{Lander2011b} who choose a particular set of variables vanishing
at the surface of the neutron star, which defines the corresponding boundary
conditons. 
Other approaches \citep{Braithwaite2006b,Lasky2011,Ciolfi2011} involve the
evolution of some parts of the neutron star's atmosphere.
For purely toroidal oscillations it is possible, however, to impose
appropriate boundary conditions and to avoid that evolution.

The {\it treatment of the crust-core interface} requires particular attention.
At this interface no boundary conditions are required and by knowing the
variables at one instant of time on both sides (crust and core) one should be
able to evolve the system. However, the stability of the employed schemes turned
out to depend sensitively on the particular treatment of the reconstruction of
the variables, which are allowed to be discontinous or to have discontinous
spatial derivatives (see below). In the following we will describe how to
proceed to achieve stable evolutions.

As anywhere else the traction is supposed to be continous
\begin{eqnarray}
 T^{r\varphi}_\mathrm{core}&=&T^{r\varphi}_\mathrm{crust}\,,\\
- b^r b^\varphi_\mathrm{core} &=& -b^r
b^\varphi_\mathrm{crust}-\frac{\mu_\mathrm{S}}{\Phi^4}
\xi^\varphi_{\,,r}\,.
\end{eqnarray}
This can be transformed by virtue of the linearized induction equation
\begin{equation}\label{eq_lin_induction}
b^\varphi =
b^r \xi^\varphi_{\,,r}+ b^\theta  \xi^\varphi_{\,,\theta}\,,
\end{equation} the continuity
of the displacement, and thus
$\xi^\varphi_{\,\mathrm{core},\theta}=\xi^\varphi_{\,\mathrm{crust},\theta}$,
into
\begin{eqnarray}
&&b^r( b^r \xi^\varphi_{\,\mathrm{core},r} + b^\theta 
\xi^\varphi_{\,\mathrm{core},\theta}) =
\nonumber\\
&&\hspace{5mm} b^r
(b^r \xi^\varphi_{\,\mathrm{crust},r}+ b^\theta 
\xi^\varphi_{\,\mathrm{crust},\theta})+\frac{\mu_\mathrm{S}}{\Phi^4}
\xi^\varphi_{\,\mathrm{crust},r},\\
&&  \xi^\varphi_{\,\mathrm{core},r} = \left( 1+
\frac{\mu_\mathrm{S}}{\Phi^4 (b^r)^2}
\right)\xi^\varphi_{\,\mathrm{crust},r}\,. \label{eq_interface}
\end{eqnarray}
In Section \ref{sec_numerics} we will show how to ensure the continuity of the
traction and maintain the correct relation between the radial derivatives of
the displacement in the core and in the crust.

Obviously the discontinuous radial derivative allows $b^\varphi$ to be
discontinuous (Eq.\,\ref{eq_lin_induction}). Hence, in general, there are
current sheets present at the crust-core interface, which
are unavoidable, and a consequence of the coupled evolution and the
assumption of ideal MHD.

\subsection{Semi-analytic model}
A very useful tool to obtain the frequencies of purely magnetic oscillations
inside a neutron star was presented by \cite{Cerda2009}. These authors 
showed that in the linear regime and in the limit of short wavelengths it is
possible to calculate the Alfv\'en continuum with a semi-analytic model. Here, 
we will only sketch the method, and for more information we refer to
\cite{Cerda2009}. In the aforementioned limit an Alfv\'en wave
will travel along magnetic field lines corresponding to
\begin{equation}
 \frac{d\mathbf{x}}{dt}=\mathbf{v}_a (\mathbf{x})\,,
\end{equation}
where $\mathbf{v}_a$ is the Alfv\'en velocity. Any displacement $Y$ traveling
along the magnetic field lines can be expressed as a function of 
\emph{magnetic-field-line-adapted} coordinates $(\chi, \zeta)$ and time,
where $\chi$  labels the magnetic field line by the radius at which it crosses
the equatorial plane, and $\zeta=t(r,\theta;\chi) / t_\mathrm{tot}(\chi) - 1/2$
is a dimensionless parameter along each field line. Here $t_\mathrm{tot}(\chi)$
is twice the total travel time of an Alfv\'en wave traveling along a magnetic
field line starting from the equatorial plane and ending at the surface or at
another point in the equatorial plane. Note that $\zeta$ used in this work
corresponds to $\xi$ in \cite{Cerda2009}, because here $\xi$ denotes
the displacement related to the 4-velocity of the fluid. For a traveling
wave $Y$ satisfies trivially the wave equation
\begin{equation}
 \frac{\partial^2 Y(\chi,\zeta,t)}{\partial t^2} =
\frac{1}{t_\mathrm{tot}(\chi)^2}\frac{\partial^2 Y(\chi,\zeta,t)}{\partial
\zeta^2}\,. 
\end{equation}

Next we assume standing waves of the form
\begin{equation}
 Y(\chi,\zeta,t) = a(\chi) \sin(\kappa\zeta + \phi_\zeta) \cos(2\pi f t +
\phi_t)\,,
\end{equation}
where $\kappa$ is the wavenumber, $a(\chi)$ the amplitude, $\phi_t$ the
temporal phase, $\phi_\zeta$ a spatial phase, and $f$ is the oscillation
frequency. The dispersion relation then is simply given by
\begin{equation}
 f=\frac{\kappa}{2\pi t_\mathrm{tot}}\,.
\end{equation}
At this point the frequencies of the oscillations are completely determined by
the magnetic field topology and the boundary conditions.

\cite{Cerda2009} did not consider an extended crust, and hence the 
boundary condition was set at the surface and corresponded to the continuous 
traction condition. This resulted in a
vanishing radial derivative of the displacement and a maximum amplitude
of the perturbation at the surface.
However, when an extended crust is present there exist two different regimes.
For low magnetic field strengths ($B\lesssim10^{15}\,$G) the standing waves show
a node at the crust-core interface \citep{Gabler2011letter}, which may be
interpreted as a reflection of the
standing wave at the crust-core interface. As we show below, this change of the
boundary condition for the semi-analytic model is necessary to calculate
the correct frequencies and to find the symmetry of the numerically obtained
QPOs.

For stronger magnetic fields ($B > 10^{15}\,$G), the oscillations
reach the surface, and we can apply the boundary condition of \cite{Cerda2009}.
In order to
take the crust into account the velocity of the perturbation can be
approximated by the eigenvalues (\ref{eq:eigenvalues}), assuming that the
perturbation $Y$ is still traveling along the magnetic field lines.
In an intermediate regime at around $10^{15}\,$G we do not expect the
semi-analytic model to be valid, because in this case the shear and magnetic
contributions to the evolution in the crust are of similar order.

\section{Equilibrium Models}\label{sec_models}
The initial models are self-consistent general relativistic
equilibrium models of magnetized non-rotating neutron stars with a purely
poloidal magnetic field \citep{Bocquet1995}. We use the numerical code
``magstar''  of the LORENE 
library\footnote[1]{http://www.lorene.obspm.fr} to compute these models,
which include the effects of the magnetic field on the matter and the
spacetime. 

In our models the magnetic field is generated by a current of the form 
${\cal J}^{\varphi} = \rho h C$, where $C$ is a constant which determines the
strength of the magnetic field. We note that different choices of the current
can lead to different internal magnetic fields,
still resembling an exterior dipole field \citep[see e.g.][]{vanHoven2011}. 
Hereafter, we will label the different models by
the surface value of their magnetic field strength at the pole.

As we assume a spherically symmetric spacetime and matter background in our
simulations we angle-average the density of the background model to
obtain a spherically symmetric model from the LORENE data. In the most extreme
cases, i.e. for a very strong magnetic field this simplification changes the
structure of the neutron star by at most about one per cent. For example the
density at different angles but constant radius varies within less than one per
cent at $B=5\times 10^{15}\,$G. Therefore, the influence on the oscillations is
in general less than that of other approximations we are applying. 

For the EoS we can choose between different
realistic barotropic models including the description of a crust. We use
four combinations of two EoS in the core matched to two distinct EoS for the
crust. For the core we chose the APR EoS \citep{Akmal1998} and
the stiffer EoS L \citep{Pandharipande1975}, while for the low density region of
the crust we select EoS NV \citep{Negele1973} and EoS DH
\citep{Douchin2001}. The recent discovery of a
2 $M_\odot$ neutron star by
\cite{Demorest2010}, excludes EoS which cannot reproduce such large masses.
The properties of the equilibrium models used in this work are
summarized in Table\,\ref{tab_referencemodel}. These models are a subset of the 
models used in \cite{Sotani2007}. Here, $r_\mathrm{s}$ and $r_\mathrm{cc}$ are
the radii of the surface of the star and of the crust-core interface,
respectively, and $\Delta r_\mathrm{crust}=r_\mathrm{s}-r_\mathrm{cc}$ is
the size of the crust. We note that table 1 of \cite{Sotani2007} shows $\Delta
r_\mathrm{crust} / r_\mathrm{cc}$,  instead of $\Delta r_\mathrm{crust} /
r_\mathrm{s}$ and thus the percentage for the relative
size of the crust  is different in their case. However, we checked that
the value of $\Delta r_\mathrm{crust}$ is the same in both cases. The
frequencies of the crustal modes with $n>0$ depend sensitively on the size of
the crust  \citep{Samuelsson2007}. Therefore, a proper definition
of the size of the crust is important. 

\begin{table}
\begin{tabular}{c  c c c c c}
EoS & mass & circunferencial&inner radius &relative size \\
&&radius $r_\mathrm{s}$&of crust& of crust \\
&[$M_\odot$]&[km]&[km]&$\frac{\Delta r_\mathrm{crust}}{r_\mathrm{s}}$ [\%]\\
\hline
APR+DH&1.4&12.10&11.22&7.2\\
&1.6&12.07&11.38&5.7\\
&1.8&12.00&11.43&4.8\\
&2.0&11.90&11.44&3.9\\
&2.2&11.63&11.31&2.8\\
APR+NV&1.4&11.94&10.85&9.1\\
&1.6&11.93&11.07&7.2\\
&1.8&11.92&11.19&6.1\\
&2.0&11.81&11.23&4.9\\
&2.2&11.56&11.12&3.8\\
L+DH&1.4&14.74&13.33&9.6\\
&1.6&14.85&13.72&7.6\\
&1.8&14.93&13.94&6.6\\
&2.0&14.99&14.13&5.7\\
&2.2&14.94&14.22&4.8\\
L+NV&1.4&13.29&11.88&10.6\\
&1.6&13.58&12.35&9.1\\
&1.8&13.86&12.76&7.8\\
&2.0&14.02&13.08&6.7\\
&2.2&14.12&13.30&5.8\\
\end{tabular}
\caption{EoS, masses, radii of the star, radii of the crust-core
interface, and sizes of the crust of the models studied in this paper (without
magnetic field).}\label{tab_referencemodel}
\end{table}

The shear oscillations of the crust are mainly determined by the
shear modulus $\mu_\mathrm{S}$ which we obtain from the
zero-temperature limit of \citet{Strohmayer1991} given by
\begin{eqnarray}
 \mu_\mathrm{S} = 0.1194 \frac{n_i (Ze)^2}{a}\,,\label{shear1}
\end{eqnarray}
where $n_i$ is the ion density, $(Ze)$ the ion charge and $a = \left[3/(4\pi
n_i)\right]^{1/3}$ the average ion spacing. This equation is derived for
a perfect bcc lattice, and the shear modulus, which has different magnitude
along different crystal axes, is averaged in order to obtain an isotropic
effective shear modulus $\mu_\mathrm{S}$ \citep[see][]{Strohmayer1991}. For
the NV EoS of the crust we use a simple fitting formula derived by
\citet{Duncan1998}
\begin{equation}
 \mu_\mathrm{S} = 1.267 \times 10^{30} \mathrm{erg\,cm}^{-3}
\rho^{4/5}_{14}\,,
\end{equation}
where $\rho_{14} = \rho/(10^{14}\mathrm{g\,cm}^{-3})$. 

To calculate the shear modulus for the DH EoS, one has to evaluate $n_i$ in
Eq.\,(\ref{shear1}) in terms of the nucleon number $A$, the proton number $Z$,
and the neutron fraction $X_n$ \citep{Piro2005}: $n_i=\rho_i m_i$ and $A \rho_
i\sim \rho ( 1 - X_n)$. The composition at a given density is given in
\cite{Douchin2001}.
The shear modulus can be estimated to be
\begin{eqnarray}
 \mu_\mathrm{S} &=&1.2 \times 10^{30}  \mathrm{erg\,cm}^{-3}
\rho_{14}^{4/3} \left(\frac{Z}{38}\right)^2 \left(\frac{302}{A}\right)
\nonumber\\ 
&&\hspace{.5cm}\times\left(\frac{1-X_n}{0.25} \right)^{4/3}\label{shear2}
\hspace{-12pt}\vspace{-5mm}\,.
\end{eqnarray}
\citet{Sotani2007} introduced the following fit to this equation
\begin{eqnarray}
 \mu_\mathrm{S} &=& 10^{30} \mathrm{erg\,cm}^{-3} (0.02123 + 0.37631
\rho_{14} +
3.13044 \rho_{14}^2 \nonumber\\ &&~~~~-4.718141 \rho_{14}^3 + 2.46792
\rho_{14}^4)\,.
\end{eqnarray}
This function provides a good approximation for densities larger than
$\rho = 5 \times 10^{11}\mathrm{g\,cm}^{-3}$, but below we will rely on the more
general expression in Eq.\,(\ref{shear2}). The main motivation to use this fit
is to allow for a direct comparison of the
results obtained in this work to the results presented in
\citet{Sotani2007} and \citet{Colaiuda2011}.

The crust-core boundary for the NV and DH EoS is defined at
$\rho_\mathrm{cc,NV}=2.4\times 10^{14} \mathrm{g\,cm}^{-3}$, and
$\rho_\mathrm{cc,DH}=1.28\times 10^{14} \mathrm{g\,cm}^{-3}$, respectively.

We employed two EoS which give neutron star models with large
shear moduli compared to other available EoS \citep{Steiner2009}.
Therefore, the results obtained in this work can be regarded as an upper
limit for the influence of the crust.
%
\section{Numerical methods}\label{sec_numerics}
This section is devoted to the numerical methods employed to analyze the
torsional shear oscillations. The current work is an extension of the study of
\citet{Cerda2009}, whose non-linear GRMHD code we use as a 
 basis for our
simulations. The code has been developed in order to
investigate various astrophysical scenarios where both magnetic fields and
strong gravitational fields play an important role in the evolution of the
system. \citep{Dimmelmeier2002a, Dimmelmeier2002b, Dimmelmeier2005, Cerda2008}.
The code uses high-resolution shock-capturing
schemes to solve the GRMHD equations for a dynamical spacetime, under the
approximation of the conformally flat condition (CFC) for the Einstein's
equations \citep{Isenberg2008,Wilson1996}.  For a spherically symmetric
spacetime the CFC metric is an exact solution of Einstein's equations,
and reduces to
the solution in isotropic coordinates. Therefore,
this numerical code
is well suited to describe the spacetime used in the simulations of the present
work.
The equations are cast in a first-order, flux-conservative
hyperbolic form, supplemented by the flux constraint transport method to ensure
the solenoidal condition of the magnetic field. 

The basic version of the code including the solution of the ideal GRMHD
equations was thoroughly tested in \citet{Cerda2008}, who
demonstrate the robustness of the code for a number of stringent tests, such as
relativistic shocks, highly magnetized fluids, equilibrium configurations of
magnetized neutron stars, and the magneto-rotational core collapse of a
realistic progenitor. One important feature is the ability
of the code to handle different classes of EoS which range from simple
analytical expressions to microphysically derived tables.
We want to emphasize that although the current project is concerned with
small-amplitude perturbations in order to apply simplifications appropriate to a
linear regime, the code can in principle handle large amplitudes and in general
is nonlinear.

While the numerical method employed in the original version of the code 
to integrate the equations is unchanged, we
modify the equations to be solved according to Eq.\,(\ref{conservationlaw}) and
(\ref{reduced_withcrust1})-(\ref{reduced_withcrust2}) in the crust. As
mentioned above we also have to solve for the spatial derivatives of
$\xi^\varphi$. For these equations, which do not represent conservation laws, 
we calculate the fluxes $-\alpha v^\varphi$ corresponding to
Eqs.\,(\ref{eq_xi_dr}) and (\ref{eq_xi_dtheta}) at the cell 
interfaces with the corresponding approximate Riemann solver. The
derivatives of the fluxes are approximated by dividing the difference of the
flux at the two cell interfaces by the grid extension.
Although this approach should be able to cope with discontinuities, as present
at the crust-core interface for the shear modulus, it 
 turns out that special
care has to be taken to achieve a converging method. The coupling between the
crust and the core will be described in detail below. We note that
when using this approach and setting the shear modulus in the crust to zero
 the results of \citet{Cerda2009} are recovered. 

Setting up the interface conditions turned out to
be a very delicate issue. The shear modulus is discontinuous at the crust-core
interface, and Riemann solvers are able to cope with discontinuities at
cell interfaces. Therefore, we define the crust-core interface to be located at
a
cell interface. In this case it is crucial to ensure that the reconstruction
procedure gives a value for
$\xi^\varphi_{\,,r} \sim \Sigma^{r}_{\,\,\varphi}$ which is consistent with the
continous traction conditions (Sec.\,\ref{sec_bc}). Any standard reconstruction
method not taking this condition explicitly into account failed and the
simulations produced spikes in the radial profiles that spoiled the evolution
or even made the whole evolution unstable.
The main cause for this behavior is the discontinuity of the shear modulus at
the crust-core interface. For intermediate and weak magnetic fields, the very
large shear modulus on one side and the vanishing shear modulus on the other
causes the different terms in the momentum equation for the radial flux
Eq.\,(\ref{flux_r}) to be much larger on the side of the crust than on the side
of the core, i.e.
\begin{equation}
\left| \frac{b_\varphi B^r} {W} \right|_\mathrm{crust}\, , \,
\left| \mu_\mathrm{S}\Sigma^{r}_{\,\,\varphi} \right|_\mathrm{crust}
>>
\left| -\frac{b_\varphi B^r}{W}\right|_\mathrm{core}.
\end{equation}
The evaluation of the flux at the crust-core
interface is numerically
problematic due to non-cancellations of the two terms on the side of the crust.
However, when taking the continous traction condition appropriately into
account, this problem does not arise and the flux at the core-crust interface is
well behaved, allowing one to perform simulations also for intermediate and weak
magnetic fields.

We are using the following numerical treatment based on the continous traction
condition $\xi^\varphi_{\,\mathrm{core}} = \xi^\varphi_{\,\mathrm{crust}}
\equiv \xi^\varphi$ and $\xi^\varphi_{\,\mathrm{core},r}=\eta
\xi^\varphi_{\,\mathrm{crust},r}$ with $\eta = 1+\mu_\mathrm{S} /
\left(\Phi^4 (b^r)^2\right)$. The derivatives at the crust-core interface can
be approximated by
\begin{eqnarray}
 \xi^\varphi_{\,\mathrm{core},r}(r_\mathrm{cc})&=&\frac{\xi^\varphi
(r_\mathrm{cc})-\xi^\varphi_{ \,\mathrm{core}}(r_\mathrm{cc} - 0.5 \Delta r )
}{0.5\Delta r}\,,\label{approx_xi_dr1}\\
\xi^\varphi_{\,\mathrm{crust},r}(r_\mathrm{cc})&=&\frac{\xi^\varphi_{
\,\mathrm{crust}}(r_\mathrm{cc} + 0.5 \Delta r ) - \xi^\varphi
(r_\mathrm{cc})}{0.5\Delta r}\,,\label{approx_xi_dr2}
\end{eqnarray}
where $\Delta r$ the grid spacing in radial direction. Both equations lead to
the following
expression for $\xi^\varphi$:
\begin{equation}
 \xi^\varphi (r_\mathrm{cc}) = \frac{\xi^\varphi_{\,\mathrm{core}}(r_\mathrm{cc}
- 0.5 \Delta r
) + \eta \xi^\varphi_{ \,\mathrm{crust}}(r_\mathrm{cc} + 0.5 \Delta r)}
{1+\eta}\,.
\end{equation}
Knowing $\xi^\varphi$ at the crust-core interface one can calculate the
radial derivatives $\xi^\varphi_{\,\mathrm{crust},r}$ and
$\xi^\varphi_{\,\mathrm{core},r}$, and finally the fluxes. For the calculations
presented in this paper we used a second-order approximation of the
derivatives instead of Eqs.\,(\ref{approx_xi_dr1}) and (\ref{approx_xi_dr2}).
%
\section{Results}\label{sec_results}
First, we show that our code reproduces the purely shear oscillations of the
crust, and we demonstrate how these crustal modes are damped when the
magnetic field is switched on. The next two subsections are concerned with the
behavior of the magneto-elastic oscillations at intermediate fields
($5\times10^{13}\,$G$<B<10^{15}\,$G), and at higher magnetic
field strengths, respectively. In the remaining subsections we address the issue
of different equilibrium models and investigate the magnetic field threshold at
which the magneto-elastic QPOs break out of the core and reach the surface of
the neutron star. For intermediate magnetic field strengths the QPOs are
confined to the core due to the interaction with the crust.
If not stated otherwise we refer to the results using an equilibrium model
obtained with the APR+DH EoS with a mass of $M=1.4\,\mathbf{M}_\odot$
(see Table\,\ref{tab_referencemodel}).

The 2-dimensional simulations are computationally very demanding.
On a single Intel Xeon processor with 3GHz a typical simulation necessary
to perform a Fourier analysis of the Alfv\'en spectrum takes of the order of a
few weeks for one equilibrium model. We are therefore restricted in the maximum
time we can evolve the models. In the most demanding situation the minimum
evolution time corresponds to 10 times the dynamical time scale which is set by
the Alfv\'en crossing time. For a magnetic field of 
$10^{14}\,$G the Alfv\'en crossing time for the fundamental
Alfv\'en oscillation is about $2\,$s. This value scales inversely 
proportional to the magnetic field.

\subsection{Recovering purely shear oscillations of the
crust}\label{sec_shearmodes}

\begin{table*}
 \tabcolsep 5pt
\begin{center}

\begin{tabular}{c  c c c c c c c}
Model& \multicolumn{7}{c}{mode frequency in Hz  } \\ \hline
& \multicolumn{5}{c}{n=0 ($\pm1$Hz)}& n = 1& n =2\\ 
&l = 2&l = 3&l = 4&l = 5&l = 6&\multicolumn{2}{c}{($\pm20$Hz)}\\\hline\hline
APR+DH 1.4 &25.4 (24.6) [25.1]&40.0 (38.9)&53.6 (52.2)&67.3 (65.1)&80.0
(77.8) &741 (761) [734]&  1190 (1270)\\
APR+DH 1.6 &24.3 (23.4) [24.0]&38.5 (37.0)&51.2 (49.6)&64.3 (61.9)&76.5
(73.9) &829 (860) [825]& 1340 (1430)\\
APR+DH 2.0 &21.9 (21.3) [21.7]&34.6
(33.6)&46.3 (45.1)&58.0 (56.3)&69.2
(67.3)& 1052 (1083) [1045]& 1842 (1810)\\
L+DH 1.6   &21.0 (20.6) [20.9] &33.1 (32.5)&44.8 (43.7)&55.6 (54.5)&66.8
(65.1) &565 (586) [567]& 917 (980)\\
L+DH 2.0   &19.5 (18.9) [19.2]&30.7 (29.9)&40.9 (40.2)&51.2 (50.1)&61.4
(59.9)&682 (713) [677]&1100 (1190)\\
\hline\hline
APR+NV 1.6 &23.9 (23.8) [23.6]& 37.5 (37.6)&50.7 (50.5)& 62.9 (63.0)&75.5
(75.3)&692 (689) [684]&1230 (1220)\\
APR+NV 2.0 &21.5 (21.4) [21.2]&33.7 (33.9)&45.4 (45.5)& 56.5 (56.7)& 67.8
(67.8) &838 (858) [827]& 1501 (1520)\\
L+NV 1.6   &22.0 (21.8) [21.8]&34.7 (34.5)&46.8 (46.3)&58.6 (57.7)&69.7
(69.0) & 526 (525) [522]& 936 (930) \\
L+NV 2.0   &19.5 (19.7) [19.6]&31.2 (31.1)&41.9 (41.7)&52.2 (52.1)&62.5
(62.2) & 615 (615) [608]& 1092 (1090)\\
\hline
\end{tabular}
\end{center}
\caption{ Frequencies of some torsional shear modes of the crust for
different EoS. Numbers following the abbreviation of the EoS give the
mass of the stellar model in solar mass units. The frequencies in round
parenthesis are from \citet{Sotani2007}, and the squared brackets give the
result of the eigenmode analysis (see Appendix \ref{ap_eigenmodes}). For the
$n=1$ and $n=2$ modes we compare to the $l=2$ case only. The frequencies of the
eigenvalue calculation for $n=0$ and $l>2$ can be obtained by multiplying the
corresponding frequencies for $l=2$ with $\sqrt{(l-1)(l+2)}/2$ .
The error ranges shown in the table header 
correspond to the frequency spacing in the Fourier analysis.}
\label{crust_modes}
\end{table*}
The purely shear oscillations for various realistic EoS in general 
relativity have been calculated for the linearized problem by 
\citet{Messios2001} and \citet{Sotani2007}. To recover their results, we
performed a series of simulations for a selection of models with zero magnetic
field strength. As initial velocity perturbation we use a simple
radial law in the form  $v\approx \sin{ (\pi / 2 *(r-r_\mathrm{cc}) /
(r_\mathrm{s}-r_\mathrm{cc}))}$ multiplied by a sum of the first ten
vector spherical harmonics for the angular dependence. With this
kind of perturbation
we ensure to excite single modes with different values for the radial and
angular mode numbers $n$ and $l$.
We evolve the system for $1$~s in the case of $n=0$ modes and $50$~ms for the
modes with $n\geq1$.
The resolution of the simulations was $120 (r)\times60 (\theta)$ points in the
domain $[0,r_\mathrm{s}]\times[0,\pi/2]$, the grid is equidistant in both
directions and we used equatorial symmetry. The chosen grid corresponds to about
20 radial zones inside the crust.

Table \ref{crust_modes} gives the oscillation frequencies of our 
dynamical simulations extracted from the Fourier analysis at points
inside the crust. The frequencies of the modes agree up to a few
percent with those of the linear approximation (given in round parenthesis).
For modes with $n\geq1$, the frequency resolution of the Fourier transform 
does not allow us to resolve modes with the same $n$ but different $l$, 
which are only separated from each other by a few Hz. Therefore, the measured
frequency is a mixture of different $l$ contributions, and its value is expected
to be slightly larger than that of \cite{Sotani2007} for $l=2$,
which are reported in the Table\,\ref{crust_modes}  as well.

Additionally, we have computed the frequencies from the associated eigenvalue
problem (Appendix\,\ref{ap_eigenmodes}). For the computation of these
frequencies we used a radial grid of about 80 zones in the crust. The results
for the $n=0,\,l=2$ and $n=1$ modes are given in squared brackets in
Table\,\ref{crust_modes}. They agree up to a similar accuracy with those that
have been obtained in the literature.

\begin{figure}
\begin{center}	
 \includegraphics[width=.47\textwidth]{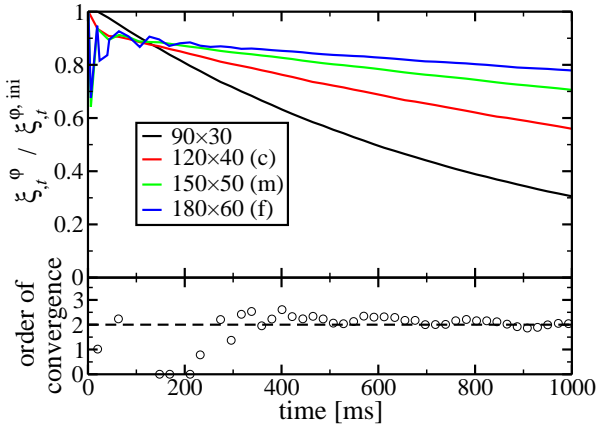}
\end{center}
\caption{Numerical damping of crustal oscillations.
The upper panel shows the evolution of the maximal amplitude of
 $\xi^\varphi_{,t}$ normalized to its initial value, at a point in the crust 
near the pole for different grid resolutions.  The numerical damping decreases 
with increasing resolution. The lower panel shows
the order of convergence (circles) computed using the three highest resolution
simulations
compared to the expected second order convergence (dashed line).}
\label{fig_damping_0}
\end{figure}
To investigate the convergence properties of our code when no magnetic field is
present, we have calculated the evolution of the $n=0$ and $l=2$ mode of
our reference neutron star model for different grid resolutions: $180\times60$,
$150\times50$, $120\times40$, and $90\times30$. The angular grid is equidistant,
while the radial grid is equidistant only in the crust, where 40 percent of the
zones are located, and coarsens towards the center of the
star. The finer mesh in the crust ensures higher accuracy without 
 significant increase of
the computational costs. In order to save further
computational power we assume equatorial symmetry.  The
mode frequencies extracted from the simulations at different resolutions agree
within the frequency resolution of the Fourier transform.
The upper panel of Fig.~\ref{fig_damping_0} shows the time evolution of the
maximum amplitude  of $\xi^\varphi_{,t}$ at the crust for different grid 
resolutions. One clearly sees that the numerical damping decreases
with increasing resolution. For this simple test case it is possible to compute
the order of  convergence using the results for the three highest resolutions,
when one assumes that  the error in the interesting variable, $f$, scales as
$\Delta^p$, where $\Delta$ is the size of the numerical cell and $p$ the order
of convergence. To compute $p$ one searches for roots of 
\begin{equation}\label{eq_Q}
\frac{f_\mathrm{coarse}-f_\mathrm{medium}}{f_\mathrm{medium}-f_\mathrm{fine}}
=\frac{\Delta_\mathrm{coarse}^p -
\Delta_\mathrm{medium}^p}{\Delta_\mathrm{medium}^p-\Delta_\mathrm{fine}^p}\,,
\end{equation}
where the subscripts denote fine, medium or coarse grid resolution. 
The lower panel of Fig.~\ref{fig_damping_0} shows that after 
a short initial transient the order of convergence, $p$, rapidly converges to
$2$, which is the expected order of convergence of our numerical scheme.

\subsection{Absorption of crustal shear modes by the Alfv\'en
continuum}\label{sec_damping}
This subsection is concerned with the absorption of purely shear modes of
the crust into the Alfv\'en continuum of the core. We will often refer to this
process as the {\it damping} of crustal shear modes. This expression may
mislead the reader to think of dissipation processes. However, damping in the
current context refers to the transfer of energy from crustal modes into the
continuum of the core, but not to dissipation of energy.
Only when referring to numerical damping we mean the usual concept
describing the loss of ordered kinetic energy by numerical dissipation.

\subsubsection{$n=0$ shear modes}
As we have shown in \cite{Gabler2011letter} purely shear oscillations are
absorbed very efficiently by the Alfv\'en continuum of the core for magnetic
field strengths $B\gtrsim5\times 10^{13}\,$G. In this case the amplitude of the
perturbations of the crust is damped by transferring their energy to the
Alfv\'en continuum. To analyze how this
damping scales with the magnetic field strength, we have performed a series of
simulations for different crustal modes with $n=0$, $l\geq2$ for different
magnetic
fields ($0$, $10^{13}$,$2\times10^{13}$, $5\times10^{13}$,
$8\times10^{13}$,$10^{14}$   and
$2\times10^{14}\,$G). For these simulations we use a grid of
$150\times100$ zones covering a domain $[0,
  r_{\mathrm{s}}] \times [0,\pi]$, which is equivalent to the medium grid of
the previous section but not assuming equatorial symmetry. Here, we use
the solution of the eigenvalue problem for the unmagnetized crust (see
Appendix\,\ref{ap_eigenmodes}) as initial perturbation for the
velocity. Symmetries are exploited whenever a perturbation is purely symmetric
or antisymmetric with respect to the equatorial plane. 

\begin{figure}
\begin{center}	
 \includegraphics[width=.47\textwidth]{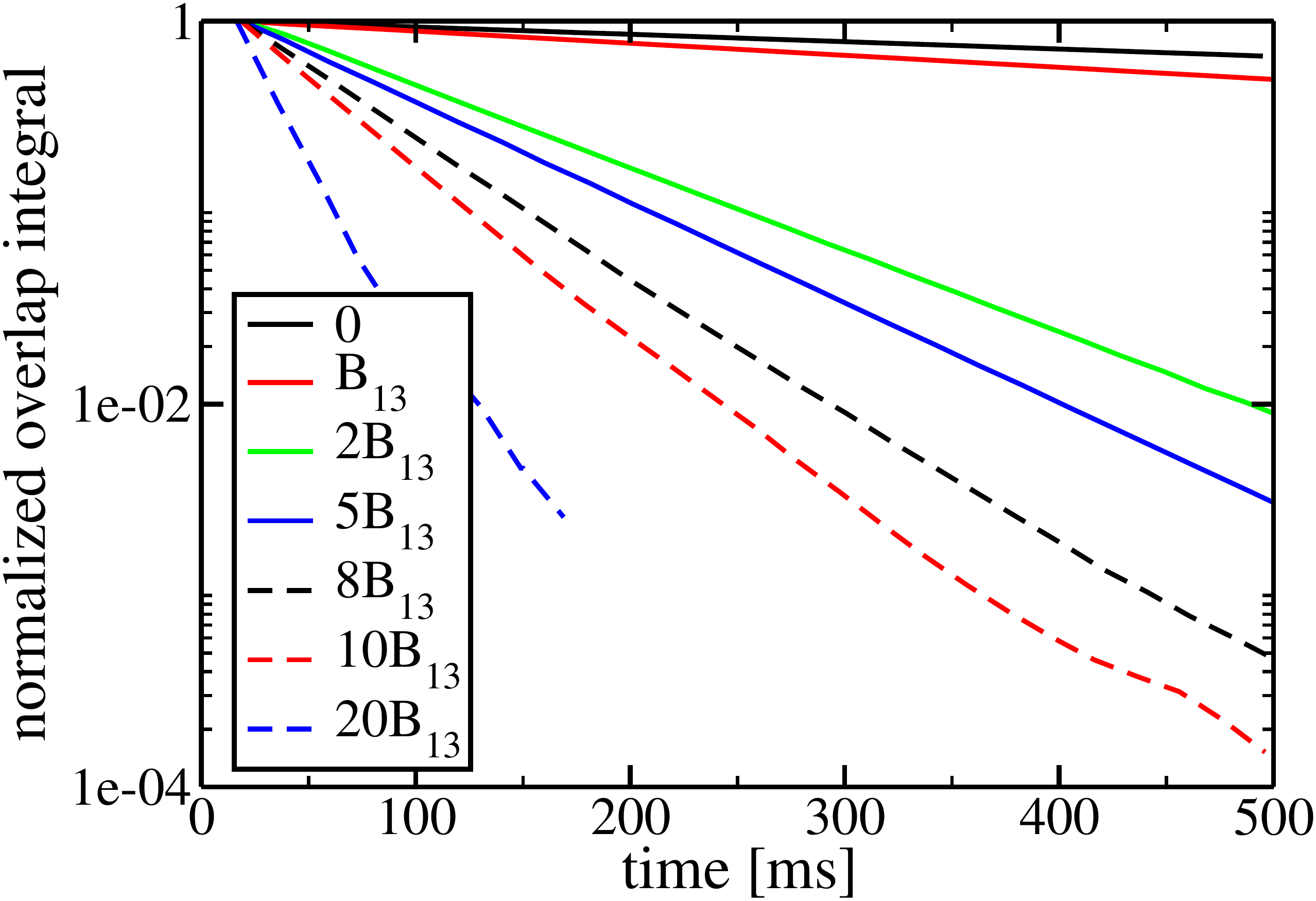}
\end{center}
\caption{The maximum of the normalized overlap integral for the $n=0$, $l=2$
eigenmode of the crust as a function of time. Stronger magnetic field
results in faster damping of the initially excited crust mode.
Orthogonality of different eigenmodes is fulfilled numerically up to about
$10^{-5}$. In the legend we introduced the abbreviation
$B_{13}=10^{13}\,$G.}\label{fig_damping_l2}
\end{figure}

In the following we will discuss our results by using so-called overlap
integrals (derived in Appendix\,\ref{ap_eigenmodes}). These overlap integrals
are the expansion coefficients of an arbitrary spatial function in the basis of
the crustal oscillation eigenmodes, i.e. they give a measure of how strong the
different eigenmodes are excited \cite[see][for an application to
radial oscillations of neutron stars]{Gabler2009}.
In Fig.\,\ref{fig_damping_l2} we show the time dependence of the maxima of the 
overlap integrals defined in Eq.\,(\ref{overlap_integral}) corresponding to
the $n=0$, $l=2$ crustal mode for different magnetic field strengths. The
stronger the field is the faster the damping of the shear mode proceeds. For
high field strength, $B> 2\times10^{14}\,$G, it is not possible to obtain
a characteristic damping time $\tau$, because the time scale is shorter than
one oscillation period. Therefore, the latter can be used as an upper bound for
$\tau$. For $2\times10^{14}\,$G we show the evolution of the overlap integral
only up to the time when global magneto-elastic oscillations start to dominate
and interfere with the purely shear modes of the crust.

Table\,\ref{tab_damping} shows the damping timescales, obtained by analyzing the
overlap integrals, for
different $n=0$, $l\geq2$ modes and different magnetic field strengths.
\begin{table}
 \tabcolsep 5pt
\begin{center}
\begin{tabular}{c | c c c c c c c}
magnetic&\multicolumn{7}{c}{$\tau$ [ms] for mode $n=0$ }\\
field [G]&$l=2$&$l=3$&$l=4$&$l=5$&$l=6$&$l=7$&$l=8$ \\ \hline
0 		&1130&1110&1030&874&654&466&313\\ 
$1\times10^{13}$	&688&835&846&764&599&441&302\\ 
$2\times10^{13}$&102&287&478&534&481&385&279\\
$5\times10^{13}$&83&85&72&54&38&63&104\\
$8\times10^{13}$&58&60&43&38&37&34&28\\
$1\times10^{14}$	&46&48&37&38&40&33&27\\ 
$2\times10^{14}$&20&21&21&23&19&21&21\\
\hline
\end{tabular}
\end{center}
\caption{Damping timescale $\tau$ in ms for different $n=0$, crustal shear
modes for different magnetic field strengths.}
\label{tab_damping}
\end{table}
The values for zero magnetic field serve as 
 a measure of the numerical
damping of the code. As expected for numerical dissipation processes,
modes with higher $l$ suffer stronger from numerical damping than lower $l$ 
\citep[see also][]{Cerda2010}. For weak magnetic fields, e.g.
$10^{13}\,$G, the damping
of high $l$ modes is dominated
by numerical dissipation, while for low $l$ modes it is caused by
the interaction with the Alfv\'en continuum of the core. For magnetic fields
stronger than $5\times10^{13}\,$G, we are confident that the damping time of
all studied modes is physical and not due to numerical dissipation. Above
$2\times10^{14}\,$G the oscillations are damped on shorter time scales than 
the respective oscillation period, and hence it is impossible to obtain accurate
damping times.

\begin{figure}
\begin{center}	
 \includegraphics[width=.47\textwidth]{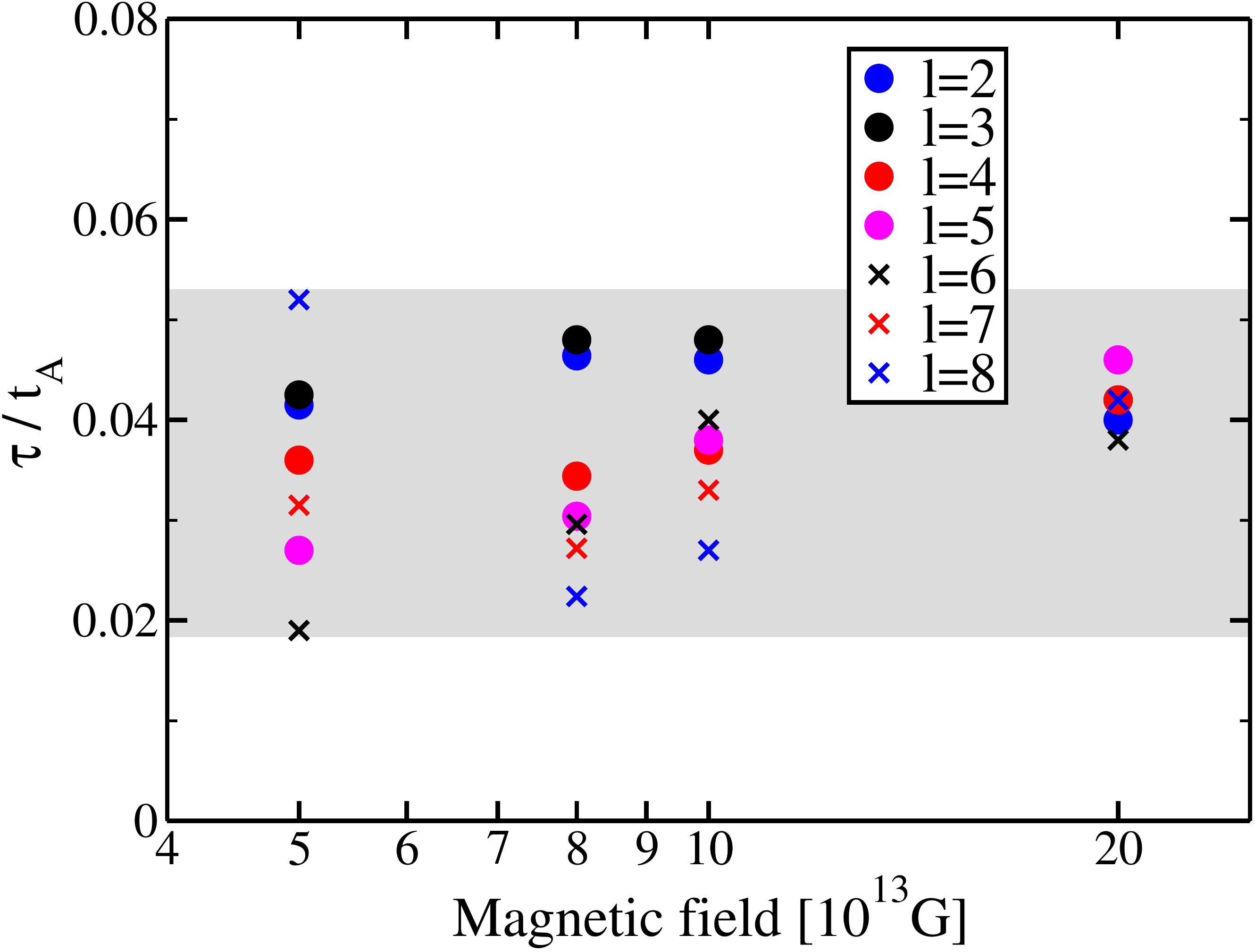}
\end{center}
\caption{The dependence of $\tau / t_\mathrm{A}$ on the magnetic
field. The stronger the magnetic field is the less is the spread of the
numerical values for different $l$ around this value. The shaded area indicates
the maximum range of variation of $\tau / t_\mathrm{A}$ around the mean value
($\sim0.04$). }\label{fig_damping_t_over_cross}
\end{figure}

Figure \ref{fig_damping_t_over_cross} shows  $\tau / t_\mathrm{A}$, where
$t_\mathrm{A}$ is
the Alfv\'en crossing time of the star at the pole.
The damping time of crustal modes due to the absorption by the Alfv\'en
continuum scales linearly with $t_\mathrm{A}$, i.e. $\tau$ decreases with
increasing
magnetic field. The mean damping time is about $0.04\,t_\mathrm{A}$. 
Deviations from this value (see Fig.\,\ref{fig_damping_t_over_cross})
depend non-trivially on the magnetic field and the
mode number $l$. The spread decreases with increasing magnetic field strength, 
being smallest for our $B=2\times10^{14}\,$G simulation. Low $l$-modes 
(filled circles in Figure\,\ref{fig_damping_t_over_cross}) show a smaller spread
around the mean value than the higher $l$-modes (crosses in the same figure).

These deviations are expected, because the damping depends on a variety of
parameters as for example the frequency of the crustal mode, the frequencies
available in
the Alfv\'en continuum of the core, and the spatial structure of the
crustal modes. Numerical effects may also affect
the damping times. These are: Firstly, the grid resolution necessary to obtain
a comparable accuracy for different modes increases with increasing mode number
$l$. Secondly, waves in the crust reaching the core-crust interface will
propagate into
the core as Alfv\'en waves. Due to the jump in the wave velocity at the
interface, the wavenumber increases, i.e. the resolution
requirements in the core are more restrictive. This holds, in particular, for
weak magnetic fields, where the jump in the wave velocity is larger, and for
large $l$ modes with higher frequencies. 

The damping time at a given magnetic field strength $B>10^{13}\,$G varies only
by at most a factor of two between different $l$, but may vary significantly
with the magnetic field strength. This indicates that the damping is dominated
by the magnetic field and does not depend sensitively on the mode structure
itself, i.e. there will always be a part of the continuum
that is able to drain the energy from the crustal oscillations.

A more detailed discussion of the damping of the different crustal modes is
beyond the scope of this work. 
Our results do not favor a crustal-mode interpretation of
the observed QPOs in SGRs, because any crustal
shear mode is damped sufficiently fast for magnetic field strengths well below
the typical magnetar field strengths $\sim5\times10^{13}\,$G. Although there
might exist SGRs with weaker magnetic fields \citep{Rea2010}, QPOs have
only been observed so far in magnetars with the strongest fields.

\subsubsection{$n>0$ shear modes}
The higher radial overtones ($n>0$) of the shear modes have frequencies above
$500\,$Hz (see Table\,\ref{crust_modes}) and are usually used to explain the
QPOs of SGR 1806-20
with the highest frequencies of $625$ and  $1840\,$Hz. As these modes have at
least one node inside the crust computing their evolution demands much
higher spatial resolution than necessary for the $n=0$ modes. Thus, it was
practically impossible for us to follow their evolution over several Alfv\'en
crossing times to the same accuracy as for the $n=0$ modes. However this would
have been necessary to draw more reliable conclusions about damping times or
interaction of these modes with the Alfv\'en continuum in the core.
Nevertheless, we can make some qualitative statements.

The time needed for a shear wave ($v_\mathrm{S}\sim1000\,$km/s) to travel
through the crust ($\Delta r_\mathrm{crust} \sim 1\,$km) corresponds to the
inverse of the
frequency of the first overtone ($n=1$), which is of the order of $f\sim1\,$kHz.
Assuming that the wave travels inside the crust along the $\theta$-direction
(travel path $\Delta r \sim 10\pi\,$km), a similar estimate results in
frequencies $ v_\mathrm{S}/\Delta r \sim 30\,$Hz which is of the order of the
frequency of the fundamental $n=0$ oscillation. Therefore, we may
conclude that the $n=0$ modes represent waves that travel predominantly parallel
to the crust-core interface, while the $n=1$ modes correspond to waves that
travel radially. This may explain the strong dependence of the $n=0$ modes on
the angular number $l$ and the weak dependence of the $n=1$ modes \citep[see
Table \ref{crust_modes} with the corresponding discussion, and][]{Sotani2007}.
Hence, we expect the $n=0$ modes, derived with isotropic shear modulus, to be
much more affected by the presence of an anisotropic magnetic field than the
$n=1$ modes. In particular, near the equator where the coupling between the
crust and the core is weaker than close to the pole (see
Eq.\,\ref{eq_interface}) the shear waves
may travel back and forth in the crust without interacting strongly.
Additionally, at the equator the magnetic field is almost parallel to the
$\theta$-direction, i.e. the direction of the Alfv\'en waves is
perpendicular to the direction of the $n=1$ shear waves. This suggests that the
$n=1$ shear waves are not influenced strongly by the presence of moderate
magnetic fields ($\lesssim10^{15}\,$G). According to these theoretical
considerations we expect the overtones of the shear modes to survive
longer than the $n=0$ modes. 

To investigate this issue, we have performed two simulations with our
reference model at magnetic field strengths of $2\times10^{14}\,$G and
$5\times10^{14}\,$G. The resolution for antisymmetric simulations
($l=\{2,4,6,..\}$) was $150\times50$ zones.
The initial perturbation consisted of the $n=1$, $l=2$ mode  of the crust only.
\begin{figure}
\begin{center}	
 \includegraphics[width=.47\textwidth]{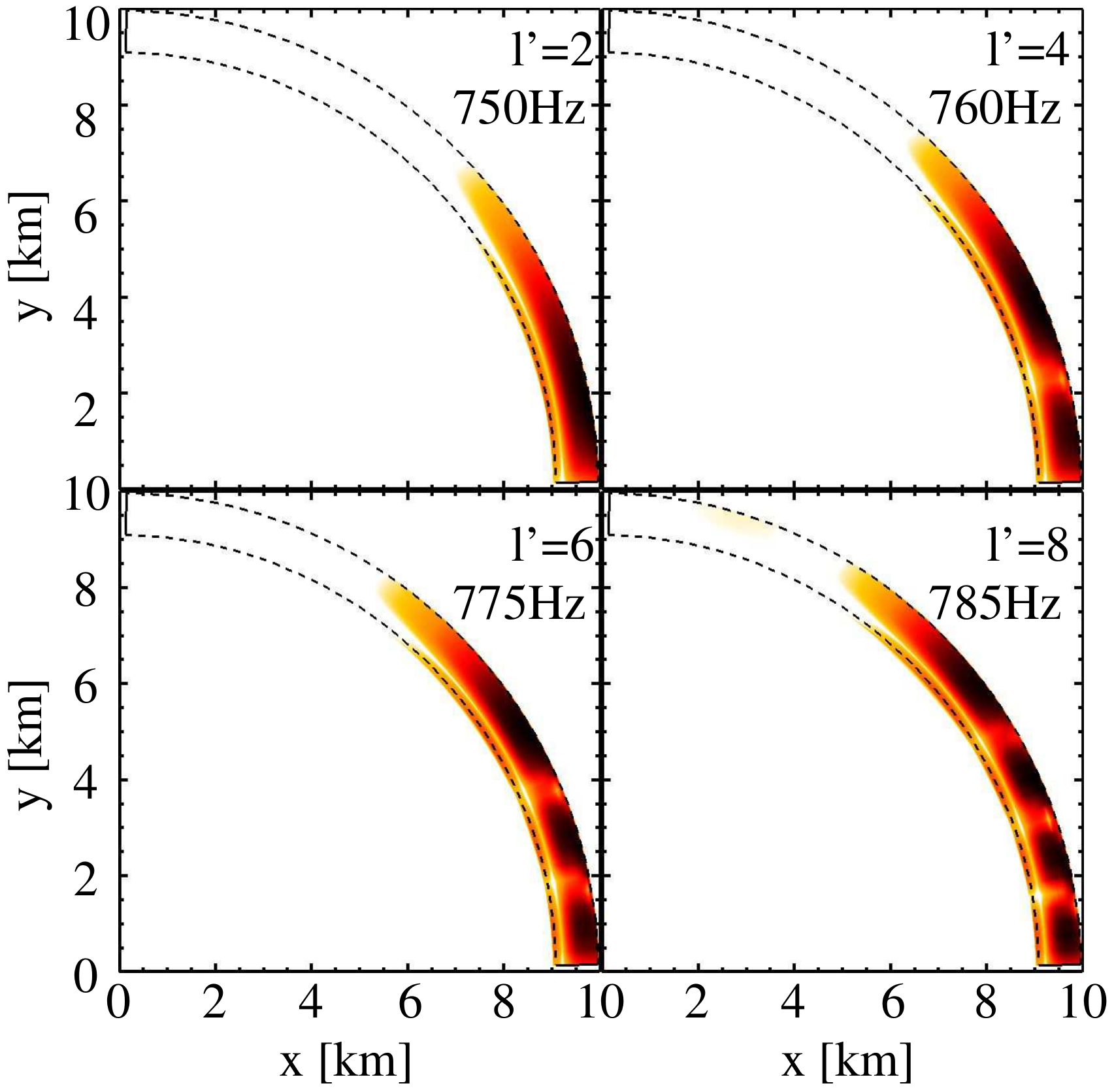}
\end{center}
\caption{The spatial structure of the first $n=1$, antisymmetric crustal shear
modes in the presence of a moderate magnetic field of $5\times10^{14}\,$G. The
oscillation patterns are strongly distorted from the typical $l$-dependence of
the spherical harmonics and only exist in regions close to the equator, where
the coupling between the core and the crust is weakest. We therefore label their
angular dependence with $l^\prime$. The dashed lines
indicate the region of the crust and the color scale ranges from white
(minimum) to red-black (maximum).}\label{fig_n1}
\end{figure}
In Figure \ref{fig_n1} we show the main contributions to the oscillations as
obtained by Fourier analyzing the time evolution of the simulation with
$5\times10^{14}\,$G. The QPO patterns
are compressed towards the
equator and strongly distorted from the typical $l$-dependence of spherical
harmonics of the purely shear eigenmodes of the crust (see Appendix
\ref{ap_eigenmodes}). To account for this difference we label these QPOs with
$l^\prime$. Naturally, the
initial data of the undistorted $l$-mode excites many distorted $l^\prime$-QPOs.
As expected from the theoretical considerations above, the strongest amplitudes
of the oscillations appear near the
equator, where the coupling to the core is weakest (see Fig.\,\ref{fig_n1}). 
The magnetic field also increases the spacing between the frequencies
of successive $n=1$, $l^\prime$ modes from $\Delta f \approx 1\,$Hz
without field to $\Delta f \approx 10\,$Hz \citep[see][for a discussion of
purely shear eigenmodes]{Sotani2007}.

\begin{figure*}
\begin{center}	
 \includegraphics[width=.46\textwidth]{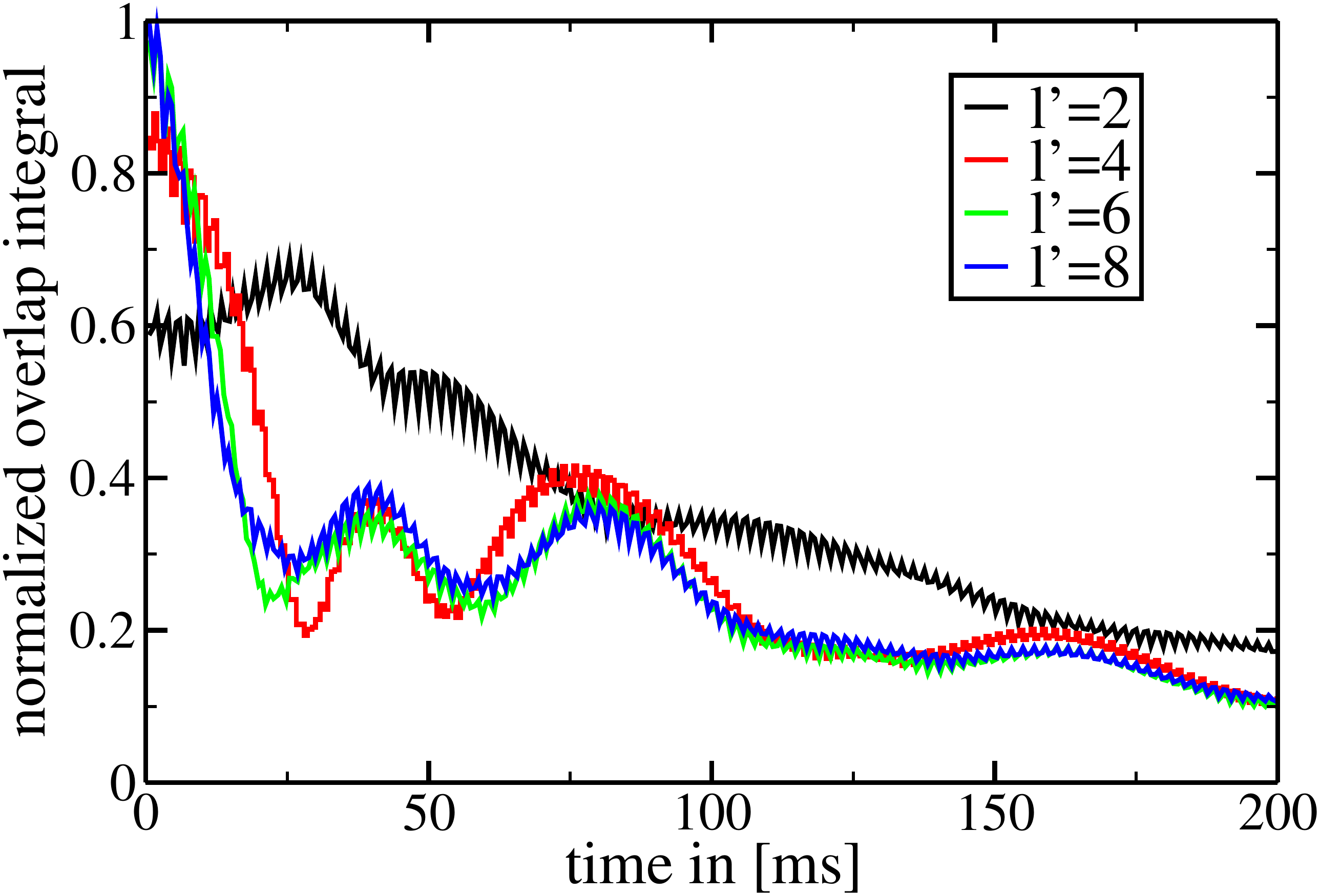}\hspace{1cm}
 \includegraphics[width=.46\textwidth]{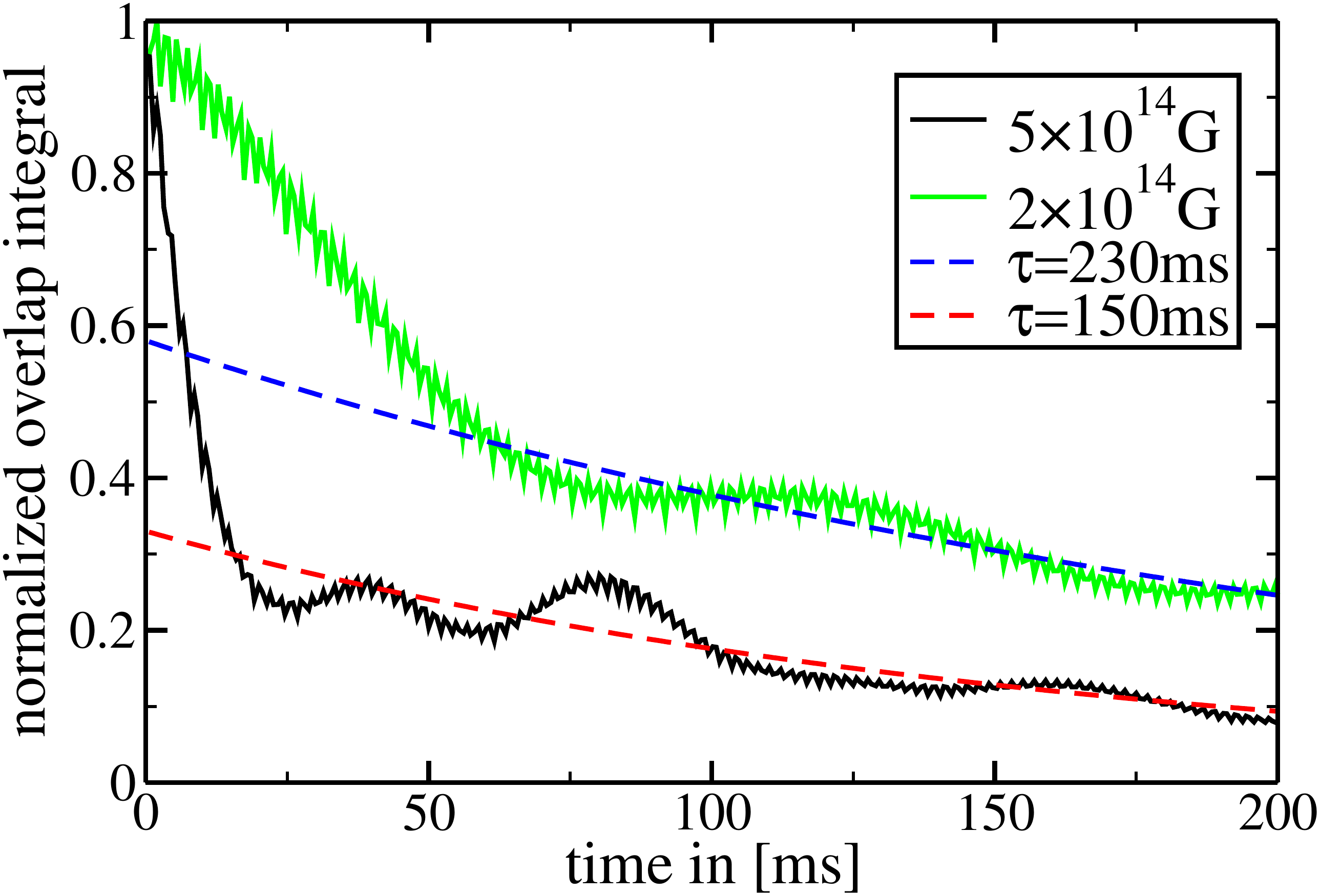}
\end{center}
\caption{{\it Left panel}: The normalized overlap integrals of the evolution for
the first few $n=1$, antisymmetric magneto-elastic modes at
$B=5\times10^{14}\,$G. The approximate mode
structure is taken from the Fourier analysis (see Fig.\,\ref{fig_n1}). Strictly
speaking there are no modes in terms of the result of the linear analysis, but
the influence of the magnetic field is insufficient to destroy the coherent
oscillations in all parts of the crust. The initial perturbation of the
$n=1$, $l=2$ crustal mode excites a large number of the
magneto-elastic modes which are all damped on a timescale of 
$\sim 150$\,ms. {\it Right panel}: The overlap integral performed for a basis
with purely radial dependence $\Xi_i (r, \theta)=R_{\lambda_r}$ (see
Appendix \ref{ap_eigenmodes}) of the $n=1$ modes. These integrals are a measure
of how strong the ensemble of all $l^\prime$, $n=1$ modes is excited. The
damping time, indicated by the fits to
an exponential (dashed lines), for $2\times10^{14}\,$G
($5\times10^{14}\,$G) of $\tau=230\,$ms ($150\,$ms) is much longer than that
of the $n=0$ modes, which is of the order of several ms only.}
\label{fig_damp_n1}
\end{figure*}

To study the behavior of the different QPOs, we calculate the corresponding
overlap integrals (Eq.\,(\ref{overlap_integral})) but taking the spatial
structure obtained from the Fourier analysis (Fig.\,\ref{fig_n1}) as basis
functions. 
The time evolution of the maxima of these overlap integrals for
$l^\prime=\{2,4,6,8\}$ can be seen in the left panel of Figure
\ref{fig_damp_n1}. Indeed all of $l^\prime$-QPOs are excited by the purely
shear $n=1$, $l=2$ eigenmode perturbation. 
Since the frequencies of the different $n=1$ modes are very similar, i.e. hard
to disentangle in any analysis of the simulations,
we average all 
$n=1$ modes to estimate the total damping time of the $n=1$ QPOs.
To this
end we calculate the overlap integral with the radial function
$\Xi_i (r, \theta)=R_{\lambda_r}$ corresponding to the pure shear eigenmodes
obtained with Eq.\,(\ref{eq_eigenmode_R}). This effectively averages over the
angular dependence, and provides a
measure of how strong the ensemble of all $l^\prime$, $n=1$ QPOs is excited. The
corresponding plots for $B=2\times10^{14}\,$G and $B=5\times10^{14}\,$G are
shown in the right panel of Fig.\,\ref{fig_damp_n1}. 
As indicated by the fitting functions
with damping times of $150$ and $230\,$ms, the damping timescale of the $n=1$
QPOs is much longer than for the $n=0$ modes at the given magnetic field
strength of a few $10^{14}\,$G.

However, one has to be very cautious at this point. With the resolution used
here, we are not able to resolve the Alfv\'en oscillations
inside the core of the neutron star which could damp the crustal
shear oscillations resonantly. At $5\times10^{14}\,$G the fundamental Alfv\'en
oscillation is about $2\,$Hz. To resolve the resonant coupling
to the crustal $n=1$ mode of roughly $500\,$Hz one would need the
250th overtone. Simulations with appropriate grid resolution would take of the
order of years. For the damping process itself this lack
of resolution should not be a problem, because the numerical method employed
should take all necessary information into account, i.e. the Riemann solver
at the crust core interface considers all the local information of possible
waves traveling into the core. The problem arises inside the core, where a low
resolution leads to an averaging out of all 
 fine-scale structure. Hence, the
energy of the Alfv\'en overtones of the continuum is transformed to resolved low
order oscillations, i.e. we cannot trust the
oscillations inside the core. However, as long as the Alfv\'en oscillations do
not reach the crust at the opposite side of the star at $t\approx
t_A=0.5/f_A=1.2\,$s (for $B=5\times10^{14}\,$G), a simulation of the crust
region should give correct results. Therefore, the present estimate of the
damping time of the $n=1$ overtones 
should be considered as a lower limit, because the main effect we
are missing is the excitation of crustal magneto-elastic QPOs by incoming
Alfv\'en oscillations of the core.

\subsubsection{Results for different equilibrium models}
All previous results were obtained for a model based on the APR+DH EoS and
a mass of $1.4\,\mathrm{M}_\odot$. Simulations using other EoS or different
masses for the equilibrium model yield qualitatively similar results. 

In table 2 of \cite{Gabler2011letter} we showed that the damping of purely
crustal shear modes occurs on the order of $100\,$ms or less for a variety of
EoS and masses at a magnetic field strength 
of $B=10^{14}\,$G. The initial models that we used in those simulations employed
a different prescription for the internal energy (ideal gas) than the
values provided in the EoS tables. As these models 
 were thus thermodynamically
inconsistent, we have recomputed them with the appropriate internal energy
values. The resulting damping times are at most 20\% longer than the values
reported in \cite{Gabler2011letter}. Some of the corrected damping times are
shown in Table \ref{tab_damping_eos}. The conclusions drawn in
\cite{Gabler2011letter} are not affected
at all by this change, because the
maximal damping time is still less then $100$ ms in all cases.

\begin{table}
\begin{center}
\begin{tabular}{c | c c c}
  EoS    &\multicolumn{3}{c}{$\tau\,$[ms] at $B=10^{14}\,$G}\\
 &                 $n=0$, $l=2$ & $n=0$, $l=3 $ &$n=0$, $l=9$\\
\hline
APR+DH 1.6	&45&47&21\\
APR+DH 2.0	&38&41&17\\
L+DH 1.6	&57&61&27\\
L+DH 2.0	&52&56&25\\
\hline 
\end{tabular}
\end{center}
\caption{Damping timescales $\tau$ due to resonant absorption of
  crustal shear modes by the Alfv\'en continuum for
  initial perturbation modes $l=2$, $l=3$, and $l=9$ for different
  combinations of equations of state at $B = 10^{14}\,$G. The number 
  in the labeling of the EoS represents the mass of the neutron star 
  model in M$_\odot$.}
\label{tab_damping_eos}
\end{table}
The variation of the damping times with the EoS at a given magnetic field is not
surprising. The relative size of the crust of
these models varies roughly by a factor of 3 \cite[see][]{Sotani2007}, and the
shear modulus of both crustal EoS is of comparable size. This explains the
smallness of the observed variations in \cite{Gabler2011letter} which
do not exceed  a factor of 5. A significantly lower shear modulus, as proposed
in \cite{Steiner2009}, would lead to even shorter damping times of the crustal
shear modes. 

The influence of the details of the EoS are not substantial
because when trying
to explain the frequencies of the QPOs observed in SGRs as shear oscillations,
the shear modulus should lie in the range we use in this work. 
Otherwise it is already impossible to reproduce the correct range of frequencies
within the crustal oscillation model. Changing the shear modulus somewhat
would have only a modest effect on the damping times. However, even for a
hypothetically exotic shear modulus, which could be one order of magnitude
larger than the actual values we use, the crustal shear oscillations would be
damped much too fast to explain long-lived QPOs. We therefore argue that we
can safely exclude shear oscillations as a viable explanation of observed
magnetar QPOs, 
for the magnetic field configurations studied here.

\subsection{QPO structure for intermediate field strength}
\label{sec_intermediate}
In \cite{Gabler2010Proceedings,Gabler2011letter} we have shown that
the crust significantly changes the structure of the QPOs for intermediate
magnetic field strengths between $5\times10^{13}-10^{15}\,$G compared to
models without crust. In this section we investigate the
behavior of the Alfv\'en QPOs of the core in this regime further.
\subsubsection{QPO structure at $4\times10^{14}\,$G}
\begin{figure}
\begin{center}	
 \includegraphics[width=.47\textwidth]{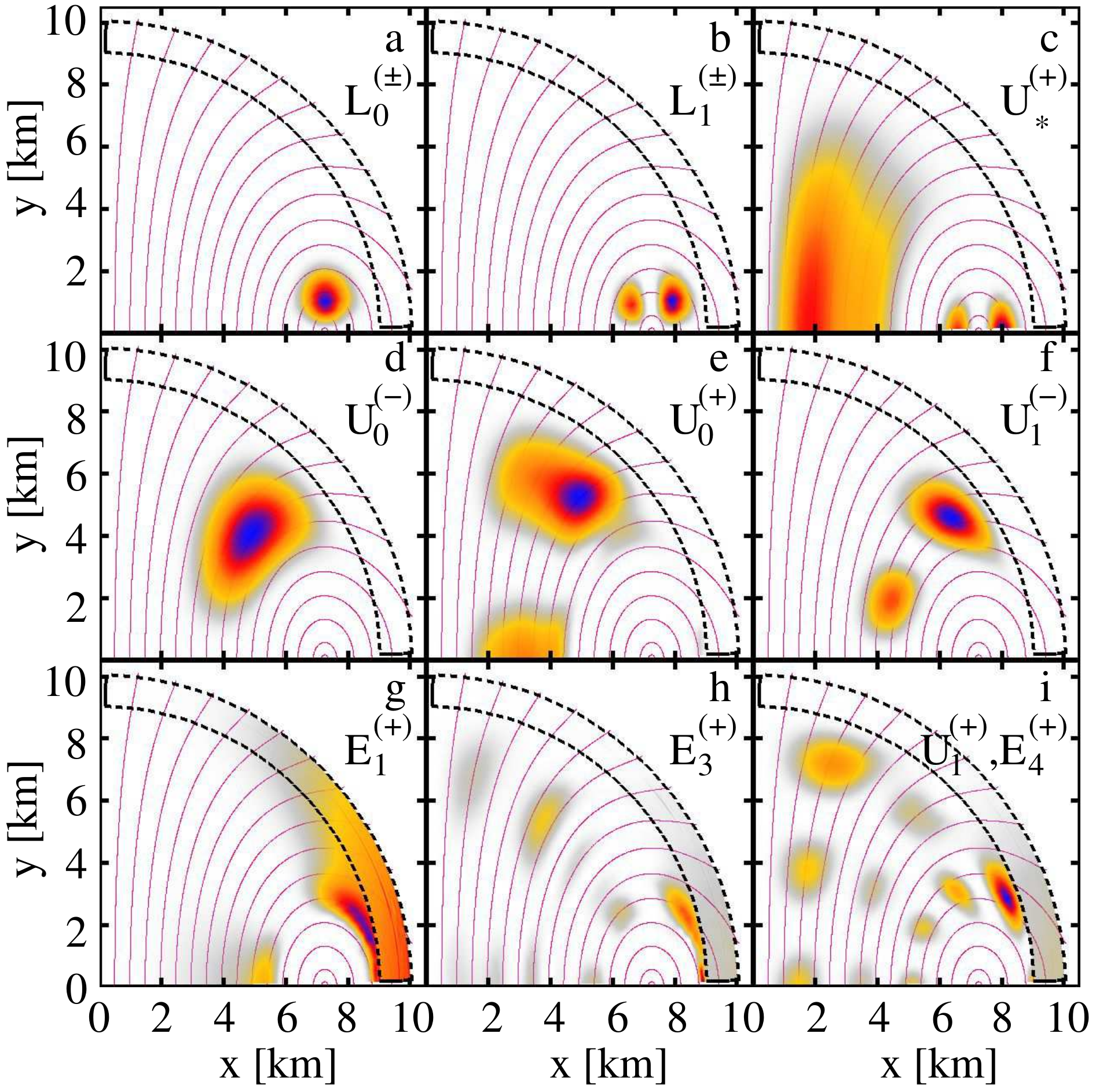}
\end{center}
\caption{The Fourier amplitude inside the neutron star for the model APR+DH 1.4
at $B=4\times10^{14}\,$G. Shown are the first two
lower QPOs $L_0^{(\pm)}$ and $L_1^{(\pm)}$, the first four upper QPOs
$U_{*}^{(+)}$, $U_0^{(-)}$, $U_0^{(+)}$, and $U_1^{(-)}$ and some selected edge
QPOs $E_1^{(+)}$, $E_3^{(+)}$, and $E_4^{(+)}$. (The figure for
$E_2^{(+)}$ was not very clear due to contamination with other QPOs.) The
plus and minus sign indicate symmetry (+) and antisymmetry (-)  with
respect to the equatorial plane.
Magenta lines indicate magnetic field lines, and black, dashed lines the
location of the crust. The color scale ranges from white-blue (minimum) to
red-black (maximum). 
}\label{fig_FFT_4_14}
\end{figure}
In this subsection we will analyze the results of two simulations with a
resolution of $100\times40$ zones and computational domain $[0,
r_\mathrm{s}]\times[0,\pi/2]$. We perform one simulation with $l=2$ initial data
and antisymmetry with respect to the
equatorial plane, and another symmetric one with $l=3$ initial data.
Both runs were evolved up to $t\approx5\,$s.

For the model APR+DH 1.4 with a magnetic field strength of $B=4\times10^{14}\,$G
we find three different families of QPOs. 
Following \cite{Cerda2009} we label the \emph{lower} QPOs as $L^{(\pm)}_n$
and  the \emph{upper} QPOs as $U^{(\pm)}_n$. A new family of QPOs appears, which
we call \emph{edge} QPOs and label them as $E^{(\pm)}_n$. To
avoid confusion with the previous work of \citet{Cerda2009}, and
because the fundamental upper symmetric QPO has
special properties as we will show below, it will be labeled as $U_*^{(+)}$ at
low magnetic field strength. The plus and minus sign in the description of the
QPOs indicate symmetry (+) or antisymmetry (-) of the QPO with respect to the
equatorial plane.

\begin{figure}
\begin{center}	
 \includegraphics[width=.47\textwidth]{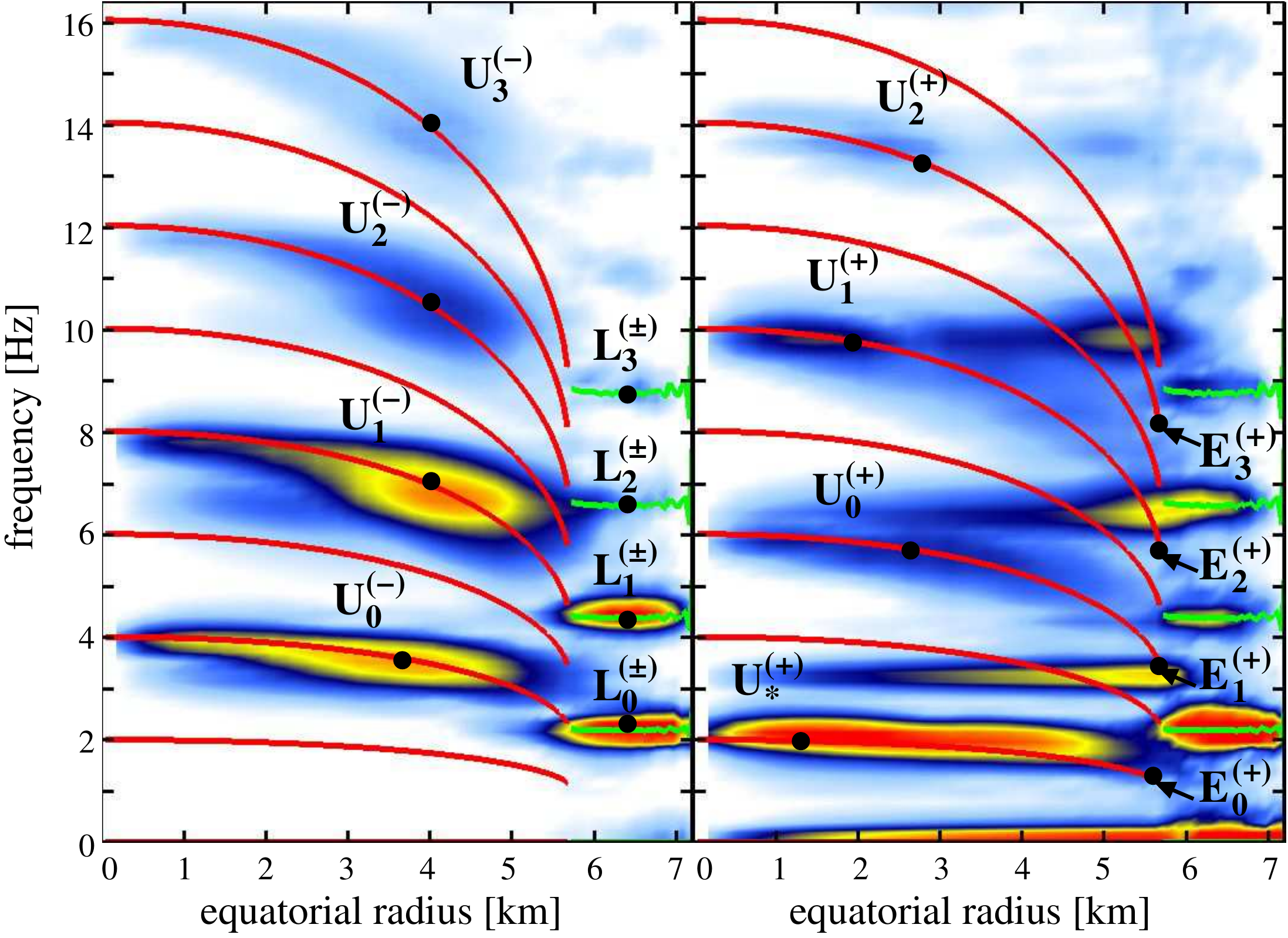}
\end{center}
\caption{The averaged Fourier amplitude along different magnetic field lines
labeled by the radius where they cross the equator for model APR+DH with
$M=1.4\,\mathrm{M}_\odot$ and $B=4\times10^{14}\,$G. Red and green lines
give the Alfv\'en continuum obtained with the semi-analytic model.
 The locations of the QPOs are indicated by black dots.
{\it Left panel}: antisymmetric simulation; {\it right panel}: symmetric
simulation. The color scale ranges from white-blue (minimum) to orange-red
(maximum). }\label{fig_FFTlines}
\end{figure}

In Fig.\,\ref{fig_FFT_4_14} we plot the distribution of the Fourier amplitude
inside the star. Lower QPOs, as shown in panels a and b, are attached
to field lines which close inside the core. In contrast, upper QPOs are located
closer towards the magnetic poles at open field lines (panels c, d-f). A
third family, the edge QPOs, are connected to the open field line inside the
core of the neutron star which just fails to close inside 
(last open field line). These QPOs can be
seen in panels g-i of Fig.\,\ref{fig_FFT_4_14}. The two QPOs $U_1^{(+)}$
and $E_4^{(+)}$ have very similar frequencies. 
Due to limited evolution time and hence limited resolution for
the Fourier transform both QPOs contribute significantly to the Fourier signal
at the corresponding frequency as can be seen in panel i.
Similarly, the figure for $E_3^{(+)}$ contains some contribution of $U_0^{(+)}$
along the field lines crossing the equator between a radius of 1 and 4 km. In
both panels the edge QPOs are
concentrated on the field lines  which cross the equatorial plane at
around 5 km.
The naming of the edge QPOs becomes clearer
in Fig.\,\ref{fig_FFTlines}, where we plot the
Fourier amplitude of the velocity, averaged per field 
line in the frequency-radius plane. The maxima indicate
the position of the QPOs. The red and green
lines are the continuum of frequencies obtained with the semi-analytic model
 introduced in \cite{Cerda2009} and
adopted here to the problem in the presence of the crust. 

The different families of QPOs mentioned above 
 and shown in Fig.\,\ref{fig_FFT_4_14} can also be identified
in Fig.\,\ref{fig_FFTlines}.
The lower QPOs are attached to the closed field lines which cross the
equatorial plane near $6.5\,$km. Since they are connected to the
closed field lines these QPOs do not have a preferred symmetry, and hence are
present in symmetric and antisymmetric simulations.
The continuum of frequencies
derived with the semi-analytic model (green lines in Fig.\,\ref{fig_FFTlines})
has a minimum at the point where we find the lower QPOs. 
We thus interpret the $L^{(\pm)}_n$ as turning point QPOs 
\citep[see][]{Levin2007}. With the exception of
the fundamental $ U^{(+)}_*$, which is located almost at the turning point 
 of the semi-analytic model, the
upper QPOs are localized in the continuum
of the open field lines (red lines in Fig.\,\ref{fig_FFTlines}) 
 that cross the equatorial plane at $2-4\,$km. 

The members of the new family of QPOs, also obtained in
\cite{Colaiuda2009}, are called edge QPOs because of their position in
Fig.\,\ref{fig_FFTlines}. These QPOs are related to those parts of the
continuum, obtained with the semi-analytic model (red lines), which do not
connect to the continuum of the closed field lines (green lines) (see also the
sketch in Fig.\,\ref{fig_intro}). 
For more details on the interpretation of the QPO structure without crust we
refer to \cite{Cerda2009} and with crust to \cite{Gabler2010Proceedings}.

\subsubsection{Differences caused by the presence of the crust}

The lower QPOs attached to the closed field lines are reproduced
qualitatively similar as in the case without crust. The only 
difference is that they 
are limited to the field lines which close inside the core and do not
extend into the crust. 

However, the upper QPOs which are
located near the pole in models without crust \citep{Cerda2009,
Colaiuda2009, Sotani2007}, can now be found at substantial distance from the
poles, i.e. at lower latitudes  \citep[see also figure 4
in][]{Gabler2011letter}.
In simulations without crust the oscillations were
associated with the maximum at the turning point
of the continuum at the pole, while if
the crust is included, we obtain the maximal amplitudes away from the pole 
and inside the continuum predicted by the semi-analytic model in the absence of
a crust.
 One possible interpretation of this new feature is that
the shear modulus in the crust alters the propagation of magneto-elastic
oscillations in the region near the pole in such a way that standing waves
cannot form at all along individual field lines or, if they form, they go
quickly out of phase with nearby field lines in this region.

At this point it is helpful to recall the problem of the reflection of
plane-parallel waves, where the reflection coefficient
depends on the jump in the propagation velocity or equivalently on the index of
refraction. The stronger the jump in the index is, the larger is the fraction of
the incident wave which becomes reflected.
Using this analogy we would expect that the smaller the difference in
propagation velocity at the crust-core interface is, the more refraction into
the crust should occur and less reflection back into the core should be
produced. If significant refraction into the crust occurs, no stable standing
waves can be maintained during the evolution. When following the
crust-core interface from the pole towards the equator the coupling between
crust and core becomes weaker, see Eq.\,(\ref{eq_interface}) where the $\theta$
dependence of the coupling factor is realized in $b^r$. The
magnetic field in the radial direction, and
hence the Alfv\'en velocity, decreases with increasing $\theta$. Therefore, the
jump in the propagation velocity increases, and thus the fraction of the
wave which is reflected will also increase along this trajectory. 
For sufficiently low magnetic fields there should always be a region near the
equator were almost perfect reflection occurs. However, when following the
crust-core interface from the equator towards the pole, one
will reach a characteristic magnetic field strength where insufficient
reflection occurs to maintain stable standing waves along a certain field line.
At this point we find the maximum amplitude of the QPO in the presence of
a crust. We will investigate the reason for this behavior further in Sec.
\ref{sec_pulse}. 

The new position of the maximum amplitude of the QPOs is thus determined by two
effects. Near the pole the magnetic field lines get out of phase due to the
interaction through the extended crust, because a significant fraction of the
oscillation is refracted. The magnetic field lines can be seen like strings
which are not attached to a rigid but rather ``moving'' boundary, i.e. the crust
which responds to the
oscillations. The resulting effect is a strong coupling of different magnetic
field lines and, consequently, an energy transfer between the lines due to the
scalar shear modulus. In this region and for magnetic field strengths studied
here, each field line seems to act as a damped oscillator. Near the
equator the magnetic field lines get out of phase due to phase mixing like in
the case without extended crust. The additional damping close to the pole makes
the corresponding upper QPOs to be shorter lived than in the case without crust.

\begin{figure}
\begin{center}	
 \includegraphics[width=.44\textwidth]{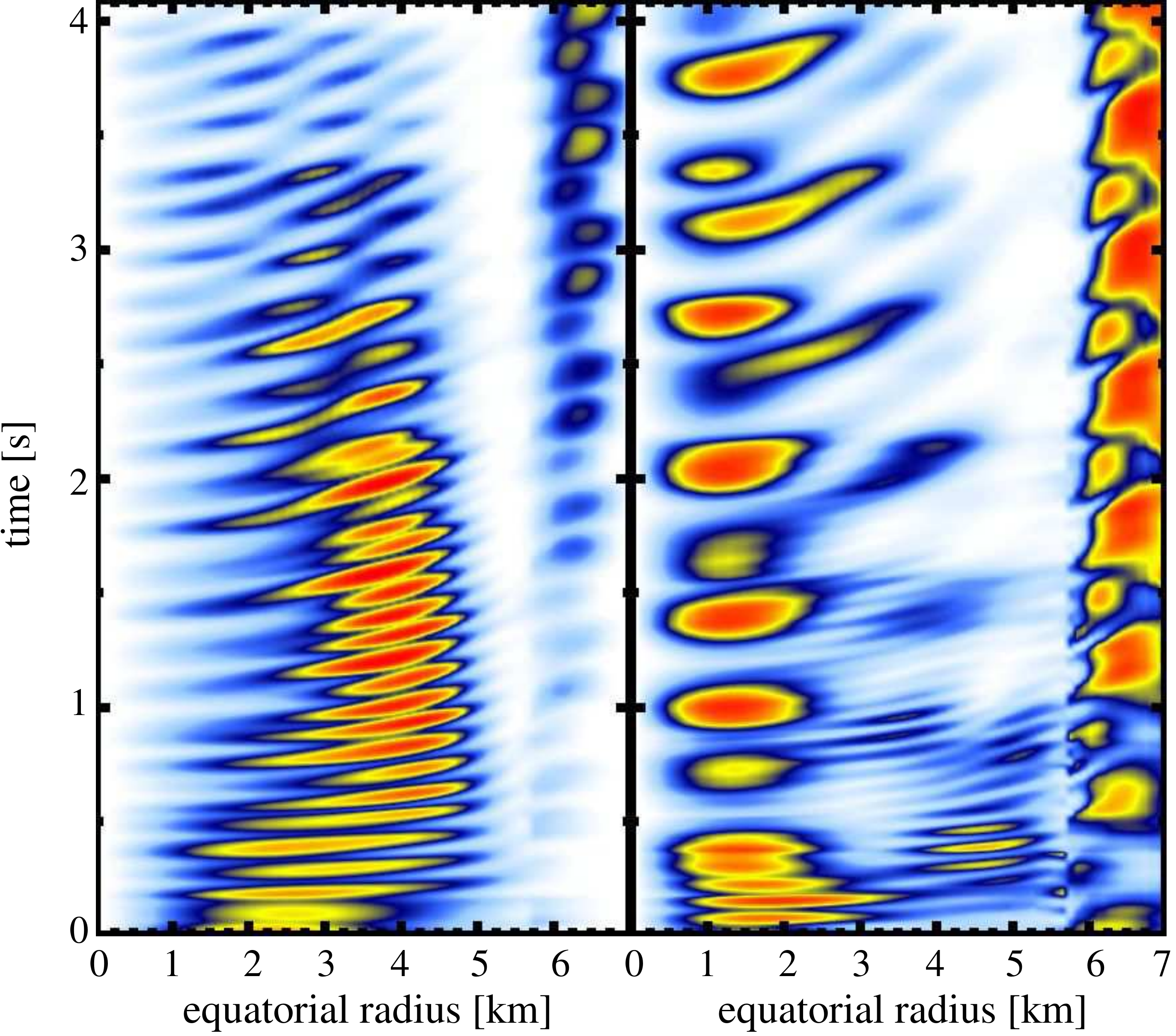}
\end{center}
\caption{Evolution of the magnetic plus kinetic energy per field line
divided by the sum of the energy over all field lines. The field lines are
labelled by their crossing point with the equator. Note that as the total
energy decreases with time the apparent increase of the energy of the lower QPOs
is due to color rescaling at every time step.
Their amplitude actually decreases, but more slowly than that
of the upper QPOs.
The \emph{left panel} shows antisymmetric and the \emph{right panel} symmetric
simulations. The color scale ranges from white-blue (minimum) to orange-red
(maximum).
}\label{fig_ene_4_14}
\end{figure}
This effect can be observed in the right panel of
Fig.\,\ref{fig_ene_4_14}, where we show the magnetic plus the kinetic energy per
field line divided by the total magnetic plus kinetic energy at the given
time as a function of time for different field lines. The initially excited
QPOs attached to the field lines between $4.5$ and $5\,$km disappear rapidly
after about $0.5\,$s. In contrast, the lower QPOs
and the fundamental symmetric QPO near the pole persist during the whole
evolution. Fig.\,\ref{fig_ene_4_14} may suggest that the lower QPO gain energy
with time. However, this apparent energy increase is not a physical effect,
because the total energy decreases with time due to numerical dissipation.
Hence, the relative amplitudes of the energy of the lower QPOs
increase, while their absolute amplitude decrease slightly because of numerical
dissipation.

Compared to simulations without crust \citep{Cerda2009} we find a new
fundamental upper QPO $U^{(+)}_*$. This QPO appears because the boundary
condition at the crust-core interface causes a reflection which results in a
node at this surface. Without crust the boundary condition at the surface of the
star implies a maximum there and the fundamental oscillation has the node at
the equator in this case. Therefore, the symmetric QPO
$U_0^{(+)}$ must have an additional node inside the core (see panel e in
Fig.\,\ref{fig_FFT_4_14} or the second row of Fig.\,\ref{fig_FFT_moving}).
$U^{(+)}_*$ situated between 1 and 2\,km (Fig.\,\ref{fig_ene_4_14}, the right
panel) decays less rapid than the other upper QPOs, $U^{(-)}_0$,
$U^{(+)}_{n\geq0}$, and $U^{(-)}_{n>0}$. This may be related to the fact,
that at $B=4\times 10^{15}\,$G, $U^{(+)}_*$ is located close to the maximum of
the continuum (see Fig.\,\ref{fig_FFTlines}). There, the gradient of the
continuum is less steep, and neighboring field lines get out of phase
less rapidly. This behavior is similar to a turning point QPO, which persists
for longer time than edge QPOs \citep{Levin2007}.

\subsubsection{Changing QPO position with increasing magnetic field}%
\begin{figure*}
$2\times10^{14}\,$G\hspace{6mm}$6\times10^{14}\,$G\hspace{9mm}$10^{15}\,
$G\hspace{7mm}$1.5\times10^{15}\,$G\hspace{6mm}$2\times10^{15}\,$G\hspace{6mm}
$3\times10^{15}\,$G\hspace{6mm}$5\times10^{15}\,$G\hspace{6mm}$8\times10^{15}\,
$G\hspace{6mm}without crust
\begin{center}	
\includegraphics[width=\textwidth]{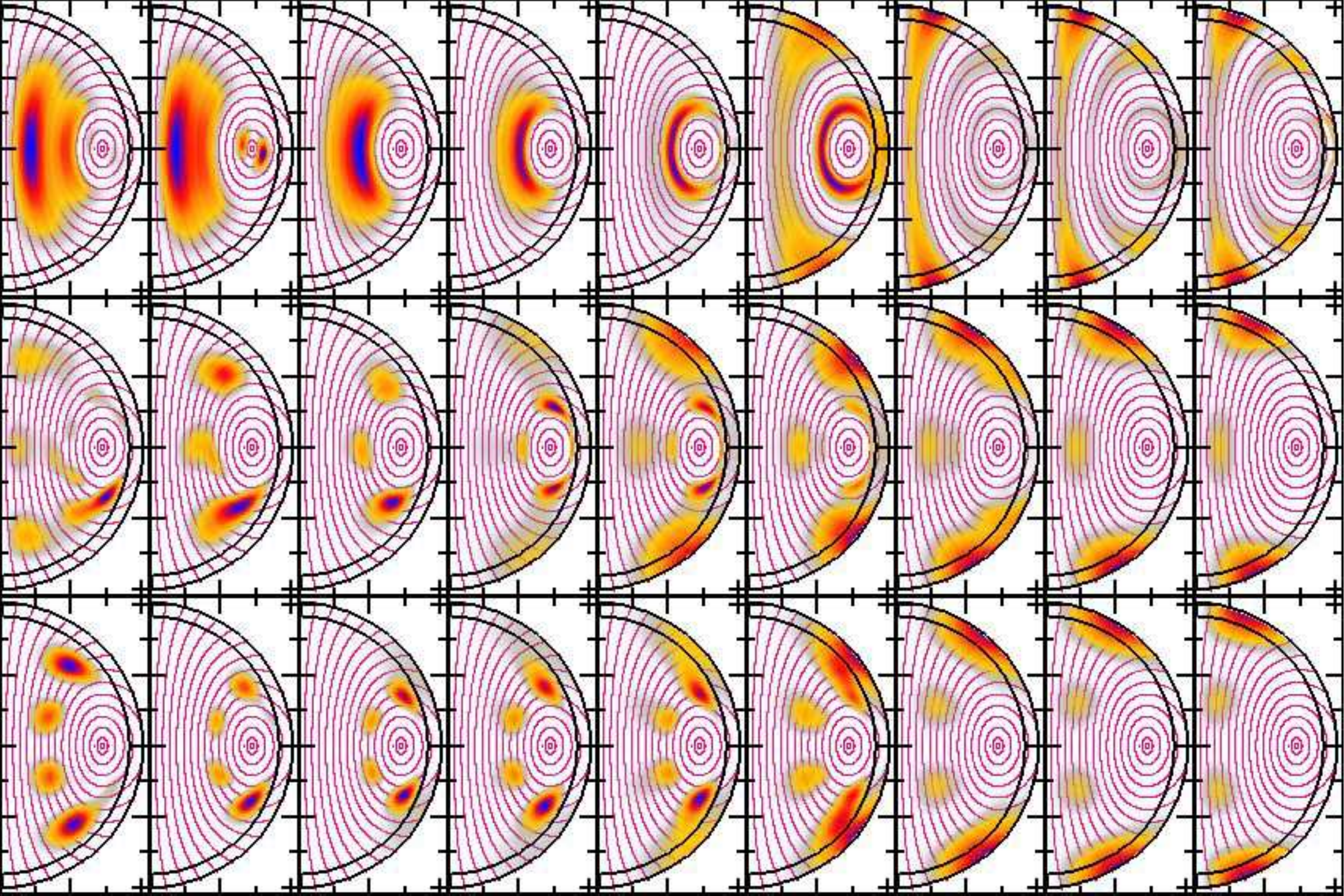}
\end{center}
\caption{Structure of upper QPOs at different magnetic field strengths for the
model APR+DH 1.4.
The panels show the fundamental symmetric $U^{(+)}_*$ (upper panels), the
symmetric $U^{(+)}_0$ (middle panels) and antisymmetric $U^{(-)}_1$ QPOs (lower
panels). The frequencies of
the QPOs shown here are given in Table \ref{tab_freqs}. The
color scale ranges from white-blue (minimum) to red-black (maximum).
}\label{fig_FFT_moving}
\end{figure*}

For magnetic field strengths between $10^{14}$ and $10^{15}\,$G the
simulations reveal that the location of the upper QPO, $U^{(\pm)}_n$
changes within the neutron star. This was not observed in the case of
 pure Alfv\'en oscillations in
\cite{Cerda2009} or \cite{Sotani2008}, where the upper QPOs were always
observed close to the pole. Fig.\,\ref{fig_FFT_moving} (first three
columns) shows the new effect, where we plot the
spatial structure of the Fourier amplitude of the QPOs. When increasing
the magnetic field from $2\times10^{14}$ to
$10^{15}\,$G the upper QPOs $U^{(\pm)}_n$ move from a location near
the pole towards the equator. 
The change in position of the $U^{(\pm)}_n$ is shown 
in Fig.\,\ref{fig_FFT_moving} only for $U^{(+)}_0$ and $U^{(-)}_1$, but holds
for all higher overtones as well and does not depend on the symmetry. For the
fundamental
$U^{(+)}_*$ the dislocation is less (upper row in Fig.\,\ref{fig_FFT_moving}).
One can understand this behavior at least partially with the help of the
semi-analytic model. Fig.\,\ref{fig_FFTlines} shows that the
frequencies and the symmetry of the QPOs are correctly predicted, if we assume
that the Alfv\'en wave is reflected at the crust-core boundary.
However, the semi-analytic model cannot explain where within the continuum the
QPOs are situated. 

Remembering the analogy with the reflection of plane-parallel waves in the
preceding subsection and bearing in mind that with increasing magnetic field
$b^r$ the relative jump in
the propagation velocity on both sides of the crust-core interface at a given
position $\theta$ decreases, more parts of an incident wave get
refracted into the crust. This means that the point where stable standing
waves can be maintained should move towards the equator. This is exactly
confirmed in Fig.\,\ref{fig_FFT_moving}, where for magnetic fields 
$\lesssim10^{15}\,$G the QPOs move from close to the pole towards the
equator, as the magnetic field strength increases.

\subsubsection{Reflection of pulses and spread in crust}\label{sec_pulse}
\begin{figure}
\begin{center}	
 \includegraphics[width=.43\textwidth]{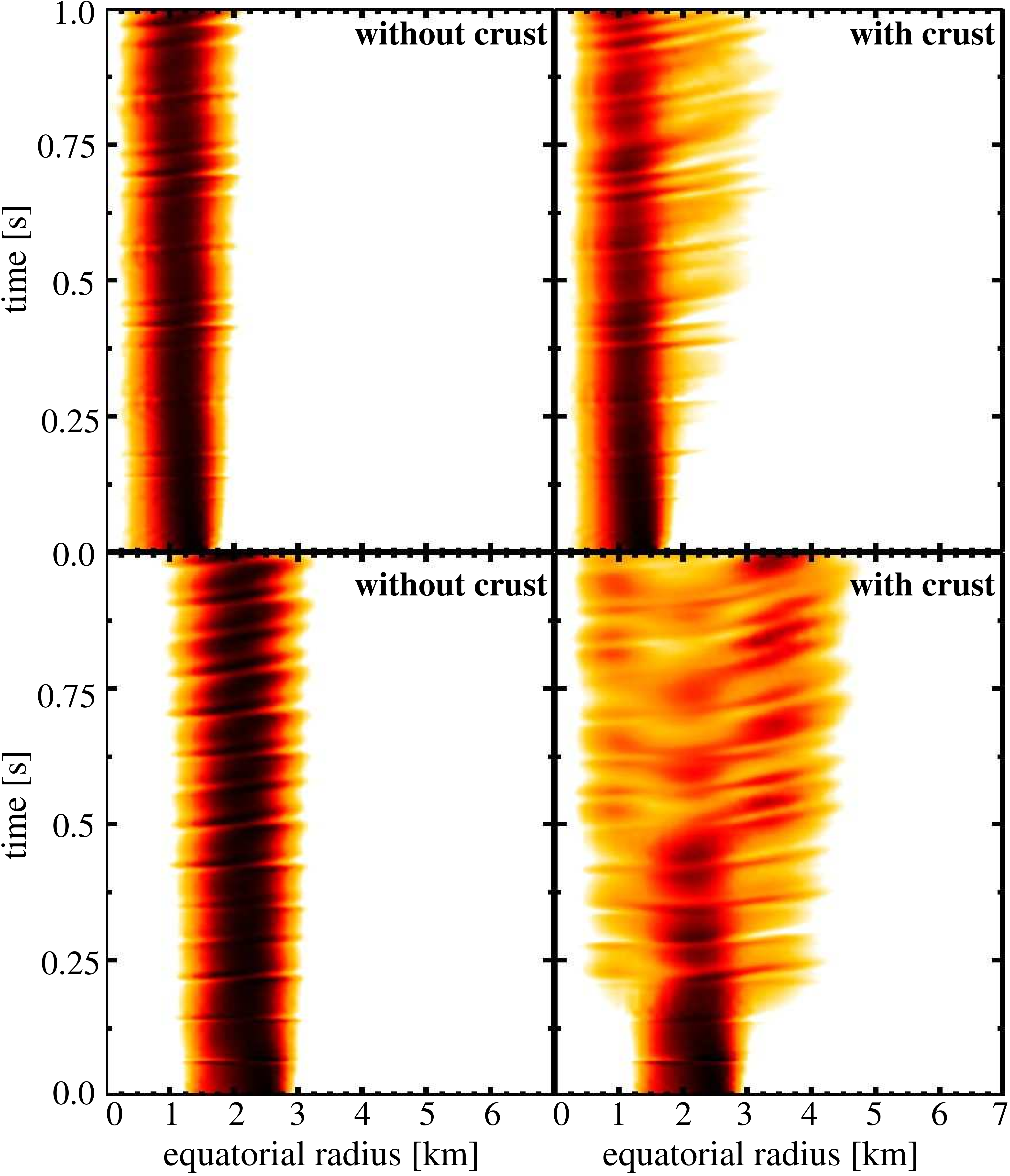}
\end{center}
\caption{Evolution of the magnetic plus kinetic energy per field line
divided by the sum of the energy of all field lines for the model APR+DH 1.4.
The field lines are
labeled by their crossing point with the equator. The pulse
reaches the surface at around $70\,$ms for the first time and the crossing time
is about $130\,$ms. \emph{Left panels:} simulation without
crust, \emph{right panels:} simulation including crust. The upper and low
panels differ by the location of the initial perturbation. Color scale
ranges from white (minimum) to red-black (maximum).
}\label{fig_pulse}
\end{figure}
In the absence of an elastic crust, Alfv\'en wave packets are supposed to travel
approximately along magnetic field lines, as the characteristic direction of
propagation of any magnetic perturbation coincides with the direction
of the magnetic field. However, when a crust is added
this picture changes. The direction of propagation 
 of magneto-elastic waves no longer coincides with the magnetic field direction
(compare the different
eigenvalues in this case given in Eq.\,(\ref{eq:eigenvalues})). 
One would therefore expect a perturbation, traveling along magnetic field
lines from the center of the star towards the surface to spread out 
 past the crust-core interface. 
Such a spread is strong for low magnetic field strengths, when 
the isotropic shear modulus dominates in the crust region and weak for high
magnetic field strengths, when the opposite is true. This behavior is shown
in
Fig.\,\ref{fig_pulse}, where we display the renormalized sum of the kinetic and
magnetic energy per field line for simulations at $5\times10^{14}\,$G. The
initial perturbation is restricted to a limited region of the star about
$4\,$km above the equator. The left panels
show the expected behavior for a pulse which travels along a field line
with no crust present. Two initial perturbations differing only by the location
of the star get reflected at the surface and travel back
towards the center. Any deviations from traveling perfectly along the initially
excited field lines is caused by a numerical coupling of different field
lines and the very weak coupling through the boundary condition at the surface
of the star. 
Taking a crust into account but imposing the same initial perturbations
the wave packets are spread whenever entering the crust, which happens at around
$70 \,$ms for the first time, and subsequently after about every $130\,$ms
(right panels). For initial data located at field lines crossing the
equator around $2$ to $3\,$km (lower right panel), the spread is more
drastic. After some reflections there are phases (around $700$
and $850\,$ms), when no significant perturbation amplitudes can be found
around the field lines which initially carried the perturbation. 

However, the scenario of an initially localized wave packet considered here
cannot rule out the existence of standing waves along individual field lines.
Nevertheless, it suggests that additional effects may be introduced by
the spreading of wave packets in the crust which probably change the Alfv\'en
continuum of our semi-analytic model, where one assumes standing waves
which get reflected at the crust-core interface (or at the surface for stronger
magnetic fields).

\subsubsection{Conservation of angular momentum}
Analyzing the convergence properties of our numerical simulations 
in the intermediate magnetic field
case is more complicated because the contributions to the stress-energy tensor
from the magnetic field and the shear are of the same order of magnitude. The
two extreme regimes have been tested above (purely shear oscillations) or in
\cite{Cerda2009} (purely Alfv\'en oscillations). 
In the present approach, the total angular momentum $J_{\rm tot}\equiv \int
S_\varphi dV$ is the only globally conserved quantity, with
$dV=\sqrt{\gamma}dr d\theta d\phi$. We
do not expect conservation of the energy of the perturbation in our
simulations, because neglecting 
the coupling to poloidal oscillations and assuming purely poloidal magnetic
fields, renders the deviations of the total energy from the energy of the
unperturbed background configuration to be of second order in the
perturbations, while our approach is accurate to first order.

The total angular momentum  $J_{\rm tot}$ 
should be conserved inside the computational volume, but the boundary condition
we have chosen (see Sec.\,\ref{sec_bc})
allows for non-vanishing flux through the surface. 
As we chose initial perturbations with the angular dependence of the vector
spherical harmonics the angular momenta in both hemispheres cancel
by construction and the total angular momentum of the star is zero. For
antisymmetric perturbations the losses/gains through the surface cancel
respectively, while for symmetric perturbations there remains a
non-zero contribution.

\begin{figure}
\begin{center}
 \includegraphics[width=.47\textwidth]{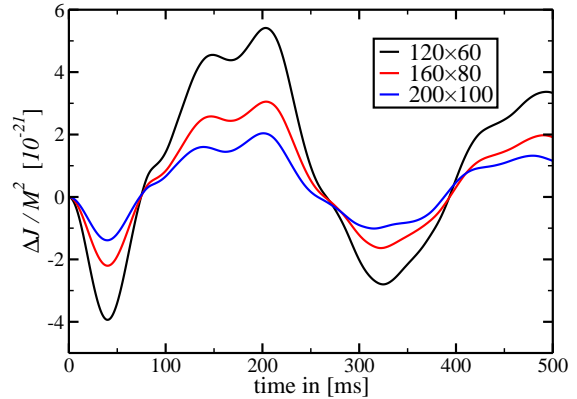}
\end{center}
\caption{Variation of the total angular momentum of the star $J_{\rm tot}/M^2$
during the evolution for three different grid resolutions:$120\times60$,
$160\times80$ and $200\times100$ zones, respectively. The model used in the
simulation is APR+DH 1.4.}
 \label{fig_momentum}
\end{figure}
Fig.\,\ref{fig_momentum} shows the variation of the total angular momentum
during the evolution for an symmetric simulation with $l=3$ initial data at
three different grid resolutions. When analyzing the differences between the
curves (see Section\,\ref{sec_shearmodes}) we obtain the order of convergence
of $1.95$,  which is near the expected second-order convergence.
To estimate the absolute magnitude of the resulting angular momentum error
we compare our perturbation to a rigidly rotating sphere with the
same total angular momentum. Taking the typical total
angular momentum during the simulation, and comparing it to that of a rigidly
rotating sphere $J=2/5 M R^2 \Omega = 2/5 M R v_\mathrm{rot}$, where
$\Omega=v_\mathrm{rot} / R$ is the rotation frequency, we obtain a
maximal velocity which is only a fraction of the perturbation
used in the simulations $v_\mathrm{rot} / v_\mathrm{pert} \sim 10^{-10}$.
Thus, the total angular momentum introduced by our perturbation is very small.
Moreover, the total angular momentum
variations converge to zero, i.e. compared to the numerical errors the losses of
angular momentum through the surface is a small effect not affecting our
simulations. 

\subsection{QPO structure for high magnetic field strength}
\label{strong_fields}

\begin{figure*}
\begin{center}
 \includegraphics[width=.9\textwidth]{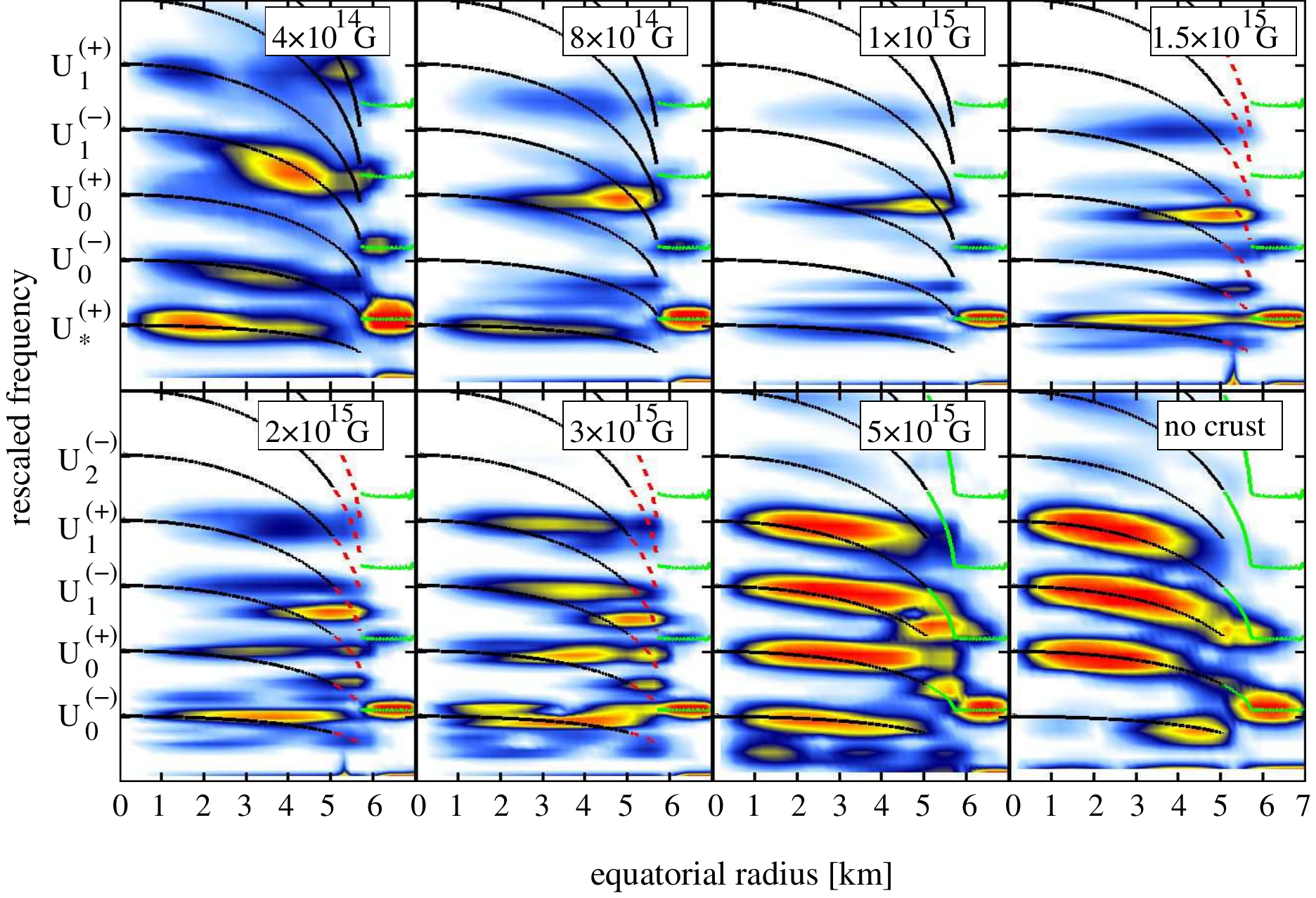}
\end{center}
\caption{The averaged Fourier amplitude along different field lines labelled by
the radius where they cross the equatorial plane for different magnetic field
strengths. In the last panel we show the
expected behavior for a model without crust. The
y-axis represents the frequency rescaled to the fundamental QPO frequency
predicted by the semi-analytical model (see Table\,\ref{tab_freqs}). The color
scale ranges from white-blue
(minimum) to orange-red (maximum). Black and green lines represent the continuum
when taking the crust into account. Dashed, red lines indicate regions where
the semi-analytic model is not supposed to work. We do not know where the waves
get reflected and thus cannot identify whether to consider the lines as open or
closed. }
 \label{fig_FFT_average}
\end{figure*}
For strong magnetic fields we are interested in the structure of magneto-elastic
QPOs, but not in the damping of crustal modes, as the latter are damped
already at much lower field strengths (see Sec.\,\ref{sec_damping}). Therefore,
we can reduce the grid resolution, i.e. the computational costs. In the strong
field case we thus use a uniform radial grid with 100 zones and an angular grid
with $80$ zones in the interval $[0,\pi]$.

For very strong magnetic fields, $B\gtrsim5\times10^{15}\,$G, the
maxima of the Fourier transform align towards the polar axis. With increasing
magnetic field the influence of the shear inside the crust becomes negligible,
and the QPO
pattern approaches that expected for the purely magnetic limit \citep[see figure
3 in][]{Cerda2009}, in agreement with the semi-analytic model. Another
effect
caused by the anisotropy of the shear modulus is a more wide-spread spatial
structure of the QPOs compared to the case without
crust. This can be inferred from the last two columns of
Fig.\,\ref{fig_FFT_moving}, where QPOs are still quite
extended inside the crust ($B=8\times10^{15}\,$G), and in the model without
crust.

Between the two extremes, the QPOs are confined in the core ($B\lesssim
10^{15}\,$G). For strong magnetic fields ($B\gtrsim10^{15}\,$G) there is a
transition from the QPO structure observed in Section \ref{sec_intermediate} to
the purely magnetic case (see Fig.\,\ref{fig_FFT_moving} from the 3 - 6
column). Between $10^{15}$ and $2\times10^{15}\,$G the QPOs begin to have
significant amplitudes in large parts of the crust. 

This transition becomes clearer in Fig.\,\ref{fig_FFT_average}, where we plot
the Fourier amplitude for individual field lines averaged over the
length of the line and labelled by their crossing point with the equatorial
plane. The solid lines represent the
continuum as obtained by the semi-analytic model, where we assume
reflection at the crust-core interface in the upper row and reflection at the
surface of the star in the lower row. The obtained
frequencies are very similar in both cases, because the travel time of the
waves is dominated by the time spent in the core. We note a change of the
structure of the QPOs with the different boundary conditions. We have already
noted in Fig.\,\ref{fig_FFT_moving} that in the case
of weak magnetic fields the QPOs move from being near
the pole towards
the equator, for increasing magnetic field strength below $10^{15}\,$G.
The same can be observed in the first two panels of Fig.\,\ref{fig_FFT_average}
for $U_0^{(-)}$, $U_1^{(-)}$, and $U_1^{(+)}$. 
Between $10^{15}$ and $2\times10^{15}\,$G a transition occurs from 
reflection at the crust-core interface (for weaker magnetic fields) and
reflection at the surface of the star (for stronger fields). Therefore,
we do not expect any of the two approximations to agree with the semi-analytic
model.  Neither can we assume that the oscillations
get reflected only at the crust-core boundary nor only at the stellar surface.
When increasing the magnetic field beyond
$2\times10^{15}\,$G the maximum amplitude of the QPO aligns again
towards the polar axis, and the numerically obtained pattern approaches the
expected structure, which is similar to the case without the crust. The main
differences are the presence of the fundamental $U_*^{(+)}$ at finite frequency
and a broader maximum because of the anisotropic shear contribution.
The QPOs are moving from the pole towards
the equator according to the frequency predicted by the semi-analytic model for
reflection at the crust-core interface, but when the effects of the
magnetic
field become comparable to those of the shear modulus in the crust, the QPOs
reach the end of the continuum and \emph{jump} from the symmetric
(antisymmetric) branch
to the antisymmetric (symmetric) branch of the next part of the continuum.

What happens to $U_*^{(+)}$, which has no possibility to jump to? For strong
magnetic fields, $B\gtrsim2\times10^{15}\,$G $U_*^{(+)}$ has different features.
First, there is no node along the field lines, due to the
change of the boundary conditions, which require a maximum at the surface.
Second, as can be seen in the last few columns of the upper row of
Fig.\,\ref{fig_FFT_moving}, there are nodes perpendicular to the field
lines at a given frequency, and we even find two different
contributions to the $U_*^{(+)}$ at slightly different frequencies.
Between $2\times10^{15}\,$G and $5\times10^{15}\,$G there is one node
perpendicular to the field lines, while for
$B\geq6\times10^{15}\,$G the QPO has predominantly two nodes in that
direction. Note that both features are always present for
$B\gtrsim2\times10^{15}\,$G. This splitting can also be seen in the
leftmost three panels of the lower row in Fig.\,\ref{fig_FFT_average}. The
corresponding features are located at frequencies below the fundamental,
$U_0^{(-)}$. The same panels also show, that
with increasing magnetic field strength, the relative Fourier amplitudes and the
frequencies of those features decrease with respect to the fundamental
frequency.

\begin{table*}
  \begin{center}
\begin{tabular}{c | c c c c c c c c c c}
polar magnetic& frequency of fundamental
$U_0$&$U_*^{(+)}/U_0$&$U_0^{(-)}/U_0$&$U_0^{(+)}/U_0$&$U_1^{(-)}/U_0$&$U_1^{(+)}
/U_0$&
$U_2^{(-)}/U_0$&$\delta U_n^{(\pm)}/U_0$\\
field [G]&with semi-analytic model [Hz]& \\ \hline
$2\times10^{14}$&1.0	&1.0&1.9&2.9&3.6&5.0&5.7&$\pm$0.24\\
$4\times10^{14}$&2.0	&1.0&1.8&2.8&3.5&4.9&5.5&$\pm$0.20\\
$6\times10^{14}$&3.0	&1.0&1.8&2.8&3.1&4.7&5.0&$\pm$0.10\\
$8\times10^{14}$&4.0	&1.0&1.8&2.5&3.5&4.3&5.1&$\pm$0.20\\
$10^{15}$	&5.0	&0.9&1.3&2.4&3.3&4.2&5.3&$\pm$0.10\\\hline
$1.5\times10^{15}$&7.4	&0.8&1.1&2.2&3.2&4.1&5.1&$\pm$0.10\\
$2\times10^{15}$&9.8	&0.7&1.0&2.0&3.0&4.0&5.1&$\pm$0.10\\
$3\times10^{15}$&14.7	&0.5&0.9&2.0&3.0&4.0&5.1&$\pm$0.10\\
$5\times10^{15}$&24.4	&0.5&1.0&2.0&3.0&4.0&5.1&$\pm$0.10\\
$8\times10^{15}$&39.1	&0.3&0.9&2.0&3.0&4.1&5.1&$\pm$0.10\\
\hline
\end{tabular}
\end{center}
\caption{The relation of the frequencies of the lowest QPOs for the model
APR+DH 1.4, obtained by analyzing the local maxima of the Fourier amplitudes, to
the fundamental frequency $U_0$, obtained by the semi-analytic method. Note that
the fundamental of the semi-analytic model changes symmetry at a magnetic field
of
about $10^{15}\,$G, such that $U_0\simeq U_*^{(+)}$ for $ B\leq10^{15}\,$G and
$U_0\simeq U_0^{(-)}$ for $B>10^{15}\,$G. 
The last column shows the uncertainty in the separation of two
successive frequencies in the Fourier spectrum. It gives an estimate of the
error caused by choosing the position of the local maxima, but it does not
include other numerical errors.}
\label{tab_freqs}
\end{table*}

The picture of the transition from one asymptotic behavior (reflection at the
crust-core interface) to the other (reflection at the surface) gets supported by
the frequencies obtained for the different QPOs shown in Table\,\ref{tab_freqs}.
There we compare the frequencies corresponding to the maxima of the Fourier
amplitude with those of the fundamental oscillation obtained with the
semi-analytic model. Thus, we use the version with reflection at the
crust-core interface up to $10^{15}\,$G, while we set the boundary at the
surface of the star for stronger fields. For magnetic fields up to
$B\lesssim8\times10^{14}\,$G,
$U_*^{(+)}$ has a similar frequency as the fundamental obtained with the
semi-analytic model. The frequencies of the other QPOs approximately behave as
$1\,(U_*^{(+)}):3\,(U_0^{(+)}):5\,(U_1^{(+)})$ and
$2\,(U_0^{(-)}):4\,(U_1^{(-)}):6\,(U_2^{(-)}$). For stronger
magnetic fields $B\gtrsim3\times10^{15}\,$G the frequencies ratios approach
$1\,(U_0^{(-)}):3\,(U_1^{(-)}):5\,(U_2^{(-)})$ and
$2\,(U_0^{(+)}):4\,(U_1^{(+)})$. 
The two asymptotic integer relations between successive overtones
and their order is what is expected from the semi-analytic model in the
two regimes.
In the intermediate regime  $8\times10^{14}\lesssim B \lesssim3\times10^{15}
\,$G the frequencies change smoothly from one relation to the other.

Deviations from exact integer
ratios may have different reasons.
First, the time of numerical integration is limited. Therefore, the
spectral resolution of the Fourier analysis is limited, too. 
Second, for the lowest magnetic field shown here, $B\approx2\times10^{14}\,$G,
not all upper QPOs
have reached their position near the polar axis, i.e. their frequencies
still lie in the continuum, resulting in lower frequencies. 
Third, in particular in the transition regime it is sometimes difficult to
identify where the maximum of a QPO is located. The interesting QPO
may be excited only very weakly by our initial data, and/or some other QPO may
be excited more strongly at a similar frequency. This occurs more frequently for
higher overtones, because there the different continua overlap (see
Fig.\,\ref{fig_FFTlines} or \ref{fig_FFT_average}).
\subsection{Crustal modes in the gaps of the Alfv\'en continuum?}

\cite{vanHoven2011} and \cite{Colaiuda2011} have pointed out the possibility of
crustal modes, which may have frequencies outside of
the continuum of the core. These modes would be only very weakly
coupled to core oscillations, because no
Alfv\'en wave of any field line could match the necessary frequency.
In the models we have studied here, we find gaps in the continuum only between
the lowest overtones of Alfv\'en oscillations (see
Fig.\,\ref{fig_FFTlines}), e.g. for model APR+DH 1.4 already the continua of
$U_1^{(-)}$ and $U_1^{(+)}$ overlap
and there are only gaps between $U_1^{(-)}$ and $U_0^{(+)}$, $U_0^{(+)}$ and
$U_0^{(-)}$ and below $U_0^{(-)}$. For the other models shown in the
Table \ref{crust_modes} the number of gaps is limited to a maximum of about
three for models with the crust EoS NV and to two for models with the
DH EoS. Note that we only consider the continua of the open field lines
indicated by the black lines in Fig.\,\ref{fig_FFTlines}, because they are
decoupled from the continua related to the closed field lines (green lines). To
have the fundamental $n=0$, $l=2$ oscillation of the crust in one of the gaps,
very strong magnetic fields $B>10^{15}\,$G are required.

\begin{figure}
\begin{center}	
 \includegraphics[width=.4\textwidth]{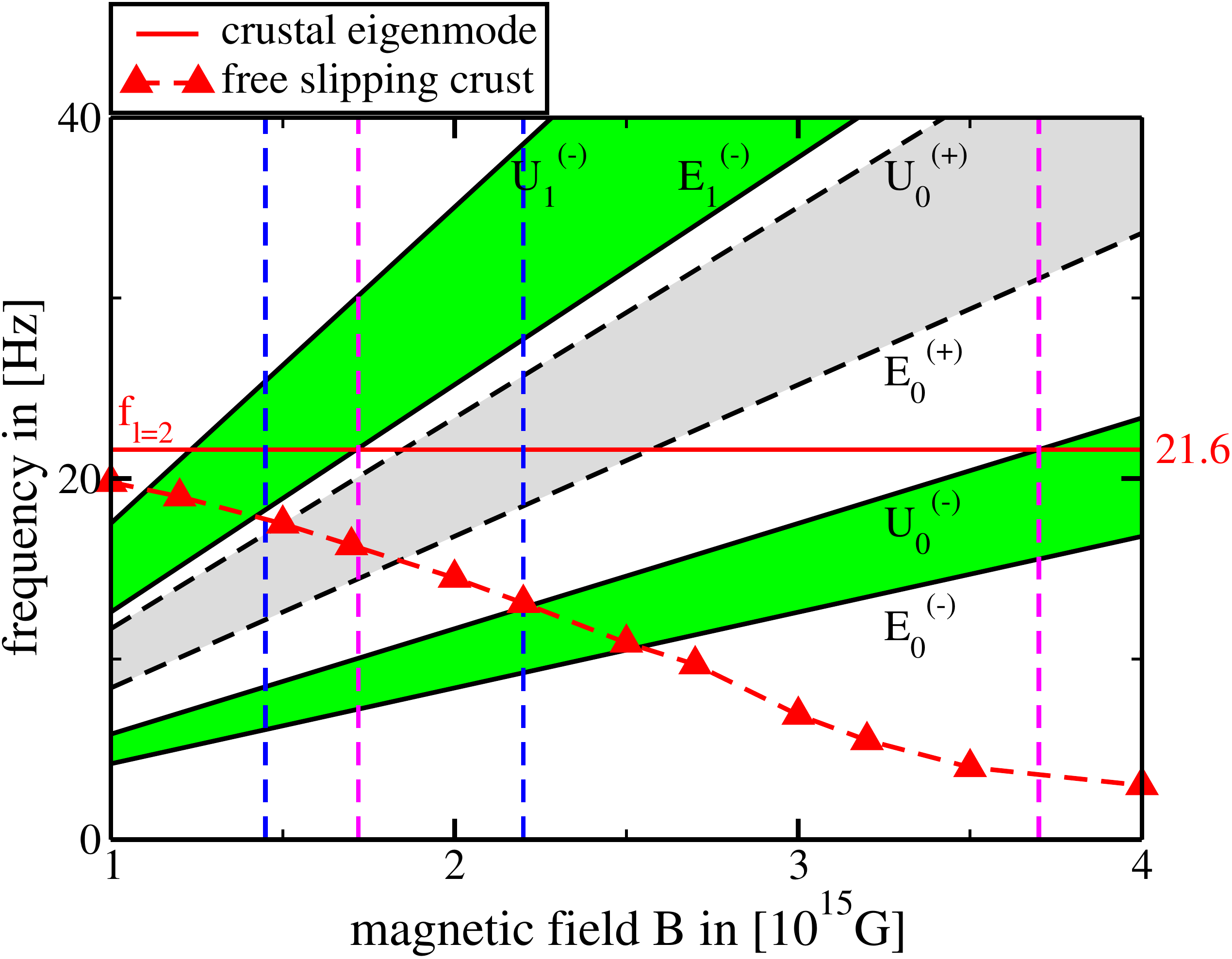}
\end{center}
\caption{Frequency continua of model L+DH 1.4 at magnetic fields between
$B=10^{15}\,$G and $B=4\times10^{15}\,$G.
Black, solid lines indicate the edges of the continua of the open field lines as
indicated by the names $E_n^{(\pm)}$ and $U_n^{(\pm)}$. The horizontal red
line gives the frequency of the $l=2$ crustal
shear mode and the dashed red lines gives the frequency obtained by
simulations with a free slipping crust (see text). The green shaded areas show
continua having the same symmetry with
respect to the equator as the $l=2$ shear mode. The grey area shows the
continuum with opposite symmetry.  Magenta (blue) dashed lines show the
range of magnetic field strengths where the frequency of the $n=0$, $l=2$
crustal mode lies in the gap between allowed Alfv\'en continua for purely shear
modes (for free slipping crust with magnetic field).} \label{fig_gap_freq}
\end{figure}

As an example we choose the EoS L+DH 1.4, because the corresponding model has
broader gaps than the APR+DH EoS. Furthermore, models with lower mass have
shorter Alfv\'en crossing times, and therefore weaker magnetic fields are
necessary to have the shear mode in the gap between the lowest overtones of
the continua (see Table\,\ref{tab_matchfreq}). 
The spectral structure of this model is displayed in
Fig.\,\ref{fig_gap_freq}, where we show the edges of the continua of the open
field lines as a function of the magnetic field strength. The shaded areas
between two
edges represent the corresponding continuum, where crustal modes can be absorbed
resonantly. The red line indicates the frequency $21.6\,$Hz of the purely shear,
$n=0$, $l=2$ mode of the crust.
Using model L+DH 1.4, we performed simulations with initial data with the
crustal $n=0$, $l=2$ mode, allowing only for antisymmetry with
respect to the equatorial plane.  The oscillations of the continuum associated
with symmetric QPOs, as for example $U_0^{(+)}$, are thus not allowed and cannot
be
excited.
These forbidden oscillations are indicated by the grey shaded region in Fig.
\ref{fig_gap_freq}. We also confirmed that without imposing equatorial symmetry
only Alfv\'en QPOs having the same symmetry as the
corresponding crustal mode can be excited with significant amplitude during the
evolution.
Therefore, the antisymmetric $n=0$, $l=2$ shear mode of the crust lies in the
gap between
$U_0^{(-)}$ and $E_1^{(-)}$ for magnetic fields between $1.74$ and
$3.7\times10^{15}\,$G (Fig. \ref{fig_gap_freq}).
Following previous works by \cite{vanHoven2011} or \cite{Colaiuda2011} the
crustal mode in the gap should not be damped significantly, because there is no
oscillation at the resonant frequency available in the continuum.

\begin{figure}
\begin{center}	
 \includegraphics[width=.4\textwidth]{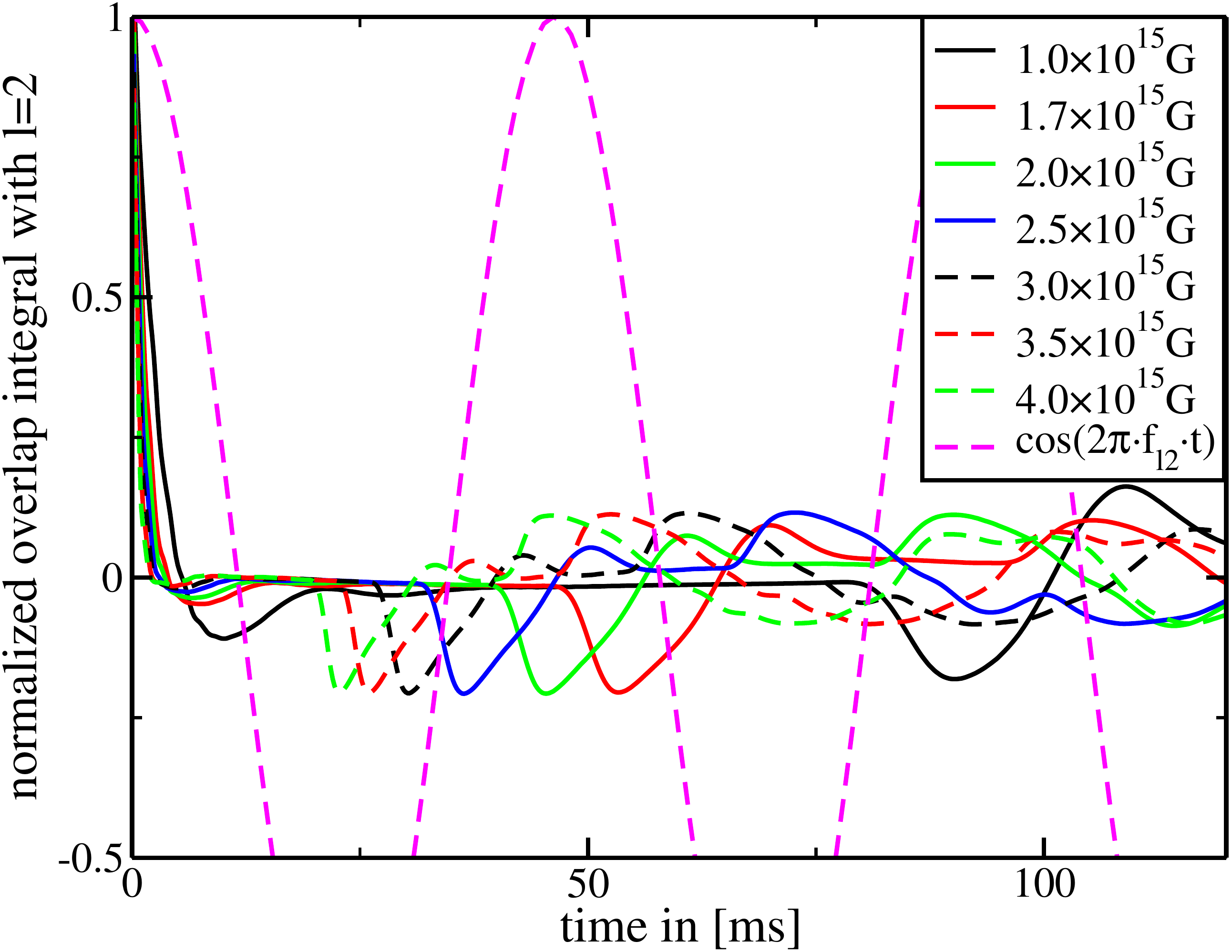}
\end{center}
\caption{Evolution of the overlap integral of the $n=0$, $l=2$ mode of the crust
at different magnetic field strengths: $B=$\{1.0, 2.0, 2.5, 3.0, 3.5,
4.0\}$\times10^{15}\,$G. The dashed
magenta line shows the expected oscillation of the purely shear $l=2$ mode of
the crust. All other lines show strong damping of the excited
$l=2$ mode at early times ($t<20\,$ms). Contributions to the overlap integral at
later times $t>20\,$ms originate from magneto-elastic oscillations.}
\label{fig_damping_LDH14}
\end{figure}
However, we still find very strong damping of this crustal
mode in our simulations, as indicated by the overlap integral of the
$n=0$, $l=2$ shear mode in Fig. \ref{fig_damping_LDH14}. 
Clearly the mode is damped after a few msec for all considered magnetic field
strengths. Later contributions originate from coupled magneto-elastic pulses
travelling through the whole star. Therefore, the time
when they contribute to the overlap integrals depends inversely on the magnetic
field strength, a behavior which is completely different from that of discrete
oscillations for purely crustal shear modes.
These observations indicate that we do not have a weakly coupled system of
two sub-systems (the crust and the core)
but we are dealing with coupled,
global magneto-elastic oscillations.
To compare with what one would expect for an undamped purely crustal
mode, we plotted the magenta, dashed line in Fig.\,\ref{fig_damping_LDH14}.

We checked the influence of the magnetic field on the frequency of the
purely shear mode
by performing a series of simulations, where
we apply an artificial boundary condition at the crust-core interface, i.e.
we use the same condition $\xi^\varphi_{,r}=0$ that would apply in the absence
of the magnetic field \citep[see also][]{Sotani2007}. This
allows the crust to slip freely on top of the core, and the oscillations inside
the crust cannot be damped into the core. The frequency of
the corresponding $l=2$ crustal mode is influenced as displayed in
Fig.\,\ref{fig_gap_freq}.
This is expected, because global, poloidal
magnetic fields of the order of $B\sim10^{15}\,$G and stronger begin to have
measurable effects \citep{Duncan1998, Messios2001, Sotani2007, Sotani2008b,
Shaisultanov2009}. Intuitively, one would expect the shear mode frequencies to
increase with the magnetic field strength, because the magnetic tension could
be interpreted as effectively augmenting the shear modulus
\citep{Messios2001,Sotani2007}.  
In contrast to \cite{Sotani2007} we find a decrease of the $l=2$ mode
frequency with increasing dipolar magnetic field strength (see
Fig.\,\ref{fig_gap_freq}). In their work \cite{Sotani2007} neglected couplings
between $l$- and $l\pm2$-modes 
due to the magnetic field, which led to discrete modes. However,
by analyzing the evolution of the corresponding overlap integrals we
observe strong excitation of the $l=4$ mode by the $l=2$
mode in our simulations. This strong coupling may explain the
 opposit change of frequency than expected from the study by \cite{Sotani2007}.
For a different magnetic field configuration also \cite{Messios2001} and
\cite{Sotani2007} find that the frequency of the fundamental $l=2$ crustal mode
decreases with increasing magnetic field strength.

If the frequency of the $l=2$ crustal mode changes according to our simulations
with the free slipping crust, the magnetic field strengths for which the mode
lies in the continuum gap is limited to $1.4\times10^{15}$ to
$2.2\times10^{15}\,$G. In this regime we have performed four simulations at
$1.5\times10^{15}$, $1.7\times10^{15}$, $2.0\times10^{15}$, and
$2.2\times10^{15}\,$G (triangles in Fig.\,\ref{fig_gap_freq}). In all of them
we find the strong damping of the crustal mode.

A second problem of matching of crustal frequencies into the gaps of
the Alfv\'en continuum emerges at the field strengths ($\gtrsim 10^{15}\,$G) at
which we find the crustal frequencies in the continuum gaps, as there is no
clear way of how to compute the Alfv\'en frequencies. In this
transition regime,
reflection neither occurs predominantly at
the crust-core interface nor at the surface, therefore, we do not
claim that the continuum shown in Fig. \ref{fig_gap_freq} is perfectly 
valid at all magnetic field strengths (compare the panels for
$1.5\times10^{15}\,$G to those of $3\times10^{15}\,$G in Fig.
\ref{fig_FFT_average}).  However, by performing simulations for 10
different magnetic field strengths between $10^{15}$ and
$4\times10^{15}\,$G for the current model L+DH 1.4, we can ensure that the $l=2$
crustal frequency lies in the gap between the Alfv\'en continua at least for one
of the models. In none of the above
simulations we find a different behavior than the one reported, i.e. we do not
observe any of the crustal shear modes. For crust models with
lower shear modulus, we expect the transition to occur at lower magnetic
field strength. The magnetic field required to fit a crustal shear mode into the
gap between successive continua decreases with decreasing shear modulus.
However, a smaller shear modulus also lowers the relative importance
of the shear terms compared to the magnetic field.

In the example shown above, we may have just missed to match the
frequency of the crustal mode to a gap of the continuum, as the frequency may
have been changed due to the presence of the strong magnetic field, and
because the continuum is probably not reliably predicted by the
semi-analytic model. To this end we performed a large number ($>50$) of
simulations at different magnetic field strengths $>10^{15}\,$G and different
equilibrium models but we never found any crustal shear mode at such strong
magnetic fields. 
 
Generalizing the dipolar magnetic field configuration, which is our main model 
simplification in the current context, would
probably increase the complexity of the continuum, making it even harder to find
gaps. However, there might arise new effects due to an entanglement of the
magnetic field \citep[see][]{vanHoven2011}. 

Crustal modes in the gaps of the Alfv\'en continua of the core have been
reported by \cite{Colaiuda2011}, but our results suggest that these QPOs are
rather oscillations of the continuum. 
It is possible that the mode recycling
technique (using the amplitude of the FFT at the frequency of the crustal shear
modes) employed by \cite{Colaiuda2011} may give an excitation of the continuum
at this frequency. 
\cite{vanHoven2011b} also observe gap modes. In their approach
several 1-dimensional eigenmodes of the magnetic field in the core are coupled
to 2-dimensional eigenmodes in the crust. However, they do not
prove that their assumption that the crustal dynamics can be described with
an eigenvalue problem is valid also in the presence of a magnetic field. Thus,
we expect that the ansatz of \cite{vanHoven2011b} is valid only approximately
for weak magnetic fields ($B<10^{15}\,$G).
Such an eigenmode approach cannot describe the complicated coupling behavior,
including e.g. travelling wave packets, which would require the inclusion of a
much larger number of coupled oscillators than those considered by
\cite{vanHoven2011b}.

\subsection{Threshold for the outbreak of the QPOs through the crust}
In Section\,\ref{strong_fields} we noticed that for weak magnetic fields,
$B\lesssim 10^{15}\,$G, the QPOs are largely confined to the fluid
core, and
that there exists a threshold magnetic field strength beyond which QPOs can be
observed with significant amplitudes at the surface of the star.
To quantify when magneto-elastic QPOs have a significant amplitude in
the crust of the  neutron star, we studied their maximum amplitude
at the surface. 

\begin{figure}
\begin{center}	
 \includegraphics[width=.47\textwidth]{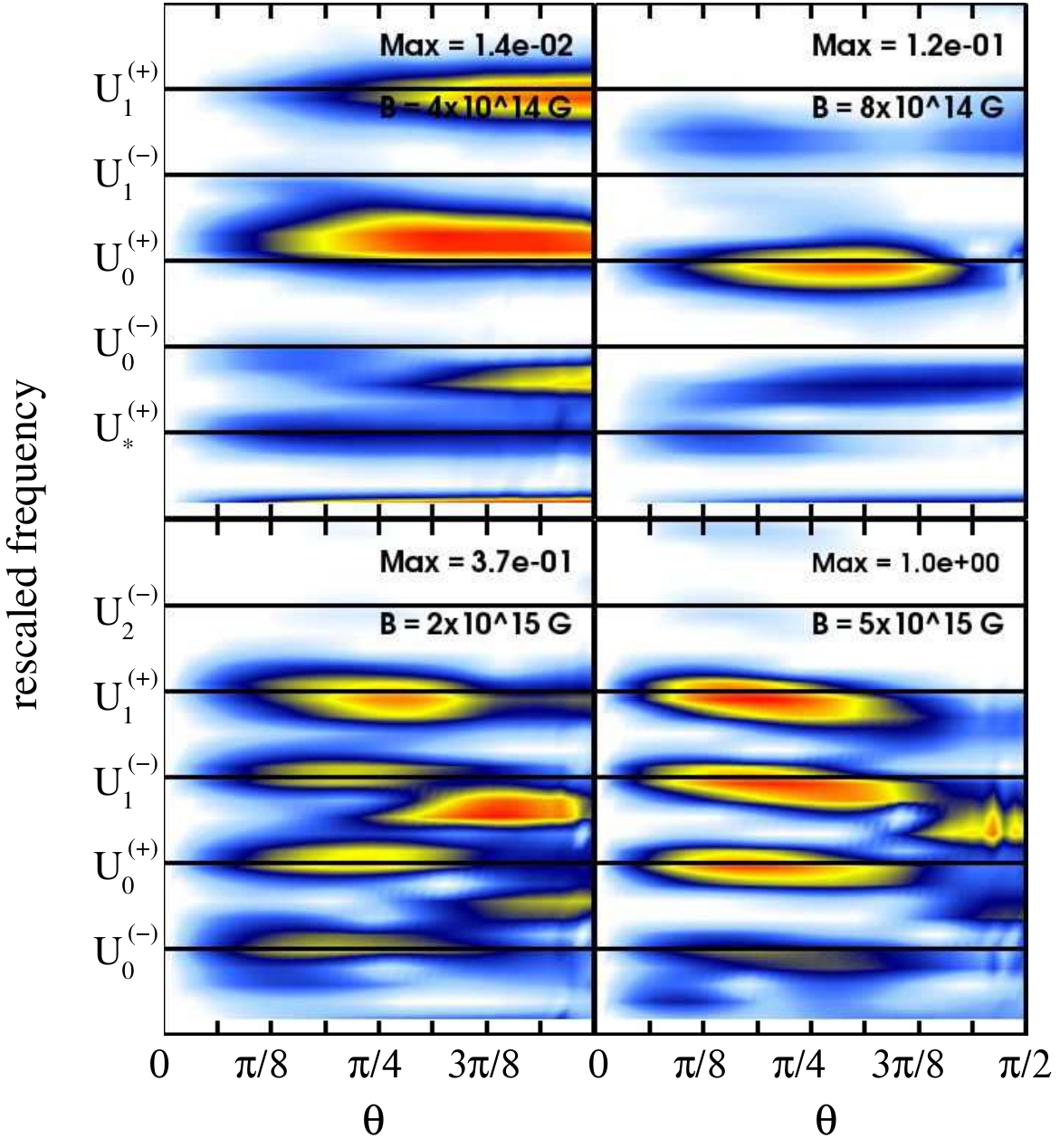}
\end{center}
\caption{Normalized amplitude of the Fourier transform at the surface of the
star (between pole and equator) as a
function of the frequency for the stellar model APR+DH with 1.4 solar masses.
The horizontal lines indicate the frequency for the
upper QPOs predicted by the semi-analytic model. The color scale ranges from
white-blue (minimum) to orange-red (maximum).} \label{fig_FFTsurface}
\end{figure}
In Fig.\,\ref{fig_FFTsurface} we plot the amplitude of the
Fourier transform at the surface for different magnetic field strengths as a
function of the polar angle $\theta$ and the frequency. 
The color scaling is rescaled in each panel
to enhance the main contributions at each magnetic field
strength. We consider APR+DH 1.4 as our reference model.
First we note, that for very strong fields ($>5\times10^{15}\,$G), 
the frequencies approach those predicted by the semi-analytic model,
because the influence of the crust should decrease for increasing magnetic field
strength. There is some low frequency oscillation which
corresponds to $U_*^{(+)}$ (compare with Fig.\,\ref{fig_FFT_average}), and
there are some additional, strong Fourier modes near to $\pi/2$ resulting from
the edge modes, which are stronger than in the case without crust.

However, when decreasing the strength of the magnetic field from
$5\times10^{15}\,$G to $2\times10^{15}\,$G, the Fourier amplitude of the QPOs
at the surface decreases, e.g. the maximum decreases from  $1.0$ to
$0.37$ in units normalized to the maximum amplitude at
$B=5\times10^{15}\,$G. As we have already seen in
Fig.\,\ref{fig_FFT_moving},
this is expected, because for decreasing field strength the crust will
shield the QPOs. We further see that the additional structure in the
Fourier amplitude at about the fundamental frequency of $U_0^{(-)}$ and the
strong feature
just below the frequency of $U_1^{(-)}$ close to the equator are
increasing in amplitude relative to the upper QPOs. We note here
that  in the transition region of magnetic field strengths
the correspondence between the frequencies of the
semi-analytic model and the simulated QPOs should be taken with
caution  and we do not expect a simple QPO structure as in the two
limiting cases.

When decreasing the magnetic field strength towards $8\times10^{14}\,$G a
dominant feature originates from $U_0^{(+)}$. In a sequence of similar
plots for different magnetic fields not shown here, one can follow the
slight change in frequency until reaching approximately the value predicted by
the semi-analytic model for reflection at the crust-core interface. Comparing
with Fig.\,\ref{fig_FFT_moving} the corresponding spatial structure
of the mode has still some amplitude inside the crust, i.e. our interpretation
in terms of the $U_0^{(+)}$ QPO makes sense. The general trend is that the
magneto-elastic upper QPOs exhibit a decreasing amplitude near the surface for
decreasing magnetic field strength.

For the weakest magnetic field ($4\times10^{14}\,$G), the upper QPOs have no
strong amplitudes, because they are confined to the fluid core. The 
dominant contribution at the surface results
from the edge modes in this case (see Fig.\,\ref{fig_FFT_4_14}). 
The QPO with the largest amplitude inside the crust is the edge QPO $E_2^{(+)}$
(see Fig.\,\ref{fig_FFTsurface}) 
with a frequency slightly above $U_0^{(+)}$ (see also the right
panel of Fig.\,\ref{fig_FFTlines}). Other
contributions to the Fourier signal at the surface stem from the two edge QPOs
$E_1^{(+)}$ and $E_3^{(+)}$ 
at frequencies just below $U_0^{(-)}$ and close to
$U_1^{(+)}$, respectively. However, the edge
QPOs are damped much faster than turning-point QPOs, and the amplitude in the
Fourier analysis for the same initial data is about two orders of magnitude
smaller than for a reference model without crust. Therefore we doubt
that edge QPOs are the explanation for the observed frequencies in SGR
for magnetic fields $B\lesssim5\times10^{14}\,$G. 

\begin{table}
  \begin{center}
\begin{tabular}{c c c c}
Model&$\frac{\mathrm{FFT
(10^{15}\,G)}}{\mathrm{FFT\,{(no\,crust)}}}$&
Model&$\frac{\mathrm{FFT\,(10^{15}\,G)}}{\mathrm{FFT\,(no\,crust)}}$\\
\hline
APR+DH 1.4&0.07&APR+NV 1.4&0.06\\
APR+DH 1.8&0.11&APR+NV 1.8&0.005\\
APR+DH 2.2&0.33&APR+NV 2.2&0.04\\
L+DH 1.4&0.16&L+NV 1.4&0.016\\
L+DH 1.8&0.36&L+NV 1.8&0.03\\
L+DH 2.2&0.15&L+NV 2.2&0.009\\ 
\hline
\end{tabular}
\end{center}
\caption{The maximal magnitude of the Fourier transform at
the surface of the star for a dipole magnetic field strength 
of $10^{15}$\,G.
Simulations for different EoS, and with and without crust are considered 
and compared.}
\label{threshold}
\end{table}
In Table \ref{threshold} we give the maximum of the Fourier amplitude at the
surface for a simulation with crust at $10^{15}\,$G with respect to the
corresponding simulation without crust. The closer this value is to $1.0$ the
stronger is the amplitude of the QPO at the surface. For very low values the
crust shields the QPOs efficiently, i.e. they are confined to the core of the
neutron star. For the DH crustal EoS there is already a considerable amount of
oscillations penetrating the crust and reaching the surface. We therefore argue
that magneto-elastic oscillations break through the crust around
$10^{15}\,$G. However, for the NV EoS the amplitudes for models with crust
never reach 10\% of the values for models without crust, as the crust is
more extended in this case, and QPOs can break through the crust only for even
stronger magnetic fields. Nevertheless, the threshold of
$B\sim10^{15}\,$G should be a good approximation for all models.
Interestingly, this result is comparable with estimates of
the magnetic field strengths for SGRs showing giant flares.

\begin{table}
  \begin{center}
\begin{tabular}{c c c}
Model&$B_\mathrm{30Hz} $[$10^{15}\,$G]&$B_\mathrm{28Hz} $[$10^{15}\,$G]\\ \hline
APR+DH 1.4&$5.0$&4.6\\
APR+DH 1.8&$7.1$&6.6\\
APR+DH 2.2&$11.0$&10.3\\
L+DH 1.4&$4.1$&3.8\\
L+DH 1.8&$5.5$&5.1\\
L+DH 2.2&$6.4$&6.7\\
\hline
APR+NV 1.4&5.1&4.8\\
APR+NV 1.8&7.4&6.9\\
APR+NV 2.2&11.3&10.6\\
L+NV 1.4&4.6&4.4\\
L+NV 1.8&6.2&5.8\\
L+NV 2.2&8.1&7.6\\
\hline
\end{tabular}
\end{center}
\caption{ Mean surface magnetic field strength required to match the
frequency of the fundamental Alfv\'en QPO to $30\,$Hz observed in SGR 1806-20
and $28\,$Hz of SGR 1900+14.}
\label{tab_matchfreq}
\end{table}
%
\section{Summary and discussion}\label{sec_discussion}
We have presented results from 2-dimensional, general-relativistic, 
magneto-hydrodynamical simulations of neutron stars with an extended solid
crust, in continuation of our initial results communicated as a letter
\citep{Gabler2011letter}. Performing a comprehensive set of simulations for
several neutron star models and EoS, we have been able to confirm our previous
findings regarding the QPO structure for  three different regimes of the
magnetic field strength. Our main results can be summarized as follows:

\begin{itemize}

\item
For weak magnetic fields, $B\lesssim5\times10^{13}\,$G, purely shear
oscillations of the crust dominate the evolution in the latter. For intermediate
magnetic field strengths, $5\times10^{13}\lesssim B\lesssim10^{15}\,$G, the
$n=0$ crustal modes are damped very efficiently into the core of the neutron
star on timescales of a fraction ($\sim 0.04$)
of the Alfv\'en crossing time of the star. For example, a model with
$B=10^{14}\,$G has
a damping timescale of  $\tau<100\,$ms. This effectively rules out purely shear
oscillations of the crust as a possible explanation for the observed QPOs in
SGRs for the poloidal magnetic field configurations studied here,
since the observed QPOs survive for tens of seconds at 
estimated magnetic field strengths
$B\gtrsim6\times10^{14}\,$G. We find a damping timescale dependence
on the magnetic field which scales as $\sim B^{-1}$,
dominated by the ability
of the Alfv\'en continuum to absorb the energy of crustal modes.

\item
In comparison to the $n=0$ modes, the $n>0$ modes of the crust
have damping timescales of order hundreds of milliseconds at around
$5\times10^{14}\,$G. The spatial structure of the $n>1$ modes 
becomes significantly distorted in the presence of such a strong magnetic 
field.  However, predictions for even stronger magnetic
fields, $B>5\times10^{14}\,$G, are currently not possible, because the grid
resolution needed to couple the $n>0$ shear modes to Alfv\'en oscillations in
the core is too high to perform simulations in reasonable time.  
We thus cannot safely exclude the $n>0$ crustal shear modes as
possible explanation for the high frequency QPOs observed in SGR 1806-20 at
$625$ and $1840\,$Hz, if the magnetic field is a purely dipolar one.

\item
For magnetic fields, $B\gtrsim5\times10^{14}\,$G, we find no sign
of the existence of discrete crustal shear modes. This is in contrast 
to the results of the model of \cite{vanHoven2011}, who proposed that weakly
damped global QPOs, resembling purely crustal shear modes in the zero
magnetic field limit, may survive if the corresponding shear mode frequency lies
in-between adjacent continua. In
the most promising of our models we find gaps between the first four continua,
but even when choosing the equilibrium model so that the fundamental crustal
mode frequency falls in
one of the existing gaps, we found that the mode
was damped very efficiently by the coupling to the Alfv\'en continuum
in the core. 
Furthermore, at dipolar magnetic field strengths $B\gtrsim10^{15}\,$G,
necessary to fit the frequencies of crustal shear modes into a gap, we expect
that the
magnetic field begins to have significant influence on the shear oscillations of
the crust \citep{Messios2001,Sotani2008b,Shaisultanov2009}.
Our findings strongly support the interpretation that for high magnetic
field strengths magnetar oscillations are a
strongly-coupled magneto-elastic system, where a 
division into purely
crustal modes and Alfv\'en oscillations is no longer valid.

\item
In the intermediate magnetic field regime ($5\times10^{13}\,$G$\lesssim B
\lesssim 10^{15}\,$G) the QPOs are largely confined to the core of
the neutron star. We find three families of QPOs: upper, edge and lower
QPOs. Their spatial structure coincides very well with the
expectations from our semi-analytic model, if we assume that the oscillations
are reflected at the crust-core boundary. 
Together with the strong damping of
the crustal shear modes, the reflection of Alfv\'en QPOs at the crust-core
interface leads to very small oscillation amplitudes in the crust. 
Moreover, when changing the magnetic field
strength, the position of the maximum of the corresponding upper QPO within 
the star changes significantly due to the interaction with the crust. 

\item
We have also determined the dipolar magnetic field strength at which the 
magneto-elastic QPOs break through the
crust and reach the surface with significant amplitudes. 
This happens around $B\sim10^{15}\,$G for the DH crust EoS and 
at slightly stronger magnetic fields for the NV crust EoS. 
This difference can be understood easily, because the NV EoS leads to 
thicker crusts and larger shear moduli, in particular near the crust-core
interface. 

\item
Between $10^{15}\,$G$\lesssim B
\lesssim 5 \times 10^{15}\,$G the QPOs are still transforming
from being confined to the core to being able to reach the surface  and the
QPOs have spatial structures different from those at
much lower or much higher magnetic field strengths.
\cite{Colaiuda2011} report global, discrete Alfv\'en modes 
in gaps between continua, at magnetic fields around $4\times10^{15}\,$G. 
In our model, this is still in the above transition regime,  
and the reported oscillations are not discrete Alfv\'en modes,
but rather an effect of the transition between the two limiting regimes.

\item
 For dipolar magnetic field strengths $B
\gtrsim 5 \times 10^{15}\,$G the 
magneto-elastic oscillations have an almost Alfv\'en-like
character in the whole star and the role of the shear modulus 
in the crust is diminished, recovering the results of 
\cite{Sotani2008} and \cite{Cerda2009}.

\end{itemize}

The model we have presented allows for a 
tentative interpretation of the observed QPOs in
terms of the predominantly Alfv\'en QPOs which reach the surface. The 
 main family of QPOs we expect to  play a role in the observed QPOs
are the upper (turning-point) QPOs. 
As first pointed out by \cite{Sotani2008b}, 
the interpretation
of the frequencies $30$, $92$ and $150\,$Hz in SGR 1806-20 in terms of the
 $U_0^{(-)}$, $U_1^{(-)}$ and $U_2^{(-)}$  QPOs is very tempting. 
Here we note that the $18\,$Hz
oscillation may be interpreted as the first edge QPO $E_0^{(-)}$, which for 
model APR+DH has a frequency of about
$0.57\times f_{U_0^{(-)}} \approx 17\,$Hz \citep[this
could also correspond to the frequency at $16.9\,$Hz found by][]{Hambaryan2011}.
Similarly, the $36.4\,$Hz oscillation 
could be interpreted as the $E_1^{(-)}$ QPO.
However, one should be cautious with these latter
identifications, since edge QPOs are not as long-lived as turning-point 
QPOs.

 If we require that the fundamental upper QPO matches the observed 
30 Hz QPO in SGR 1806-20 or the 28 Hz QPO in SGR 1900+14, we obtain 
mean dipolar magnetic field strengths at the surface in the possible 
range $ 3.8 \times 10^{15}\,$G$\lesssim B \lesssim 1.1 \times 10^{16}\,$G
(for the particular choices of EoS and masses)
as reported in Table \ref{tab_matchfreq}. This is a rather narrow 
range and it is at only somewhat larger magnetic field strengths than
simple estimates for magnetic fields in known magnetars.

We conclude by highlighting that our model provides two 
different constraints on the magnetic field strength. 
The first constraint is that the 
dipolar magnetic field strength must be larger than $B\sim10^{15}\,$G 
for QPOs to break out of the crust, which is a lower limit on the magnetic
field. This constraint is independent of a particular identification of observed
QPOs and only depends on the assumed magnetic field structure of a pure dipole.
The second constraint comes from the matching of the lowest-frequency 
observed QPO, that appears at near-integer multiples,  
with the fundamental $U_0^{(-)}$ QPO. This constrains the 
mean dipolar magnetic field strength at the surface of the neutron
star to be in the range of $ 3.8 \times 10^{15}\,$G$\lesssim B \lesssim 1.1
\times 10^{16}\,$G for
the sample of EoSs and various masses that we assumed here.

The above constraints, favoring somewhat stronger magnetic fields
than estimated for known magnetars, hint at a possible deviation of
the actual magnetic field structure from a global dipole. This 
is not surprising, as it is known that a purely dipolar or purely toroidal 
field is not a stable magnetic field configuration in compact stars
\citep[see][and references therein]
{Braithwaite2006b,Lasky2011,Ciolfi2011,Kiuchi2011}. We may thus have an
observational indication that the 
structure of the magnetic field in magnetars is in fact more complicated 
than a pure dipole. Other physical effects which may cause the
magnetic field required to match the magneto-elastic QPOs to the range of
observed frequencies to be weaker, are superfluidity of the neutrons and
superconductivity of the protons in the core.

 We are planning to investigate the effects produced by changing 
the magnetic field configuration and those caused by superfluidity and
superconductivity of the neutrons and protons in the core, which are
expected to influence the Alfv\'en speeds and the
overall dynamics \citep{Passamonti2011}. For the former, we need to investigate
different dipolar configurations, mixed poloidal-toroidal configurations,
relax the assumption of axisymmetry and include the coupling to polar
oscillations. Moreover, the coupling to an exterior magnetosphere, where the
X-ray emission is modulated, also has to  be included in a complete model.

\section*{Acknowledgements}
We thank A. Colaiuda for fruitful discussions about the interpretation of
the simulations and F. Lamb for drawing our attention
to the possibility of different realizations of dipolar magnetic fields.
This work was supported by the Collaborative Research Center on Gravitational
Wave Astronomy of the Deutsche Forschungsgemeinschaft (DFG SFB/Transregio 7),
the Spanish {\it Ministerio de Educaci\'on y Ciencia} (AYA 2010-21097-C03-01)
and the {\it Generalitat Valenciana} (PROMETEO-2009-103),
a DAAD exchange grant for MG for an extended visit at the
Aristotle University of Thessaloniki
and by CompStar, a Research Networking Programme of the European Science 
Foundation. The computations were performed at 
the {\it Servicio de Inform\'atica de la Universidad de Valencia}.

\appendix
\section{Alternative numerical method}\label{appendix_linear}
In this appendix we derive a wave equation to describe the coupled
crust-core oscillations of magnetars. To this end we expand all dynamical
variables $f (\mathbf{r},t)$ into a
static unperturbed part $\hat f(\mathbf{r})$, denoted by a caret, and a
time-dependent perturbation $\delta f(\mathbf{r},t)$:
\begin{equation}
 f (\mathbf{r},t) = \hat f (\mathbf{r}) + \delta f (\mathbf{r},t)\,.
\end{equation}
For clarity we will omit the arguments $\mathbf{r}$ and $t$ from now
on.

In the subsequent subsections we derive the equation for the displacement,
describe the numerical implementation and compare the performance of this
alternative method with the standard approach using Riemann solvers on both
sides of the crust-core interface. 

\subsection{Linear wave equation}
To derive the linearized wave equation governing the evolution of the
displacement in the crust of a magnetized neutron star, we project
the conservation equation of energy-momentum
\begin{eqnarray}
h^\mu_{\,\beta} T^{\beta\nu}_{\,\,\,;\nu} &=& 0\,, \label{Ap_conservation}\\
(\rho h + b^2 ) u^\mu_{\,;\nu} u^\nu &=& - h^{\mu\nu} \left ( P + \frac{1}{2}
b^2 \right)_{;\nu} \nonumber\\
&&+ h^\mu_{\,\beta} (b^\beta b^\nu -
2\mu_\mathrm{S}\Sigma^{\beta\nu})_{;\nu}\,,
\end{eqnarray}
with $h^{\mu\nu} = g^{\mu\nu} + u^\mu u^\nu$ and apply the following
simplifications: (i) we linearize in the perturbations $\delta f$ and (ii)
neglect all metric perturbations (Cowling approximation, $\delta g = 0$).
Then Eq.\,(\ref{Ap_conservation}) reads
\begin{eqnarray}
&&\hspace*{-.5cm}(\hat \rho \hat h + \hat b^2) \delta u^\mu_{\,;\nu} \hat u^\nu
= \nonumber\\
&& -\left(\delta \rho \hat h + \hat \rho
\delta h + 2 \hat b_\beta \delta b^\beta\right) \hat u^\mu_{\,;\nu} \hat
u^\nu - (\hat \rho \hat h + \hat b^2) \hat u^\mu_{\,;\nu} \delta u^\nu
\nonumber\\
&&+\left(\hat u^\mu \delta u_\beta + \delta u^\mu \hat u_\beta\right) \left[
\hat b^\beta
\hat b^\nu - \hat g^{\beta\nu} \left( \hat P + \frac{1}{2} \hat b^2\right)
\right]_{;\nu} \nonumber\\
&&+\hat h^\mu_{\,\beta}\left[ \hat b^\beta \delta b^\nu + \delta b^\beta
\hat b^\nu - \hat g^{\beta\nu} (\delta P + \hat b_\beta \delta b^\beta)
\right]_{;\nu}\nonumber\\
&&- 2 \hat h^\mu_{\,\beta}\left[\delta  \mu_\mathrm{S} \hat \Sigma^{\beta\nu}
+ \hat \mu_\mathrm{S} \delta \Sigma^{\beta\nu} \right]_{;\nu}\,.
\end{eqnarray}
Next we restrict ourselves to (iii) axisymmetry ($f_{,\varphi} = 0$),
and (iv) axial perturbations ($\delta
u^t=\delta u^r=\delta u^\theta = 0$, $\delta b^r = \delta b^\theta = 0$, and
$\delta \mu_{\mathrm{S}}=\delta h = \delta \rho = \delta P =0$).
Furthermore, we consider (v) the spherical symmetric, non-rotating
background described by the line element 
$ds^2 = -\hat \alpha^2 dt + \hat \Phi^4 (dr^2 + r^2 d\theta^2 + r^2
\sin(\theta)^2 d\varphi^2)$,
and (vi) purely poloidal background fields ($\hat b^\varphi=0$). Applying all
these simplifications one arrives at the following equation for $\delta
u^\varphi$:
\begin{eqnarray}\label{eq_appendix_perturbation}
&&(\hat \rho \hat h+\hat b^2) \hat u^t \delta u^\varphi_{\,,t}=\hat b^r \delta
b^\varphi_{\,,r} +
\hat b^\theta \delta
b^\varphi_{\,,\theta} - 2 \hat \mu_\mathrm{S} \delta
\Sigma^{\varphi\nu(s)}_{\,\,;\nu} \nonumber\\
&&\hspace{.8cm}+\delta b^\varphi \left[
\left(\frac{2\hat \Phi_{,r}}{\hat \Phi}+\frac{2}{r} +
\frac{\hat\alpha_{,r}}{\hat\alpha}  \right)
\hat b^r+2 \cot(\theta) \hat b^\theta 
\right]\,.
\end{eqnarray}
Because of the dependence of $\Sigma^{\mu\nu}$ on the displacement
$\xi^\varphi$, we express all other quantities in terms of the latter.
Recalling the definition of the corresponding velocity
$\xi^\varphi_{,t} = \delta u^\varphi / \hat u^t$, see  Eq.\,(\ref{def_xidot}),
the perturbed magnetic field $\delta b^\varphi$ remains the only
missing ingredient. 

To find an expression relating $\delta b^\varphi$ to $\xi^\varphi$, we
contract the Faraday equation 
\begin{equation}\label{eq_appendix_faraday} 
\left( u^\mu b^\nu - u^\nu b^\mu \right)_{;\mu} = 0,
\end{equation}
 with $u_\nu$ and obtain
\begin{equation}
 u_\nu b^\nu_{\,;\mu} u^\mu = u^\nu_{\,;\mu} u_\nu b^\mu - b^\mu_{\,;\mu}\,,
\end{equation}
where we have used $u_\nu u^\nu = -1\,$ and $u_\nu b^\nu=0\,$. From $u_\nu
b^\nu_{\,;\mu} u^\mu = 0\,$ \citep[see][]{Papadopoulos1982} it follows that
\begin{equation}\label{eq_appendix_bmu;mu}
 b^\mu_{\,;\mu} = u^\nu_{\,;\mu} u_\nu b^\mu\,.
\end{equation}
When linearizing Eq.\,(\ref{eq_appendix_faraday})
and using (\ref{eq_appendix_bmu;mu}) Faraday's equation
becomes:
\begin{eqnarray}
 \hat u^\mu \delta b^\nu_{\,;\mu}  &=&-\hat b^\nu_{\,;\mu} \delta u^\nu +
\hat h^{\nu\beta} (\delta u_{\beta;\lambda} \hat b^\lambda + \hat
u_{\beta;\lambda} \delta b^\lambda)  \nonumber\\
&&+ \hat u^\nu \delta u^\beta \hat b^\lambda \hat u_{\beta;\lambda} - \delta
u^\mu_{\,;\mu} \hat b^\nu- \hat u^\mu_{\,;\mu}\delta b^\nu \nonumber\\
&&+\hat u^\nu \hat b^\beta (\delta u_{\beta;\lambda} \hat u^\lambda +
\hat u_{\beta;\lambda} \delta u^\lambda) \nonumber\\
&&+ \hat u_{\beta;\lambda} \hat u^\lambda (\hat b^\beta \delta u^\nu +
\delta b^\beta \hat u^\nu)\,.
\end{eqnarray}
Taking the $\varphi$-component of this equation, we arrive at a relation
between $\delta b^\varphi$ and the spatial derivatives of the displacement
$\xi^\varphi$ which reads
\begin{eqnarray}
\hat u^t  \delta b^\varphi_{\,,t}&=&\frac{\hat \alpha_{,r}}{\hat \alpha}
\hat b^r\delta u^\varphi + \hat b^r\delta u^\varphi_{\,,r}+ \hat b^\theta\delta
u^\varphi_{\,,\theta}\nonumber\\
&=& \hat b^r \hat u^t \left ( \frac{\hat \alpha_{,r}}{\hat \alpha}
\xi^\varphi_{\,,t} +
\xi^\varphi_{\,,t,r}  \right)+ \hat b^\theta \hat u^t
\xi^\varphi_{\,,t,\theta} \nonumber\\
&& +\hat b^r
\hat u^t_{\,,r} \xi^\varphi_{\,,t}\,,
\end{eqnarray}
or
\begin{eqnarray}
\delta b^\varphi_{\,,t}&=&\left(\hat b^r \xi^\varphi_{\,,r}+ \hat b^\theta 
\xi^\varphi_{\,,\theta}\right)_{,t}\,,
\end{eqnarray}
where we have used that $\hat u^t = \hat \alpha^{-1}\,$.
Plugging this relation into Eq.\,(\ref{eq_appendix_perturbation}) we obtain
\begin{eqnarray}
\hat A_0 \xi^\varphi_{\,,tt} &=&  \hat A_1 \xi^\varphi_{\,,r} + \hat A_2
\xi^\varphi_{\,,\theta} +
\left(\hat b^\theta \hat b^\theta + \frac{\hat \mu_\mathrm{S}}{r^2
\hat \Phi^4}\right)
\xi^\varphi_{\,,\theta\theta} 
\nonumber\\
&&+ \left( 2 \hat b^r \hat b^\theta \right) \xi^\varphi_{\,,\theta r}+ \left(
\hat b^r \hat b^r +
\frac{\hat \mu_\mathrm{S}}{\hat \Phi^4}\right)
\xi^\varphi_{\,,rr},\label{eq_appendix_linear}\label{eq:linear}
\end{eqnarray}
with
\begin{eqnarray}
\hat A_0&=&\left( \hat \rho \hat h+\hat b^2 \right)  (\hat u^t)^2\,,\\
\hat A_1 &= & \hat A_3 \hat b^r +\hat b^r \hat b^r_{\,,r} + \hat b^\theta
\hat b^r_{\,,\theta} +
\left(\frac{\hat \mu_\mathrm{S}}{\hat \Phi^4}\right)_{,r} \nonumber\\
&&+\left( \frac {\hat \alpha_{,r} }{\hat \alpha}+ 10\frac{ \hat
\Phi_{,r}}{\hat \Phi} +
\frac{4}{r} \right) \frac{\hat \mu_\mathrm{S}}{\hat
\Phi^4}\,,\\
\hat A_2 &= & \hat A_3 \hat b^\theta+ \hat b^r \hat b^\theta_{\,,r} +
\hat b^\theta \hat b^\theta_{\,,\theta} +
\left(\frac{\hat \mu_\mathrm{S}}{r^2 \hat \Phi^4}\right)_{,\theta} \nonumber\\
&& + 3\cot(\theta) \frac{\hat \mu_\mathrm{S}}{r^2 \hat \Phi^4}\,,
\end{eqnarray}
and
\begin{eqnarray}
\hat A_3 &=&\left( 4\frac {
\hat \Phi_{,r}}{\hat \Phi} + \frac{2}{r} + \frac {\hat \alpha_{,r}}{
\hat \alpha}\right)
\hat b^r+2 \cot(\theta) \hat b^\theta\,.
\end{eqnarray}

The solution of the coefficient determinant of the second-order derivatives of
Eq.\,(\ref{eq_appendix_linear}) 
\begin{eqnarray}
 0&=&a_{ij} |x^i||x^j|\frac{\partial^2 \xi}{\partial x^i \partial x^j} = c_{ij}
\frac{\partial^2 \xi}{\partial x^i \partial x^j}, 
\\
 c_{ij}&=&
 \left( \begin{array}{c c c}
 \hat \alpha^2 \hat A^0&0&0\\
 0&\hat\Phi^4 \hat b^r \hat b^r + \frac{\hat
 \mu_\mathrm{S}}{r^2}&\hat \Phi^4 r \hat b^r \hat b^\theta \\
 0&\hat \Phi^4 r \hat b^\theta \hat b^r & \hat\Phi^4 r^2 \hat b^\theta
\hat b^\theta +\mu_\mathrm{S}
 \end{array}
 \right),
\end{eqnarray}
leads to the following eigenvalues:
\begin{eqnarray}
 \lambda_1 &=& 1\,,\\
 \lambda_2 &=& \frac{\hat \mu_\mathrm{S} + \hat b^2}{\hat \alpha^2 \hat
A_0}\,,\\
 \lambda_3 &=& \frac{\hat \mu_\mathrm{S}}{\hat \alpha^2 \hat A_0}\,,
\end{eqnarray}
where the eigenvector corresponding to $\lambda_2$ ($\lambda_3$) is oriented
along (perpendicular to) the magnetic field lines, i.e. they
correspond to the linearized version of Eq.\,(\ref{eq:eigenvalues}). All
eigenvalues are real, and hence Eq.\,(\ref{eq_appendix_linear}) is hyperbolic.
%
\subsection{Numerical implementation of the linearized wave equation in the
crust}
As when using Riemann solvers on both sides of the crust-core
interface, the equations and the numerical scheme in the fluid core of the
neutron star remain unmodified. In order to evolve Eq.\,(\ref{eq:linear})
numerically in the crust we split it into two equations for $\xi^\varphi$ and
$\xi^\varphi_{\,,t}$, and then perform an explicit Runge-Kutta integration.
Since
we are now evolving two systems with different variables in the core and in the
crust, we have to impose interface conditions.

The evolutions computed with this method rapidly become unstable when increasing
the magnetic field strength. It was therefore necessary to add some artificial
dissipation. We used a fourth-order Kreiss-Oliger term $\epsilon_D
\mathcal{D}_4 f$, where  $\mathcal{D}_4 f$ is the fourth-order numerical
derivative of any function $f$.
The minimal coefficient found to give stable evolutions
is $\epsilon_D=10^{-2}$. We checked the code to ensure that this additional
term does not influence the results of the simulations significantly.
%
\subsection{Comparison of the two methods} 
\begin{figure}
\begin{center}	
 \includegraphics[width=.46\textwidth]{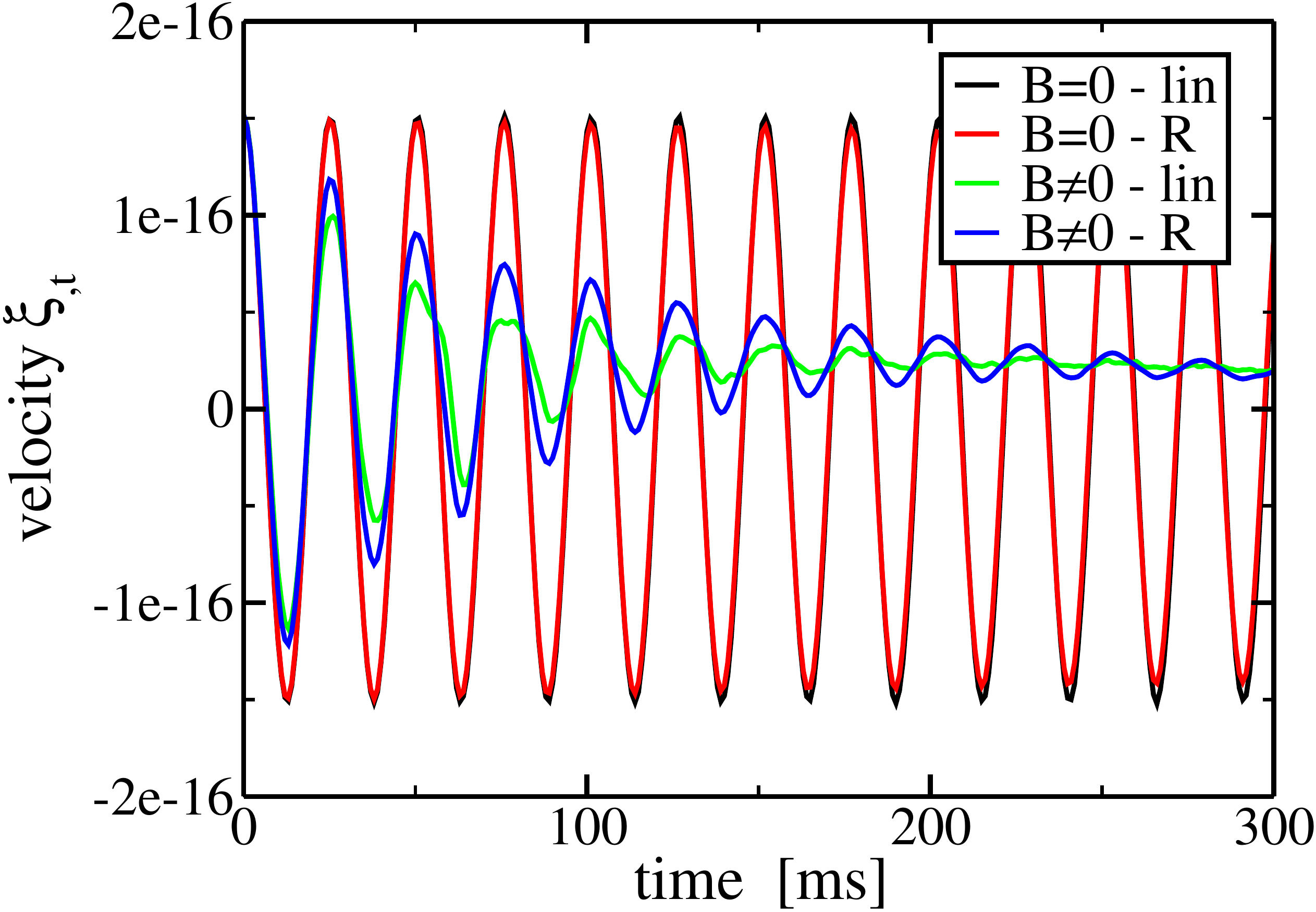}
\end{center}
\caption{Evolution of the velocity $\xi_{,t}$ at some point in the crust
near the pole for $B=5\times10^{13}\,$G and without a magnetic field. The two
different numerical methods are denoted by \emph{lin} for the linear method and
\emph{R} for the Riemann solver approach. The linear method is less dissipative
for zero magnetic field, while the opposite holds when the magnetic field
is turned on.}\label{compare}
\end{figure}

To compare the results of the crustal mode damping (see
Section\,\ref{sec_damping}) obtained with the two numerical methods presented in
Section\,\ref{sec_theory}, \ref{sec_numerics} and in Appendix
\ref{appendix_linear} we plot the evolution of the
velocity at some point in the crust near the pole (Fig.\,\ref{compare}). 
Without magnetic field, the linear method is less dissipative, which
is probably related to the set-up of the interface conditions at the
crust-core interface in this particular case. While we use the general
conditions described in Section\,\ref{sec_numerics} for the Riemann solver
approach, it is possible to use a simplified expression for the linear method. 
Because there is no magnetic field the expression of continuous traction at
the crust-core interface leads to $\xi^\varphi_{\,,r}=0$ as at the surface. 
This provides a source of dissipation for the Riemann solver method, but none
for the linear approach. In the presence of magnetic fields the picture
changes, and the evolution computed with the Riemann solver has less dissipation
of crustal modes than the linear method (see Fig.\,\ref{compare}). When using
different coefficients in the Kreiss-Oliger term, the curves for the linear
method are indistinguishable. We can therefore rule out 
that the artificial dissipation  dominates the numerical damping observed for
the linear approach.

Because of its superior behavior in the more generic case we generally used the
Riemann solver method to obtain the numerical results. We checked that the
linear approach agrees on the extracted frequencies
with the Riemann solver for all regimes of the magnetic field strength
considered by us. 
%

\section{Eigenmodes in the crust}\label{ap_eigenmodes}
Without magnetic field Eq.\,(\ref{eq_appendix_linear}) simplifies to
\begin{eqnarray}
\hat A_0 \xi^\varphi_{\,,tt} &=&  \hat A_1 \xi^\varphi_{\,,r} + \hat A_2
\xi^\varphi_{\,,\theta} +
\frac{\hat \mu_\mathrm{S}}{r^2\hat \Phi^4} \xi^\varphi_{\,,\theta\theta} 
 + \frac{\hat \mu_\mathrm{S}}{\hat \Phi^4}
\xi^\varphi_{\,,rr}\nonumber\\
&=&\nabla_\nu \left(\mu_{\mathrm{S}} g^{\nu\nu}
\xi^\varphi_{\,,\nu}\right)\label{eq_appendix_eigenmodes}\,,
\end{eqnarray}
with
\begin{eqnarray}
\hat A_0&=& \hat \rho \hat h  (\hat u^t)^2\,,\\
\hat A_1 &= & \left(\frac{\hat \mu_\mathrm{S}}{\hat \Phi^4}\right)_{,r}
+\hat C \frac{\hat \mu_\mathrm{S}}{\hat \Phi^4}\,,\\
\hat A_2 &= & \left(\frac{\hat \mu_\mathrm{S}}{r^2 \hat
\Phi^4}\right)_{,\theta} + 3\cot(\theta) \frac{\hat \mu_\mathrm{S}}{r^2 \hat
\Phi^4}\,,
\end{eqnarray}
and
\begin{eqnarray}
\hat C &=& \frac {\hat \alpha_{,r} }{\hat \alpha}+ 10\frac{ \hat
\Phi_{,r}}{\hat \Phi} + \frac{4}{r}\,.
\end{eqnarray}
Assuming a harmonic time dependence $\xi^\varphi(\mathbf{r}, t) =
\xi^\varphi(\mathbf{r}) e^{i\omega t}$ and a separation of variables
$\xi^\varphi(r, \theta) = R(r)\Theta(\theta)$ leads to
\begin{eqnarray}
 -\omega^2 \hat A_0 &=&\left[ \left(\frac{\hat \mu_\mathrm{S}}{r^2 \hat
\Phi^4}\Theta^\prime(\theta) \right)_{,\theta} + 3 \cot(\theta) \frac{\hat
\mu_\mathrm{S}}{r^2 \hat \Phi^4} \Theta^\prime(\theta)\right]
\frac{1}{\Theta(\theta)}\nonumber \\ 
&&+\left[ \left(\frac{\hat \mu_\mathrm{S}}{\hat
\Phi^4}R^\prime(r)\right)_{,r} + \hat C \frac{\hat \mu_\mathrm{S}}{\hat
\Phi^4} R^\prime(r)\right] \frac{1}{R(r)} \,,
\end{eqnarray}
where a prime denotes the derivative with respect to the corresponding variable
$r$ or $\theta$. The angular and radial part have to fulfill
this equation independently, such that:
\begin{eqnarray}
 0&=&\Theta^{\prime\prime}(\theta) + 3 \cot{\theta} \Theta^\prime(\theta) +
\lambda_\theta^2 \Theta (\theta) \,,\\
0&=&  \frac{\hat \mu_\mathrm{S}}{\hat
\Phi^4}R^{\prime\prime}(r) + \left[\left(\frac{\hat \mu_\mathrm{S}}{\hat
\Phi^4}\right)_{,r} + \hat C \frac{\hat \mu_\mathrm{S}}{\hat
\Phi^4}\right] R^\prime(r) + \lambda_r^2  R(r)\,, \label{eq_eigenmode_R}
\end{eqnarray}
where $ \lambda_r^2 = \left(\omega^2 \hat A_0 + \lambda_\theta^2\right)\,$.
Both equations are of singular Sturm-Liouville type, and therefore their
solutions $R_{\lambda_r}$ and $\Theta_{\lambda_\theta}$ form a complete,
orthonormal set $\Xi_i (r, \theta) = R_{\lambda_r} \Theta_{\lambda_\theta}$.
The solution of the angular part consists of the angular part of the vector
spherical harmonics $\Psi_n(\theta)$, which is related to the Legendre
polynomials $P_n(\theta)$ by 
\begin{equation}
 \Psi_n (\theta) \sim \frac{\partial P_n (\theta)}{\partial \theta}\,.
\end{equation}
The equation to obtain $R(r)$ is solved numerically with a shooting method.
Because the eigenfunctions $\Xi_i (r, \theta)$ form a complete set, it is
possible to expand any displacement in terms of the former:
\begin{equation}
 \xi^\varphi (r,\theta,t) = \sum_i A_i(t) \,\, \Xi_i (r, \theta)\,,
\end{equation}
where the eigenmode coefficients $A_i(t)$ are given by the inner product
\begin{eqnarray}
 A_i (t) &=& \langle \xi^\varphi (r,\theta,t), \Xi_i (r,\theta) \rangle
\nonumber\\
&=& \int_{r_\mathrm{cc}}^{r_\mathrm{s}} \xi^\varphi (r,\theta,t)
\,\, \Xi_i (r,\theta) w_\theta w_r dr d\theta\,.\label{overlap_integral}
\end{eqnarray}
The corresponding weighting functions $w_r$ and $w_\theta$ are, according to the
Sturm-Liouville theory
\begin{eqnarray}
 w_\theta &=& \sin(\theta)^3\,,\\
w_r &=& \frac{\alpha(r)^{-1} \Phi(r)^{10}\,\, r^4 \rho(r) h(r)
}{\alpha(r_\mathrm{cc})^{-1} \Phi(r_\mathrm{cc})^{10}\,\, r_\mathrm{cc}^4
\rho(r_\mathrm{cc}) h(r_\mathrm{cc}) } \,.
\end{eqnarray}
We calculate the overlap integrals defined in Eq.\,(\ref{overlap_integral})
with a fourth-order Simpsons rule algorithm.

\bibliographystyle{mn2e}
\bibliography{magnetar.bib}

\begin{thebibliography}{}

\bibitem[\protect\citeauthoryear{{Akmal}, {Pandharipande} \&
  {Ravenhall}}{{Akmal} et~al.}{1998}]{Akmal1998}
{Akmal} A.,  {Pandharipande} V.~R.,    {Ravenhall} D.~G.,  1998, \prc, 58, 1804

\bibitem[\protect\citeauthoryear{{Blaes}, {Blandford}, {Goldreich} \&
  {Madau}}{{Blaes} et~al.}{1989}]{Blaes1989}
{Blaes} O.,  {Blandford} R.,  {Goldreich} P.,    {Madau} P.,  1989, \apj, 343,
  839

\bibitem[\protect\citeauthoryear{{Bocquet}, {Bonazzola}, {Gourgoulhon} \&
  {Novak}}{{Bocquet} et~al.}{1995}]{Bocquet1995}
{Bocquet} M.,  {Bonazzola} S.,  {Gourgoulhon} E.,    {Novak} J.,  1995, \aap,
  301, 757

\bibitem[\protect\citeauthoryear{{Bonazzola}, {Villain} \&
  {Bejger}}{{Bonazzola} et~al.}{2007}]{Bonazzola2007}
{Bonazzola} S.,  {Villain} L.,    {Bejger} M.,  2007, Classical and Quantum
  Gravity, 24, 221

\bibitem[\protect\citeauthoryear{{Braithwaite} \& {Nordlund}}{{Braithwaite} \&
  {Nordlund}}{2006}]{Braithwaite2006b}
{Braithwaite} J.,  {Nordlund} {\AA}.,  2006, \aap, 450, 1077

\bibitem[\protect\citeauthoryear{{Carter} \& {Quintana}}{{Carter} \&
  {Quintana}}{1972}]{Carter1972}
{Carter} B.,  {Quintana} H.,  1972, Royal Society of London Proceedings Series
  A, 331, 57

\bibitem[\protect\citeauthoryear{{Carter} \& {Samuelsson}}{{Carter} \&
  {Samuelsson}}{2006}]{Carter2006}
{Carter} B.,  {Samuelsson} L.,  2006, Classical and Quantum Gravity, 23, 5367

\bibitem[\protect\citeauthoryear{{Cerd{\'a}-Dur{\'a}n}}{{Cerd{\'a}-Dur{\'a}n}}{2010}]{Cerda2010}
{Cerd{\'a}-Dur{\'a}n} P.,  2010, Classical and Quantum Gravity, 27, 205012

\bibitem[\protect\citeauthoryear{{Cerd{\'a}-Dur{\'a}n}, {Font}, {Ant{\'o}n} \&
  {M{\"u}ller}}{{Cerd{\'a}-Dur{\'a}n} et~al.}{2008}]{Cerda2008}
{Cerd{\'a}-Dur{\'a}n} P.,  {Font} J.~A.,  {Ant{\'o}n} L.,    {M{\"u}ller} E.,
  2008, \aap, 492, 937

\bibitem[\protect\citeauthoryear{{Cerd{\'a}-Dur{\'a}n}, {Stergioulas} \&
  {Font}}{{Cerd{\'a}-Dur{\'a}n} et~al.}{2009}]{Cerda2009}
{Cerd{\'a}-Dur{\'a}n} P.,  {Stergioulas} N.,    {Font} J.~A.,  2009, \mnras,
  397, 1607

\bibitem[\protect\citeauthoryear{{Ciolfi}, {Lander}, {Manca} \&
  {Rezzolla}}{{Ciolfi} et~al.}{2011}]{Ciolfi2011}
{Ciolfi} R.,  {Lander} S.~K.,  {Manca} G.~M.,    {Rezzolla} L.,  2011, \apjl,
  736, L6+

\bibitem[\protect\citeauthoryear{{Colaiuda}, {Beyer} \& {Kokkotas}}{{Colaiuda}
  et~al.}{2009}]{Colaiuda2009}
{Colaiuda} A.,  {Beyer} H.,    {Kokkotas} K.~D.,  2009, \mnras, 396, 1441

\bibitem[\protect\citeauthoryear{{Colaiuda} \& {Kokkotas}}{{Colaiuda} \&
  {Kokkotas}}{2011}]{Colaiuda2011}
{Colaiuda} A.,  {Kokkotas} K.~D.,  2011, \mnras, 414, 3014

\bibitem[\protect\citeauthoryear{{Demorest}, {Pennucci}, {Ransom}, {Roberts} \&
  {Hessels}}{{Demorest} et~al.}{2010}]{Demorest2010}
{Demorest} P.~B.,  {Pennucci} T.,  {Ransom} S.~M.,  {Roberts} M.~S.~E.,
  {Hessels} J.~W.~T.,  2010, \nat, 467, 1081

\bibitem[\protect\citeauthoryear{{Dimmelmeier}, {Font} \&
  {M{\"u}ller}}{{Dimmelmeier} et~al.}{2002a}]{Dimmelmeier2002a}
{Dimmelmeier} H.,  {Font} J.~A.,    {M{\"u}ller} E.,  2002a, \aap, 388, 917

\bibitem[\protect\citeauthoryear{{Dimmelmeier}, {Font} \&
  {M{\"u}ller}}{{Dimmelmeier} et~al.}{2002b}]{Dimmelmeier2002b}
{Dimmelmeier} H.,  {Font} J.~A.,    {M{\"u}ller} E.,  2002b, \aap, 393, 523

\bibitem[\protect\citeauthoryear{{Dimmelmeier}, {Novak}, {Font},
  {Ib{\'a}{\~n}ez} \& {M{\"u}ller}}{{Dimmelmeier}
  et~al.}{2005}]{Dimmelmeier2005}
{Dimmelmeier} H.,  {Novak} J.,  {Font} J.~A.,  {Ib{\'a}{\~n}ez} J.~M.,
  {M{\"u}ller} E.,  2005, \prd, 71, 064023

\bibitem[\protect\citeauthoryear{{Douchin} \& {Haensel}}{{Douchin} \&
  {Haensel}}{2001}]{Douchin2001}
{Douchin} F.,  {Haensel} P.,  2001, \aap, 380, 151

\bibitem[\protect\citeauthoryear{{Duncan}}{{Duncan}}{1998}]{Duncan1998}
{Duncan} R.~C.,  1998, \apjl, 498, L45

\bibitem[\protect\citeauthoryear{{Duncan} \& {Thompson}}{{Duncan} \&
  {Thompson}}{1992}]{Duncan1992}
{Duncan} R.~C.,  {Thompson} C.,  1992, \apjl, 392, L9

\bibitem[\protect\citeauthoryear{{El-Mezeini} \& {Ibrahim}}{{El-Mezeini} \&
  {Ibrahim}}{2010}]{El-Mezeini2010}
{El-Mezeini} A.~M.,  {Ibrahim} A.~I.,  2010, \apjl, 721, L121

\bibitem[\protect\citeauthoryear{{Gabler}, {Cerd\'a-Dur\'an}, {Font}, M\"uller
  \& {Stergioulas}}{{Gabler} et~al.}{2010}]{Gabler2010Proceedings}
{Gabler} M.,  {Cerd\'a-Dur\'an} P.,  {Font} J.,  M\"uller E.,    {Stergioulas}
  N.,  2010, Journal of Physics Conference Series, 283, 012013

\bibitem[\protect\citeauthoryear{{Gabler}, {Cerd{\'a} Dur{\'a}n}, {Font},
  {M{\"u}ller} \& {Stergioulas}}{{Gabler} et~al.}{2011}]{Gabler2011letter}
{Gabler} M.,  {Cerd{\'a} Dur{\'a}n} P.,  {Font} J.~A.,  {M{\"u}ller} E.,
  {Stergioulas} N.,  2011, \mnras, 410, L37

\bibitem[\protect\citeauthoryear{{Gabler}, {Sperhake} \& {Andersson}}{{Gabler}
  et~al.}{2009}]{Gabler2009}
{Gabler} M.,  {Sperhake} U.,    {Andersson} N.,  2009, \prd, 80, 064012

\bibitem[\protect\citeauthoryear{{Glampedakis} \& {Andersson}}{{Glampedakis} \&
  {Andersson}}{2006}]{Glampedakis2006}
{Glampedakis} K.,  {Andersson} N.,  2006, \mnras, 371, 1311

\bibitem[\protect\citeauthoryear{{Hambaryan}, {Neuh{\"a}user} \&
  {Kokkotas}}{{Hambaryan} et~al.}{2011}]{Hambaryan2011}
{Hambaryan} V.,  {Neuh{\"a}user} R.,    {Kokkotas} K.~D.,  2011, \aap, 528,
  A45+

\bibitem[\protect\citeauthoryear{{Isenberg}}{{Isenberg}}{2008}]{Isenberg2008}
{Isenberg} J.~A.,  2008, International Journal of Modern Physics D, 17, 265

\bibitem[\protect\citeauthoryear{{Israel}, {Belloni}, {Stella}, {Rephaeli},
  {Gruber}, {Casella}, {Dall'Osso}, {Rea}, {Persic} \& {Rothschild}}{{Israel}
  et~al.}{2005}]{Israel2005}
{Israel} G.~L.,  {Belloni} T.,  {Stella} L.,  {Rephaeli} Y.,  {Gruber} D.~E.,
  {Casella} P.,  {Dall'Osso} S.,  {Rea} N.,  {Persic} M.,    {Rothschild}
  R.~E.,  2005, \apjl, 628, L53

\bibitem[\protect\citeauthoryear{{Karlovini} \& {Samuelsson}}{{Karlovini} \&
  {Samuelsson}}{2003}]{Karlovini2003}
{Karlovini} M.,  {Samuelsson} L.,  2003, Classical and Quantum Gravity, 20,
  3613

\bibitem[\protect\citeauthoryear{{Karlovini} \& {Samuelsson}}{{Karlovini} \&
  {Samuelsson}}{2004}]{Karlovini2004b}
{Karlovini} M.,  {Samuelsson} L.,  2004, Classical and Quantum Gravity, 21,
  4531

\bibitem[\protect\citeauthoryear{{Karlovini} \& {Samuelsson}}{{Karlovini} \&
  {Samuelsson}}{2007}]{Karlovini2007}
{Karlovini} M.,  {Samuelsson} L.,  2007, Classical and Quantum Gravity, 24,
  3171

\bibitem[\protect\citeauthoryear{{Karlovini}, {Samuelsson} \&
  {Zarroug}}{{Karlovini} et~al.}{2004}]{Karlovini2004}
{Karlovini} M.,  {Samuelsson} L.,    {Zarroug} M.,  2004, Classical and Quantum
  Gravity, 21, 1559

\bibitem[\protect\citeauthoryear{{Kiuchi}, {Yoshida} \& {Shibata}}{{Kiuchi}
  et~al.}{2011}]{Kiuchi2011}
{Kiuchi} K.,  {Yoshida} S.,    {Shibata} M.,  2011, \aap, 532, A30+

\bibitem[\protect\citeauthoryear{{Lander} \& {Jones}}{{Lander} \&
  {Jones}}{2011}]{Lander2011b}
{Lander} S.~K.,  {Jones} D.~I.,  2011, \mnras, 412, 1730

\bibitem[\protect\citeauthoryear{{Lander}, {Jones} \& {Passamonti}}{{Lander}
  et~al.}{2010}]{Lander2010}
{Lander} S.~K.,  {Jones} D.~I.,    {Passamonti} A.,  2010, \mnras, 405, 318

\bibitem[\protect\citeauthoryear{{Lasky}, {Zink}, {Kokkotas} \&
  {Glampedakis}}{{Lasky} et~al.}{2011}]{Lasky2011}
{Lasky} P.~D.,  {Zink} B.,  {Kokkotas} K.~D.,    {Glampedakis} K.,  2011,
  \apjl, 735, L20+

\bibitem[\protect\citeauthoryear{{Lee}}{{Lee}}{2007}]{Lee2007}
{Lee} U.,  2007, \mnras, 374, 1015

\bibitem[\protect\citeauthoryear{{Lee}}{{Lee}}{2008}]{Lee2008}
{Lee} U.,  2008, \mnras, 385, 2069

\bibitem[\protect\citeauthoryear{{Levin}}{{Levin}}{2006}]{Levin2006}
{Levin} Y.,  2006, \mnras, 368, L35

\bibitem[\protect\citeauthoryear{{Levin}}{{Levin}}{2007}]{Levin2007}
{Levin} Y.,  2007, \mnras, 377, 159

\bibitem[\protect\citeauthoryear{{Levin} \& {van Hoven}}{{Levin} \& {van
  Hoven}}{2011}]{Levin2011}
{Levin} Y.,  {van Hoven} M.,  2011, \mnras, 418, 659

\bibitem[\protect\citeauthoryear{{Messios}, {Papadopoulos} \&
  {Stergioulas}}{{Messios} et~al.}{2001}]{Messios2001}
{Messios} N.,  {Papadopoulos} D.~B.,    {Stergioulas} N.,  2001, \mnras, 328,
  1161

\bibitem[\protect\citeauthoryear{{Negele} \& {Vautherin}}{{Negele} \&
  {Vautherin}}{1973}]{Negele1973}
{Negele} J.~W.,  {Vautherin} D.,  1973, Nuclear Physics A, 207, 298

\bibitem[\protect\citeauthoryear{{Pandharipande} \& {Smith}}{{Pandharipande} \&
  {Smith}}{1975}]{Pandharipande1975}
{Pandharipande} V.~R.,  {Smith} R.~A.,  1975, Physics Letters B, 59, 15

\bibitem[\protect\citeauthoryear{{Papadopoulos} \& {Esposito}}{{Papadopoulos}
  \& {Esposito}}{1982}]{Papadopoulos1982}
{Papadopoulos} D.,  {Esposito} F.~P.,  1982, \apj, 257, 10

\bibitem[\protect\citeauthoryear{{Passamonti} \& {Andersson}}{{Passamonti} \&
  {Andersson}}{2012}]{Passamonti2011}
{Passamonti} A.,  {Andersson} N.,  2012, \mnras, 419, 638

\bibitem[\protect\citeauthoryear{{Piro}}{{Piro}}{2005}]{Piro2005}
{Piro} A.~L.,  2005, \apjl, 634, L153

\bibitem[\protect\citeauthoryear{{Rea}, {Esposito}, {Turolla}, {Israel},
  {Zane}, {Stella}, {Mereghetti}, {Tiengo}, {G{\"o}tz}, {G{\"o}{\u g}{\"u}{\c
  s}} \& {Kouveliotou}}{{Rea} et~al.}{2010}]{Rea2010}
{Rea} N.,  {Esposito} P.,  {Turolla} R.,  {Israel} G.~L.,  {Zane} S.,  {Stella}
  L.,  {Mereghetti} S.,  {Tiengo} A.,  {G{\"o}tz} D.,  {G{\"o}{\u g}{\"u}{\c
  s}} E.,    {Kouveliotou} C.,  2010, Science, 330, 944

\bibitem[\protect\citeauthoryear{{Samuelsson} \& {Andersson}}{{Samuelsson} \&
  {Andersson}}{2007}]{Samuelsson2007}
{Samuelsson} L.,  {Andersson} N.,  2007, \mnras, 374, 256

\bibitem[\protect\citeauthoryear{{Schumaker} \& {Thorne}}{{Schumaker} \&
  {Thorne}}{1983}]{Schumaker1983}
{Schumaker} B.~L.,  {Thorne} K.~S.,  1983, \mnras, 203, 457

\bibitem[\protect\citeauthoryear{{Shaisultanov} \& {Eichler}}{{Shaisultanov} \&
  {Eichler}}{2009}]{Shaisultanov2009}
{Shaisultanov} R.,  {Eichler} D.,  2009, \apjl, 702, L23

\bibitem[\protect\citeauthoryear{{Sotani}, {Colaiuda} \& {Kokkotas}}{{Sotani}
  et~al.}{008a}]{Sotani2008b}
{Sotani} H.,  {Colaiuda} A.,    {Kokkotas} K.~D.,  2008a, \mnras, 385, 2161

\bibitem[\protect\citeauthoryear{{Sotani}, {Kokkotas} \&
  {Stergioulas}}{{Sotani} et~al.}{2007}]{Sotani2007}
{Sotani} H.,  {Kokkotas} K.~D.,    {Stergioulas} N.,  2007, \mnras, 375, 261

\bibitem[\protect\citeauthoryear{{Sotani}, {Kokkotas} \&
  {Stergioulas}}{{Sotani} et~al.}{008b}]{Sotani2008}
{Sotani} H.,  {Kokkotas} K.~D.,    {Stergioulas} N.,  2008b, \mnras, 385, L5

\bibitem[\protect\citeauthoryear{{Steiner} \& {Watts}}{{Steiner} \&
  {Watts}}{2009}]{Steiner2009}
{Steiner} A.~W.,  {Watts} A.~L.,  2009, Physical Review Letters, 103, 181101

\bibitem[\protect\citeauthoryear{{Strohmayer}, {van Horn}, {Ogata}, {Iyetomi}
  \& {Ichimaru}}{{Strohmayer} et~al.}{1991}]{Strohmayer1991}
{Strohmayer} T.,  {van Horn} H.~M.,  {Ogata} S.,  {Iyetomi} H.,    {Ichimaru}
  S.,  1991, \apj, 375, 679

\bibitem[\protect\citeauthoryear{{Strohmayer} \& {Watts}}{{Strohmayer} \&
  {Watts}}{2005}]{Strohmayer2005}
{Strohmayer} T.~E.,  {Watts} A.~L.,  2005, \apjl, 632, L111

\bibitem[\protect\citeauthoryear{{Thompson} \& {Duncan}}{{Thompson} \&
  {Duncan}}{2001}]{Thompson2001}
{Thompson} C.,  {Duncan} R.~C.,  2001, \apj, 561, 980

\bibitem[\protect\citeauthoryear{{van Hoven} \& {Levin}}{{van Hoven} \&
  {Levin}}{2011}]{vanHoven2011}
{van Hoven} M.,  {Levin} Y.,  2011, \mnras, 410, 1036

\bibitem[\protect\citeauthoryear{{van Hoven} \& {Levin}}{{van Hoven} \&
  {Levin}}{2012}]{vanHoven2011b}
{van Hoven} M.,  {Levin} Y.,  2012, \mnras, p.~2325

\bibitem[\protect\citeauthoryear{{Watts} \& {Strohmayer}}{{Watts} \&
  {Strohmayer}}{2007}]{Watts2007}
{Watts} A.~L.,  {Strohmayer} T.~E.,  2007, Advances in Space Research, 40, 1446

\bibitem[\protect\citeauthoryear{{Wilson}, {Mathews} \& {Marronetti}}{{Wilson}
  et~al.}{1996}]{Wilson1996}
{Wilson} J.~R.,  {Mathews} G.~J.,    {Marronetti} P.,  1996, \prd, 54, 1317

\end{thebibliography}

\end{document}